\documentclass{aa}  
\usepackage{graphicx}
\usepackage{txfonts}
\usepackage{tikz}
\usepackage{amssymb,amsmath}
\usepackage{dsfont}
\usepackage{lscape}
\usepackage{longtable}
\usepackage{float}
\usepackage{natbib}
\usepackage{color}
\begin{document} 

%
%
\title{Inner and outer rings are not strongly coupled with stellar bars
}
  \author{S. D\'iaz-Garc\'ia\inst{1,2}
          \and        
          S. D\'iaz-Su\'arez\inst{1,2}
          \and
          J. H. Knapen\inst{1,2,3}
          \and
          H. Salo\inst{4}       
          }
  \institute{Instituto de Astrof\'isica de Canarias, E-38205, La Laguna, Tenerife, Spain \\
              \email{simondiazgar@gmail.com}
         \and
             Departamento de Astrof\'isica, Universidad de La Laguna, E-38205, La Laguna, Tenerife, Spain
         \and
             Astrophysics Research Institute, Liverpool John Moores University, IC2, Liverpool Science Park, 146 Brownlow Hill, Liverpool, L3 5RF, UK             
         \and 
             Astronomy Research Unit, University of Oulu, FI-90014 Finland
             }
  \date{Received 13 March 2019; accepted 8 April 2019}
  \abstract
  {
Rings are distinctive features of many disc galaxies and their location and properties are closely related to the disc dynamics. 
In particular, rings are often associated to stellar bars, but the details of this connection are far from clear. 
We have studied the frequency and dimensions of inner and outer rings in the local Universe as a function of
 disc parameters and the amplitude of non-axisymmetries. 
We used the 1320 not highly inclined disc galaxies ($i<65^{\circ}$) from the S$^4$G survey. 
The ring fraction increases with bar Fourier density amplitude: this can be interpreted as evidence for the role of bars in ring formation. 
The sizes of inner rings are positively correlated with bar strength: 
this can be linked to the radial displacement of the 1/4 ultraharmonic resonance while the bar grows and the pattern speed decreases. 
The ring's intrinsic ellipticity is weakly controlled by the non-axisymmetric perturbation strength: 
this relation is not as strong as expected from simulations, especially when we include the dark matter halo in the force calculation. 
The ratio of outer-to-inner ring semi-major axes is uncorrelated with bar strength: this questions the manifold origin of rings. 
In addition, we confirm that 
i) $\sim 1/3$ ($\sim 1/4$) of the galaxies hosting inner (outer) rings are not barred; 
ii) on average, the sizes and shapes of rings are roughly the same for barred and non-barred galaxies; and 
iii) the fraction of inner (outer) rings is a factor of $1.2-1.4$ ($1.65-1.9$) 
larger in barred galaxies than in their non-barred counterparts. 
Finally, we apply unsupervised machine learning (self-organising maps, SOMs) to show that, 
among early-type galaxies, ringed or barred galaxies cannot be univocally distinguished based on 20 internal and external fundamental parameters. 
We confirm, with the aid of SOMs, that rings are mainly hosted by red, 
massive, gas-deficient, dark-matter poor, and centrally concentrated galaxies. 
We conclude that the present-day coupling between rings and bars is not as robust as predicted by numerical models, 
and diverse physical mechanisms and timescales determine ring formation and evolution. 
}
\keywords{galaxies: structure - galaxies: evolution - galaxies: statistics - galaxies: spiral - galaxies: fundamental parameters - galaxies: photometry}
\maketitle
%
%
\section{Introduction}
%
%
Rings are elliptical or circular closed stellar and gaseous structures in disc galaxies. 
They are intimately connected with the evolution of spiral structure, 
which may explain why many of them are pseudo-rings made of tightly wrapped spiral arms. 
Three main types of rings (nuclear, inner, and outer rings) are found in galaxies, 
named based on their position in the disc and relative to the bar \citep[e.g.][and references therein]{2015ApJS..217...32B}. 

A lens is a feature with a shallow brightness gradient interior to a sharp edge \citep[e.g.][]{1979ApJ...227..714K}. 
There are lens analogues of each type of ring \citep[][]{2011MNRAS.418.1452L}. 
A ringlens is the same kind of feature but with the appearance of a subtle enhancement at the rim, resembling a low contrast ring.

The frequency of inner and outer rings in disc galaxies reported by \citet[][]{1996FCPh...17...95B} is $\sim 40-47 \%$ and $\sim 10\%$, 
respectively, based on classifications of galaxies from the RC3 catalogue \citep[][]{1991rc3..book.....D} 
with recessional velocities lower than $3000$ km s$^{-1}$. 
\citet[][]{2014A&A...562A.121C} found a global fraction of $25\pm 1\%$ and $16\pm 1\%$ for inner and outer features (rings+ringlenses), 
respectively, based on the morphological classifications made by \citet[][]{2015ApJS..217...32B} to the 
\emph{Spitzer} Survey of Stellar Structure in Galaxies \citep[S$^4$G;][]{2010PASP..122.1397S}, 
comprised of galaxies within 40 Mpc. The distribution of inner and outer rings spans a wide range of morphological types and 
peaks for Sa galaxies, whereas ringlenses peak for earlier types (S0$^0$) 
\citep[e.g.][]{2011MNRAS.418.1452L,2014A&A...562A.121C}. 

Approximately two thirds of galaxies in the local Universe host bars 
\citep[e.g.][]{1991rc3..book.....D,2000ApJ...529...93K,2002MNRAS.336.1281W,2004ApJ...607..103L,2016A&A...587A.160D}, 
and the bar fraction in the nearby Universe has a minimum for Sb/c systems 
\citep[40-45 $\%$;][]{2015ApJS..217...32B,2016A&A...587A.160D}. 
The standard view is that most rings in disc galaxies form from interstellar gas collected near resonances, under the continuous action 
of gravity torques from a bar pattern. The bar resonances occur where the radial (epicyclic) frequency $\kappa = m (\Omega - \Omega_{\rm p})$, 
where $\Omega$ is the circular angular velocity, $\Omega_{\rm p}$ is the 
pattern speed of the bar, and $m$ is an integer. The main ring-forming resonances have a low order: $m$ = +2 and $-$2 (inner and outer Lindblad 
resonances, or ILR and OLR, respectively) and $m$=+4 and $-$4 (inner and outer 4:1 ultraharmonic resonances, or I4R and O4R, respectively). 
In numerical simulations, nuclear rings have been linked to double ILRs, 
inner (pseudo-)rings to the I4R, and outer (pseudo-)rings to the OLR and the O4R. 
For further details, the reader is referred, for example, to 
\citet[][]{1972Ap&SS..19..285M}, 
\citet[][]{1972MNRAS.157....1L}, 
\citet[][]{1975dgs..conf..419D}, 
\citet[][]{1976ApL....17..191S}, 
\citet[][]{1981ApJ...247...77S}, 
\citet[][]{1982A&A...107..101A}, 
\citet[][]{1993RPPh...56..173S}, 
\citet[][]{2000A&A...362..465R}, 
and 
\citet[][]{2017MNRAS.470.3819B}.

Because of their resonant origin, many inner rings in barred galaxies appear aligned with bars 
\citep[e.g.][]{1964rcbg.book.....D,1976ApL....17..191S}. 
However, \citet[][]{2014A&A...562A.121C} showed that nearly 50~$\%$ of inner rings 
have random orientations with respect to the bar (mainly in late-type galaxies), 
suggesting that other physical mechanisms might be responsible for their creation 
(e.g. they might be caused by spiral modes that are decoupled from the bar). 

Rings are expected to be more elongated in strongly barred galaxies than in their non-barred and weakly barred counterparts 
\citep[e.g.][]{1984MNRAS.209...93S}. Possible evidence for this was provided by \citet[][]{2010AJ....139.2465G} 
based on estimates of the tangential-to-radial force ratio \citep[e.g.][]{1981A&A....96..164C,2002MNRAS.337.1118L}, 
but their sample was fairly small (44 galaxies). 
\citet[][]{2014A&A...562A.121C} also reported that inner and outer rings tend to become more elliptical 
when the galaxy family changes from SA (non-barred) to SB (strongly barred).

The manifold theory 
\citep[see e.g.][]{2006A&A...453...39R,2006MNRAS.369L..56P,2006MNRAS.373..280V,2007A&A...472...63R,2008MNRAS.387.1264T,2009MNRAS.394...67A} 
also predicts a dependence of the de-projected ellipticity of inner and outer rings (and of the ratio between their semi-major axes) 
on the bar perturbation strength \citep[][]{2009MNRAS.400.1706A}. This theory proposes that 
stars get confined in families of orbits organised in tubes (invariant manifolds) 
that arise from the unstable $L_1$ and $L_2$ Lagrangian points close to the bar end. 

The resonance and manifold views of galactic rings are reasonable interpretations and each can be tested observationally to some extent. 
However, the existence of rings in non-barred galaxies \citep[e.g.][]{1995ApJS...96...39B} poses a challenge to these ideas, 
and this motivates us to try to identify what fundamental galaxy parameters 
lie at the heart of the formation of bars and rings.

The main goals of this paper are, firstly, to examine how inner and outer ring properties\footnote{
 We do not analyse nuclear rings because many of them are missing in the S$^4$G survey 
\citep[$\sim 1/2$ of those identified by][]{2010MNRAS.402.2462C} due to its relatively low angular resolution.} -- such as 
linear dimensions, de-projected minor to major axis ratios, 
and relative frequency -- correlate with galactic parameters such as mass, measures of bar strength, and galaxy morphological $T$-type 
in the S$^4$G survey; 
and, secondly, to use unsupervised machine learning techniques \citep[self-organising map, SOM;][]{2001som..book.....K} 
to go beyond these kinds of parameters to multiple parameters having more global significance, 
trying different visualisation techniques to check if ringed and barred galaxies can be distinguished using clustering algorithms.

The paper is organised as follows: in Sect.~\ref{sample1} we present the sample and main data used in this work, 
we briefly describe the measurements of the dimensions of ring and bars \citep[from][]{2015A&A...582A..86H}, 
and we provide a summary of the way in which bar strengths and gravitational forces were calculated \citep[from][]{2016A&A...587A.160D}. 
In Sect.~\ref{intrinsic_rings_mass} (and Appendix~\ref{app_frac}) 
we analyse the fraction of rings as a function of stellar mass, considering barred and non-barred galaxies. 
In Sect.~\ref{intrinsic_rings_ttype} we study the sizes of inner and outer rings in the Hubble sequence 
(and versus galaxy mass in Appendix~\ref{app_inner}). 
Section~\ref{rings_bars} probes the dependence of ring sizes and axis ratio on the amplitude of non-axisymmetries. 
In Sect.~\ref{rings_ratio} we test the predictions from the manifold theory on 
the effect of bar strength controlling the ratio of outer-to-inner ring semi-major axes. 
In Sect.~\ref{ml_analysis} (and Appendices~\ref{app_ml}-\ref{ttype_ml}) we analyse the outcome of training a SOM with internal 
(e.g. total stellar mass, gas fraction, star formation rate, bulge prominence, colour, or dark matter content) 
and external (e.g. torques and density of nearby galaxies) global parameters of S$^4$G galaxies. 
In Sect.~\ref{Discussion} we discuss the formation of inner and outer rings and the coupling between bars and rings. 
Finally, in Sect.~\ref{summarysection} we summarise the main results of this paper.
%
%
%
%
\section{Sample and data}\label{sample1}
%
%
\subsection{Spitzer Survey of Stellar Structure in Galaxies (S$^{4}$G)}\label{S4Gsurvey}
%
%
%
%
The S$^{4}$G \citep[][]{2010PASP..122.1397S} is a magnitude- and diameter-limited survey that 
comprises 2352 galaxies with distances $\lesssim 40$ Mpc. 
These galaxies were observed in the 3.6~$\mu$m and 4.5~$\mu$m bands with the Infrared Array Camera \citep[IRAC;][]{2004ApJS..154...10F}
installed on-board the \emph{Spitzer Space Telescope} \citep{2004ApJS..154....1W}. 

A wide range of masses ($\sim 5$ orders of magnitude) and all Hubble types ($T$) are included in the S$^{4}$G sample. 
However, there is a bias towards late-type gas-rich systems because of the distance restriction based on 21 cm 
H{\sc\,i} recessional velocities. To correct this bias, observations were extended to gas-poor early-type galaxies (ETG) with 
$T\le 0$ \citep{2013AAS...22123001S}, using optical velocities for the volume cut. Nevertheless, the survey remains incomplete: 
the S$^4$G and the early-type galaxies extension missed more than 400 late-type galaxies without 21 cm 
systemic velocity measurements listed in HyperLEDA that fulfil the original selection criteria. 
We are currently obtaining $i$-band photometry with ground-based telescopes 
for the $\sim 1/2$ of those galaxies that lack near-IR and optical archival imaging of resolution and depth 
comparable to IRAC or the Sloan Digital Sky Survey \citep[][]{2006AJ....131.2332G}. 
The current study is based on the original S$^4$G.

Total stellar masses of S$^4$G galaxies were calculated by \citet[][]{2015ApJS..219....3M} and will be extensively used in this work. 
The isophotal radii at the surface brightness 25.5~mag~arcsec$^{-2}$ ($R_{25.5}$) 
obtained from the $3.6\,\mu$m images is also taken from \citet[][]{2015ApJS..219....3M} and used as a proxy of the intrinsic disc size.
%
%
\subsection{Morphological classification}\label{morphology}
%
%
By visual inspection of the 3.6~$\mu$m images, 
\citet{2015ApJS..217...32B} carried out the morphological classification of the galaxies in the S$^4$G, 
following a revised version of the de Vaucouleurs Hubble-Sandage system \citep[][]{1959HDP....53..275D}. 
This morphological classification is a suitable training data set for computer-based classification of galaxies using machine learning techniques. 

The taxonomy by \citet{2015ApJS..217...32B} includes stellar bars as well as rings and ringlenses. 
The large depth of the S$^4$G ($\sim$~1~$M_{\odot}/$pc$^{2}$) allows the detection of nearly all outer features 
\citep[see further discussion in Sect. 2.2 in][]{2014A&A...562A.121C}. 
\citet[][]{2015ApJS..217...32B} also determined the family -- $\rm SA$ (non-barred), $\rm S\underline{A}B$ (weakest bars), $\rm SAB$, 
$\rm SA\underline{B}$, and $\rm SB$ (strongest bars) -- and the revised Hubble stage ($T$) of the galaxies that are used in this work. 
Hereafter, rings, pseudo-rings, and ringlenses are studied together, and referred to as rings, unless stated otherwise. 
%
%
\subsection{Sample of not highly inclined disc galaxies}\label{samplesample}
%
%
In this work we use the 1320 disc galaxies \citep[according to][]{2015ApJS..217...32B}\footnote{
We have excluded 18 dwarf S0s ($T=11$) and seven peculiar galaxies with uncertain morphological types ($T=99$, mostly interacting).
} 
in the original S$^4$G sample with inclinations lower than 65$^{\circ}$ \citep[according to][]{2015ApJS..219....4S}. 
Of these, 825 galaxies are barred, 465 host inner rings, and 264 host outer rings. 
%
%
%
\subsection{Measurements of dimensions of rings and bars}\label{ring_bar_measurement}
%
%
Measurements of the de-projected sizes and ellipticities of inner and outer stellar rings and bars 
are taken from \citet[][]{2015A&A...582A..86H}. Only measurements with "ok" flags ($=1-2$) are used. 
They visually marked the outline of the ridge of the rings with points over unsharp mask images\footnote{Unsharp mask images are created by 
dividing by smoothed versions of the original image, 
in which subtle galactic structures that appear against a bright and diffuse background are easier to identify.
} 
that were then fitted with an ellipse, and de-projected to the plane of the disc. 
Hereafter, we will refer to the rings' de-projected semi-major axis (SMA) as $a_{\rm ring}$
(or as $a_{\rm inner}$ and $a_{\rm outer}$ when differentiating inner and outer rings, respectively). 
Likewise, $b_{\rm ring}$ refers to the de-projected semi-minor axis, and $q_{\rm ring}=b_{\rm ring}/a_{\rm ring}$ to the intrinsic axis ratio.

We also use de-projected visual measurements of bar sizes ($r_{\rm bar}$) from \citet[][]{2015A&A...582A..86H}, 
which were performed after optimising the brightness scale of the images to make the bars stand out. 
For the bar shape, they adjusted ellipses to the 2-D light intensity distribution of the 3.6~$\mu$m images, 
and the maximum ellipticity at the bar region was calculated ($\epsilon$). 
%
%
\subsection{Gravitational potential and bar forcing}\label{forces}
%
%
\citet[][]{2016A&A...587A.160D} used the NIR-QB code \citep[][]{1999AJ....117..792S,2002MNRAS.337.1118L} 
to perform the Fourier decomposition of the de-projected 3.6~$\mu$m S$^4$G images, 
and derived the normalised Fourier amplitudes $A_{m}=I_{m}/I_{0}$, where $I_{0}$ indicates the $m=0$ surface density component. 
The maximum of $A_2$ at the bar region, named $A_2^{\rm max}$, 
is used here as a measurement of the bar strength \citep[e.g.][]{2004ApJ...607..103L}. 

From the tabulated Fourier amplitudes, \citet[][]{2016A&A...587A.160D} also calculated the gravitational potential at the equatorial plane, 
and derived tangential ($F_{\rm T}$) and radial ($F_{\rm R}$) forces. 
They calculated the radial profiles of tangential forces normalised to the mean radial force field \citep[][]{1981A&A....96..164C}:
\begin{equation}\label{torquerad}
Q_{\rm T}(r)=\frac{{\rm max}\left( |F_{\rm T}(r,\phi)| \right)}{\langle |F_{\rm R}(r,\phi)|\rangle}.
\end{equation}
The maximum of $Q_{\rm T}$ at the bar region, named $Q_{\rm b}$, 
is used as an estimate of the strength of the bar \citep[e.g.][]{2001ApJ...550..243B,2002MNRAS.331..880L}. 
In this work we also evaluate $Q_{\rm T}$ at $r_{\rm bar}$ and at $a_{\rm ring}$.

\citet[][]{2016A&A...587A.160D} calculated the stellar contribution (disc+bulge) to the circular velocity ($V_{\rm 3.6\mu m}$) from 
the mean radial force field \citep[assuming the mass-to-light ratio at 3.6~$\mu$m of 0.53 obtained by][]{2012AJ....143..139E}. 
They also obtained a first-order model of the halo rotation curve ($V_{\rm halo}$) and 
estimated the radial forces exerted by the halo\footnote{The halo velocity amplitude was obtained by 
fitting $V_{\rm 3.6\mu m}$ and $V_{\rm halo}$ to the inclination-corrected H{\sc\,i} maximum velocity 
at the optical radius. The core radius of the halo (modelled as an isothermal sphere) was estimated 
from the total $I-$band luminosity, based on the universal rotation curve model \citep[e.g.][]{1996MNRAS.281...27P,1997ASPC..117..198H}. 
Forces were calculated assuming a spherically symmetric model: $F_{\rm halo}(r)=V_{\rm halo}(r)^2/r$. 
For further details on the method the reader is referred to \citet[][]{2016A&A...587A.160D}.
}. 
Then, a first-order halo correction on the 
tangential-to-radial force profiles ($Q_{\rm T}^{\rm halo-corr}$) was implemented following \citet[][]{2004AJ....127..279B}:
\begin{equation}
Q_{\rm T}^{\rm halo-corr}(r)=Q_{\rm T}(r) \cdot \frac{F_{\rm R}(r)}{F_{\rm R}(r)+F_{\rm halo}(r)}.
\label{halo_dilution}
\end{equation}
%
%
\begin{figure*}
\centering
\includegraphics[width=0.49\textwidth]{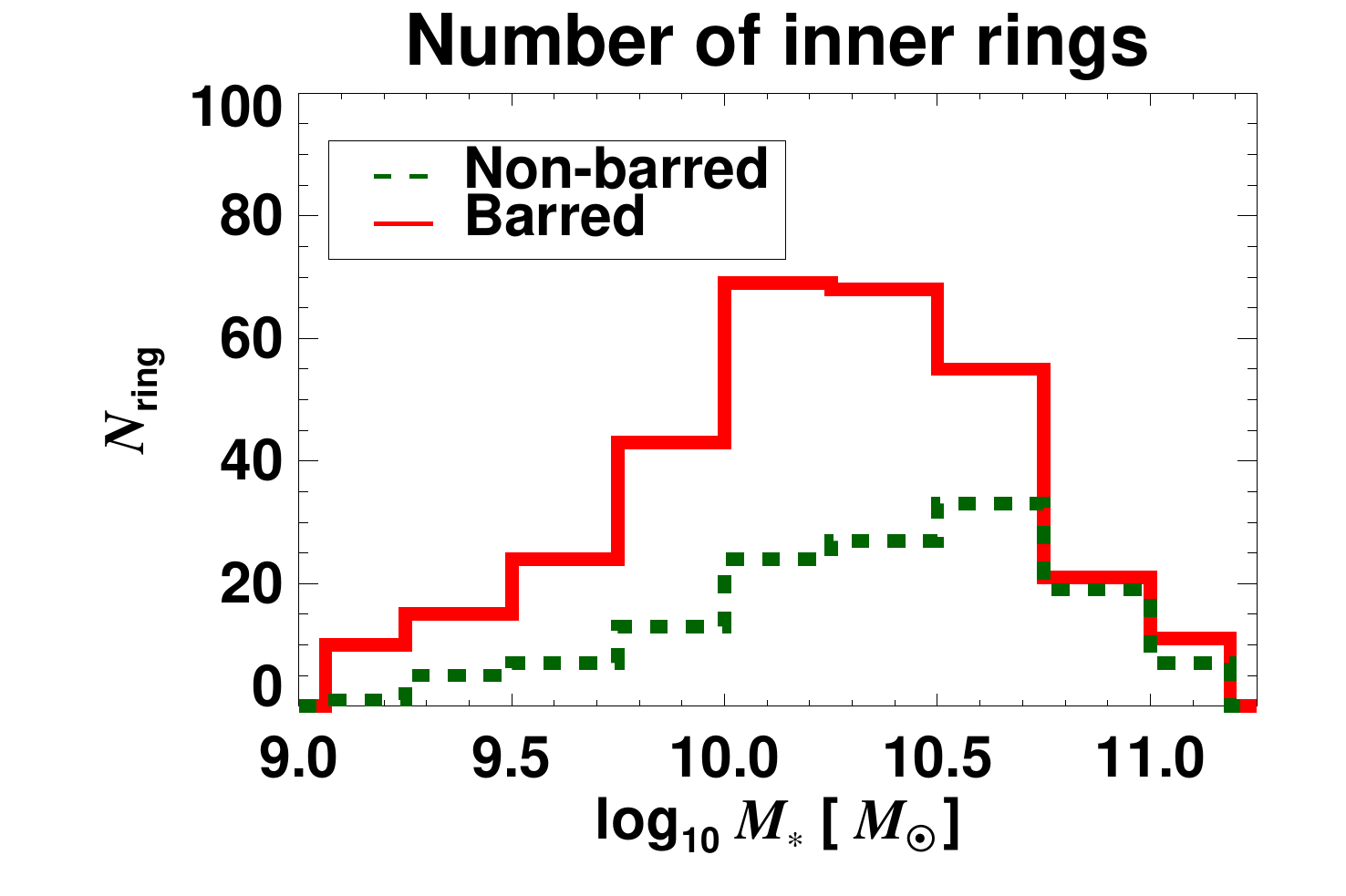}
\includegraphics[width=0.49\textwidth]{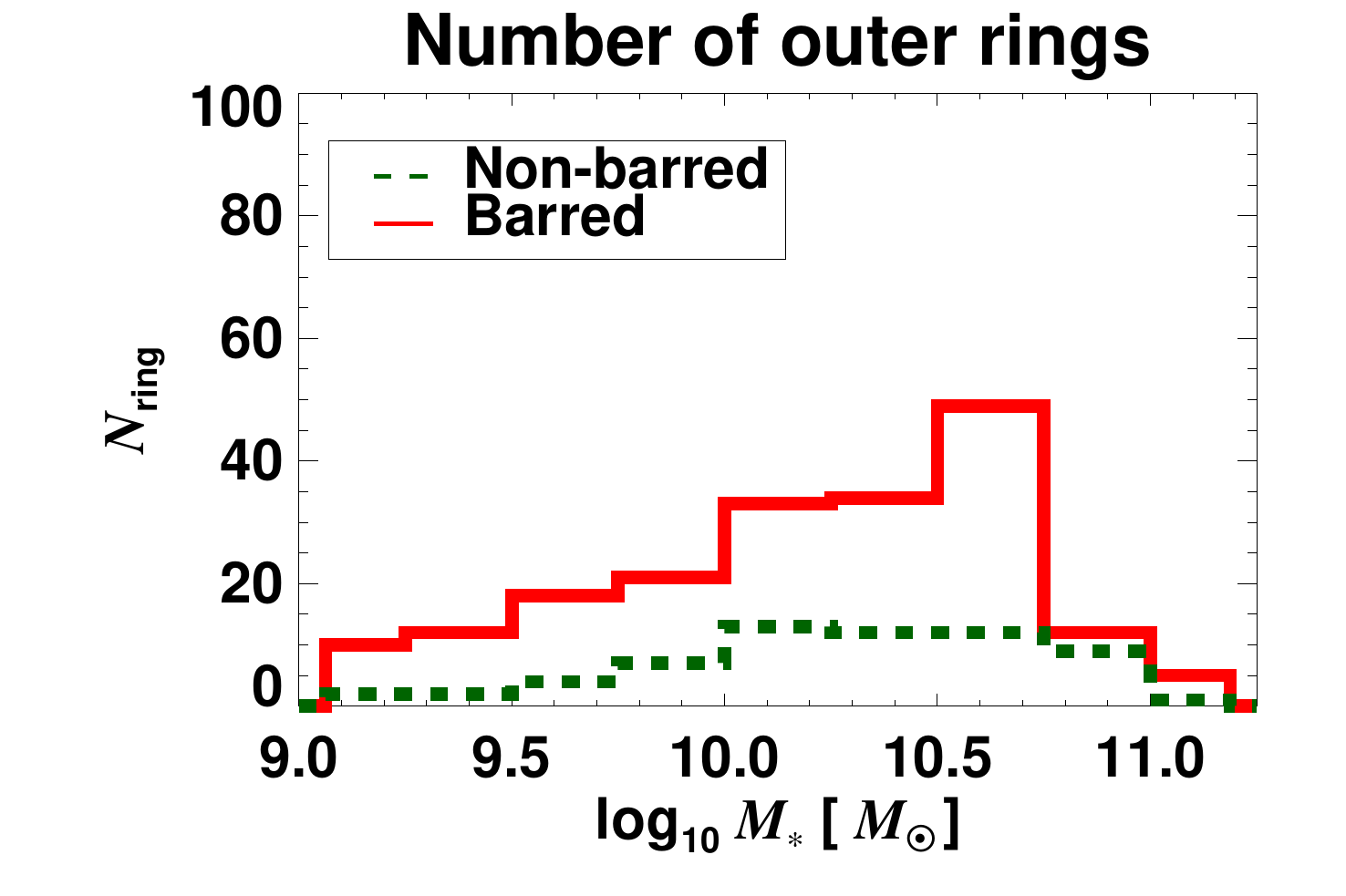}\\
\includegraphics[width=0.49\textwidth]{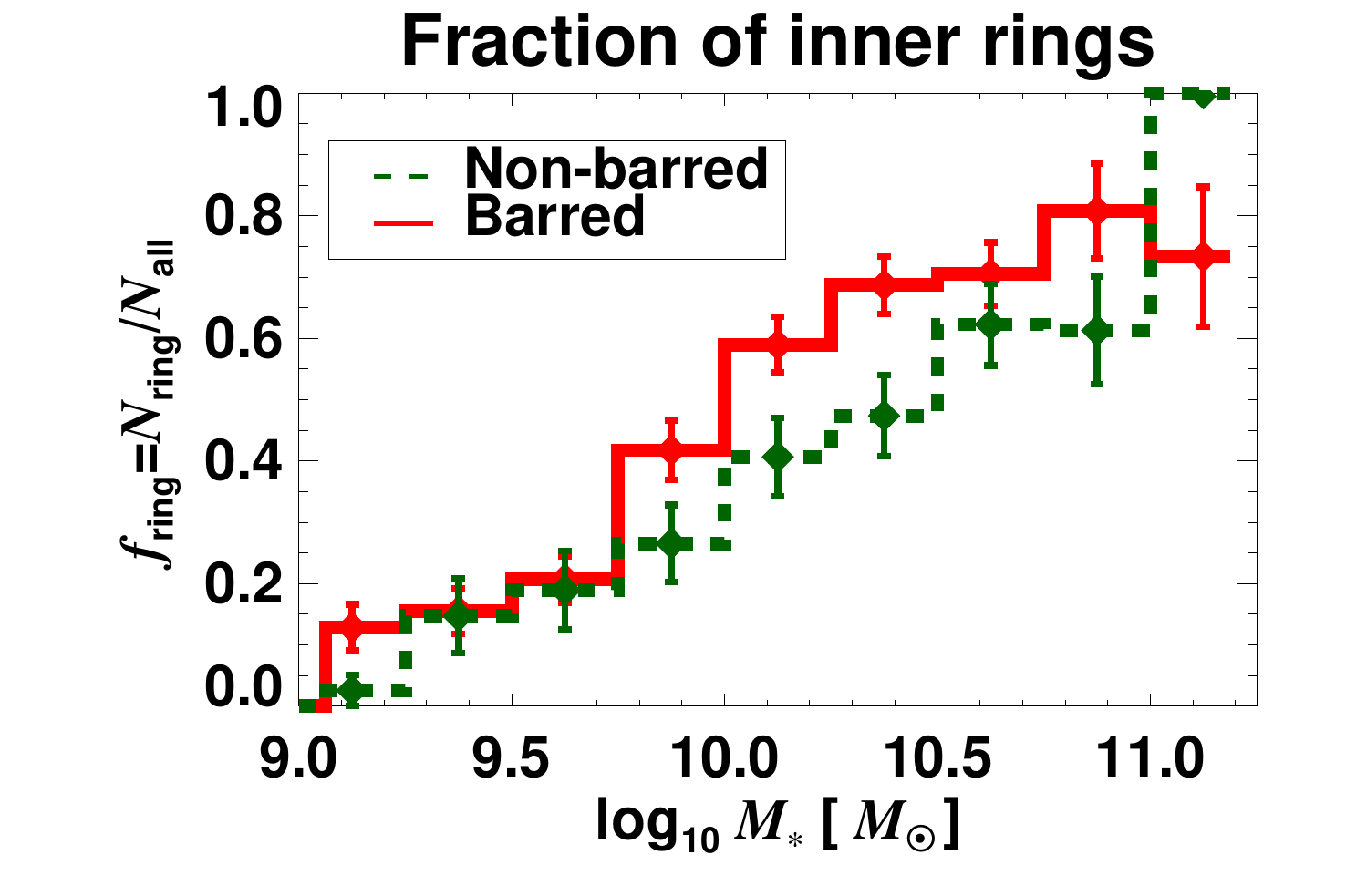}
\includegraphics[width=0.49\textwidth]{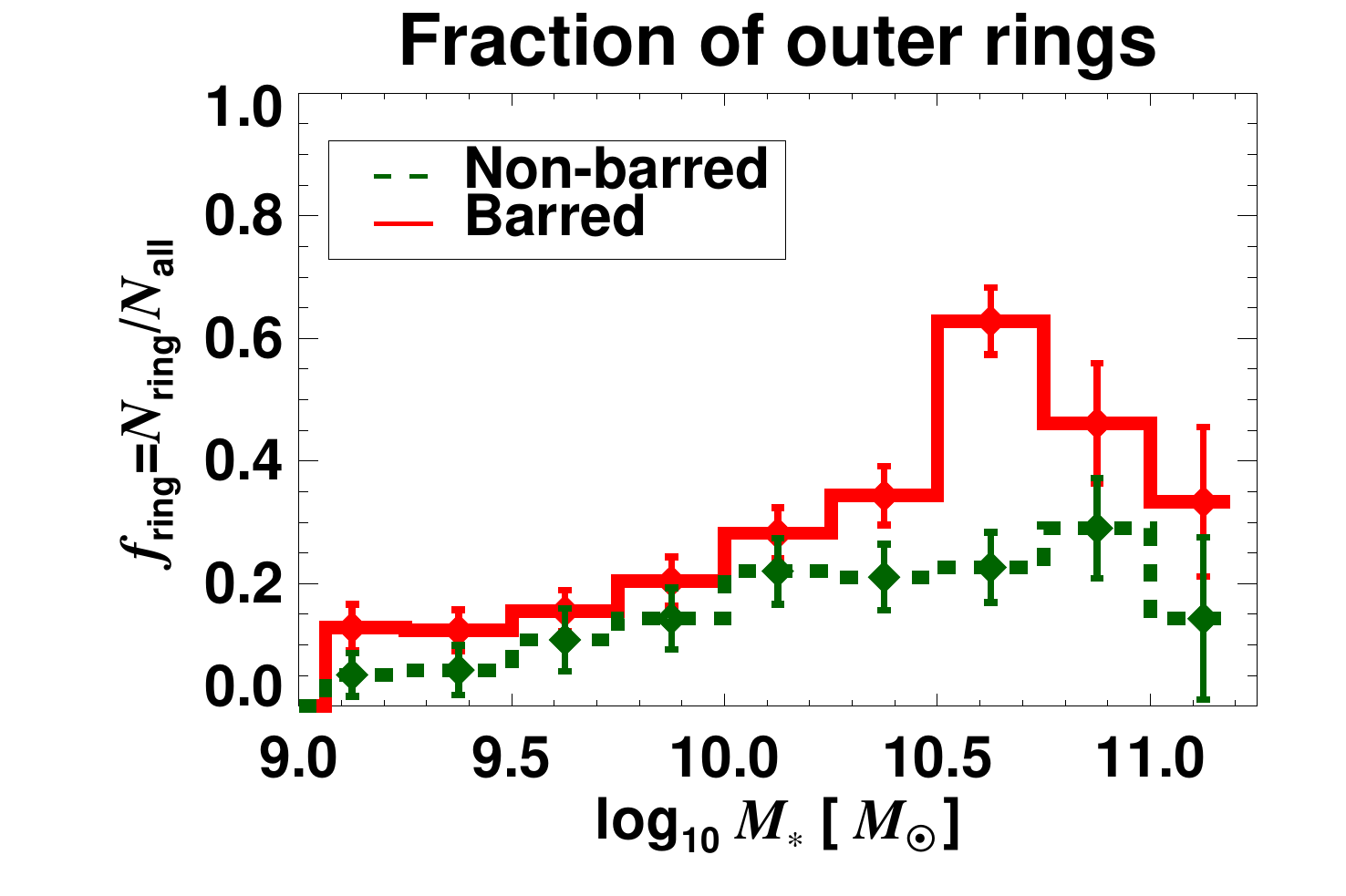}\\
\caption{
Histogram of the number (top) and fraction (bottom) of inner (left) and outer (right) 
rings as a function of the total stellar mass of the galaxies. 
The error bars correspond to the uncertainties on the fractions, estimated assuming a binomial distribution. 
}
\label{ring_frec_vs_mass}
\end{figure*}
%
%
\begin{table*}
\centering
\small
\begin{tabular}{|| c| c| c| c||}
\hline
\multicolumn{1}{||c}{\centering Inner ring fraction:} & \multicolumn{1}{|c}{\centering All} &
\multicolumn{1}{|c}{\centering Barred} & \multicolumn{1}{|c||}{\centering Non-barred} \tabularnewline
\hline
\hline
${\rm log_{10}}(M_{\ast})\in[9,9.5)$&  12.6 $\pm$   2.1 $\%$ (247)&  14.3 $\pm$   2.6 $\%$ (175)&   8.3 $\pm$   3.3 $\%$ ( 72) \\
${\rm log_{10}}(M_{\ast})\in[9.5,10)$&  28.6 $\pm$   2.6 $\%$ (304)&  30.7 $\pm$   3.1 $\%$ (218)&  23.3 $\pm$   4.6 $\%$ ( 86) \\
${\rm log_{10}}(M_{\ast})\in[10,10.5)$&  56.6 $\pm$   2.7 $\%$ (332)&  63.4 $\pm$   3.3 $\%$ (216)&  44.0 $\pm$   4.6 $\%$ (116) \\
${\rm log_{10}}(M_{\ast})\in[10.5,11)$&  68.1 $\pm$   3.4 $\%$ (188)&  73.1 $\pm$   4.3 $\%$ (104)&  61.9 $\pm$   5.3 $\%$ ( 84) \\
${\rm log_{10}}(M_{\ast})\in[11,11.5)$&  80.0 $\pm$   8.0 $\%$ ( 25)&  72.2 $\pm$  10.6 $\%$ ( 18)& 100.0 $\pm$   0.0 $\%$ (  7) \\
\hline
\end{tabular}
\caption{
The fractions of inner rings in bins of $M_{\ast}$ and for barred and non-barred galaxies. Binomial counting errors and the total number of galaxies within the bins (in parenthesis) are indicated.}
\label{inner_ring_type}
\end{table*}

\begin{table*}
\centering
\small
\begin{tabular}{|| c| c| c| c||}
\hline
\multicolumn{1}{||c}{\centering Outer ring fraction:} & \multicolumn{1}{|c}{\centering All} &
\multicolumn{1}{|c}{\centering Barred} & \multicolumn{1}{|c||}{\centering Non-barred} \tabularnewline
\hline
\hline
${\rm log_{10}}(M_{\ast})\in[9,9.5)$&  10.5 $\pm$   2.0 $\%$ (247)&  12.6 $\pm$   2.5 $\%$ (175)&   5.6 $\pm$   2.7 $\%$ ( 72) \\
${\rm log_{10}}(M_{\ast})\in[9.5,10)$&  16.4 $\pm$   2.1 $\%$ (304)&  17.9 $\pm$   2.6 $\%$ (218)&  12.8 $\pm$   3.6 $\%$ ( 86) \\
${\rm log_{10}}(M_{\ast})\in[10,10.5)$&  27.7 $\pm$   2.5 $\%$ (332)&  31.0 $\pm$   3.1 $\%$ (216)&  21.6 $\pm$   3.8 $\%$ (116) \\
${\rm log_{10}}(M_{\ast})\in[10.5,11)$&  43.6 $\pm$   3.6 $\%$ (188)&  58.7 $\pm$   4.8 $\%$ (104)&  25.0 $\pm$   4.7 $\%$ ( 84) \\
${\rm log_{10}}(M_{\ast})\in[11,11.5)$&  28.0 $\pm$   9.0 $\%$ ( 25)&  33.3 $\pm$  11.1 $\%$ ( 18)&  14.3 $\pm$  13.2 $\%$ (  7) \\
\hline
\end{tabular}
\caption{
As in Table. \ref{inner_ring_type} but for outer rings.}
\label{outer_ring_type}
\end{table*}

%
%
\section{Ring fraction as a function of stellar mass}\label{intrinsic_rings_mass}
%
%
The fraction of inner and outer rings ($f_{\rm ring}$) is strongly dependent on stellar mass (Fig.~\ref{ring_frec_vs_mass}). 
It increases with increasing $M_{\ast}$ for both barred and non-barred galaxies. 
It peaks at $M_{\ast} \approx 10^{10.5}-10^{11}M_{\odot}$. 

For inner rings, $f_{\rm ring}$ ranges from $\sim17 \,\%$ amongst faint galaxies ($M_{\ast} \le 10^{10}M_{\odot}$) to 
$\sim70 \,\%$ for the more massive ones ($M_{\ast} \ge 10^{10.5}M_{\odot}$), regardless of the presence of bars. 
We note that all of the (few) most massive non-barred galaxies host inner rings, but the statistics in this $M_{\ast}$-bin are not very robust, 
given the sub-sample size (see the histogram in the upper left panel of Fig.~\ref{ring_frec_vs_mass}). 
For outer rings, $f_{\rm ring}$ ranges from $\sim11 \,\%$ when $M_{\ast} \le 10^{10}M_{\odot}$ to $\sim42 \,\%$ when $M_{\ast} \ge 10^{10.5}M_{\odot}$.

On average, barred galaxies tend to host rings more frequently than their non-barred counterparts. 
In particular, $12.9 \pm 1.5 \%$\footnote{We calculate binomial counting errors: $df_{\rm ring}=\sqrt{f_{\rm ring}\cdot (1-f_{\rm ring})/N_{\rm gals}}$, 
where $f_{\rm ring}$ refers to the ring fraction and $N_{\rm gals}$ to the total number of galaxies.
}
of the non-barred galaxies in our sample host outer rings, in contrast to $24.2 \pm 1.5 \%$  of barred ones. 
Barred galaxies host inner rings in $39.5 \pm 1.7\%$ of the cases, which is larger than the $28.1 \pm 2.0 \%$ of non-barred galaxies hosting them. 
In Table.~\ref{inner_ring_type} and Table.~\ref{outer_ring_type} we provide the fraction of inner and outer rings, 
respectively, in bins of $M_{\ast}$\footnote{
Nine galaxies in our sample do not have measurements of $M_{\ast}$ and are not included in this analysis. 
}. 
The differences in $f_{\rm ring}$ between barred and non-barred galaxies tend to be larger for larger $M_{\ast}$, 
and more clearly so for outer rings. 
In summary, the fraction of inner (outer) rings in barred galaxies is larger than in their non-barred counterparts 
by a factor of $1.41 \pm 0.12$ ($1.88 \pm 0.25$).
%
%
\begin{figure}
\centering
\includegraphics[width=0.4\textwidth]{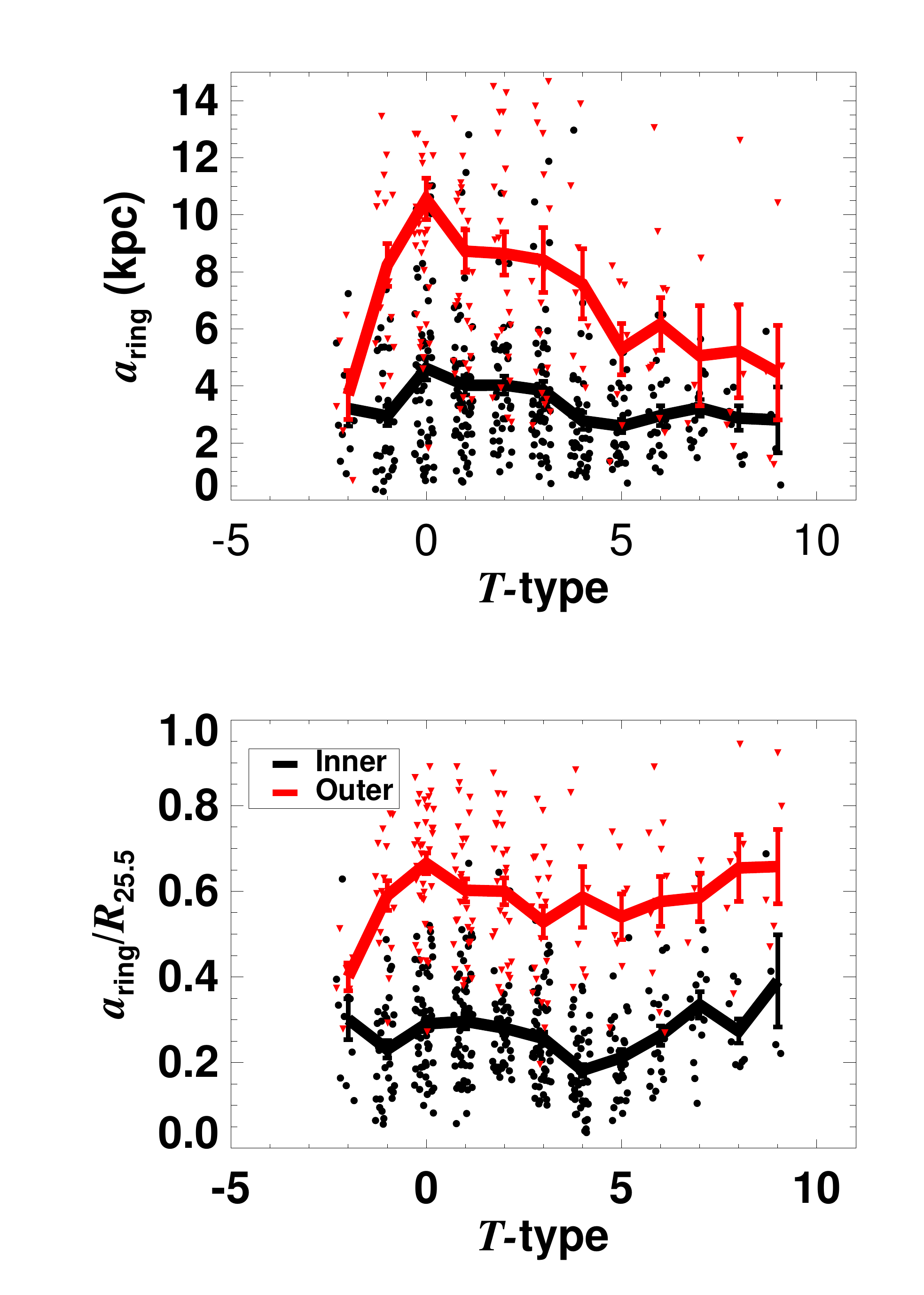}
\includegraphics[width=0.4\textwidth]{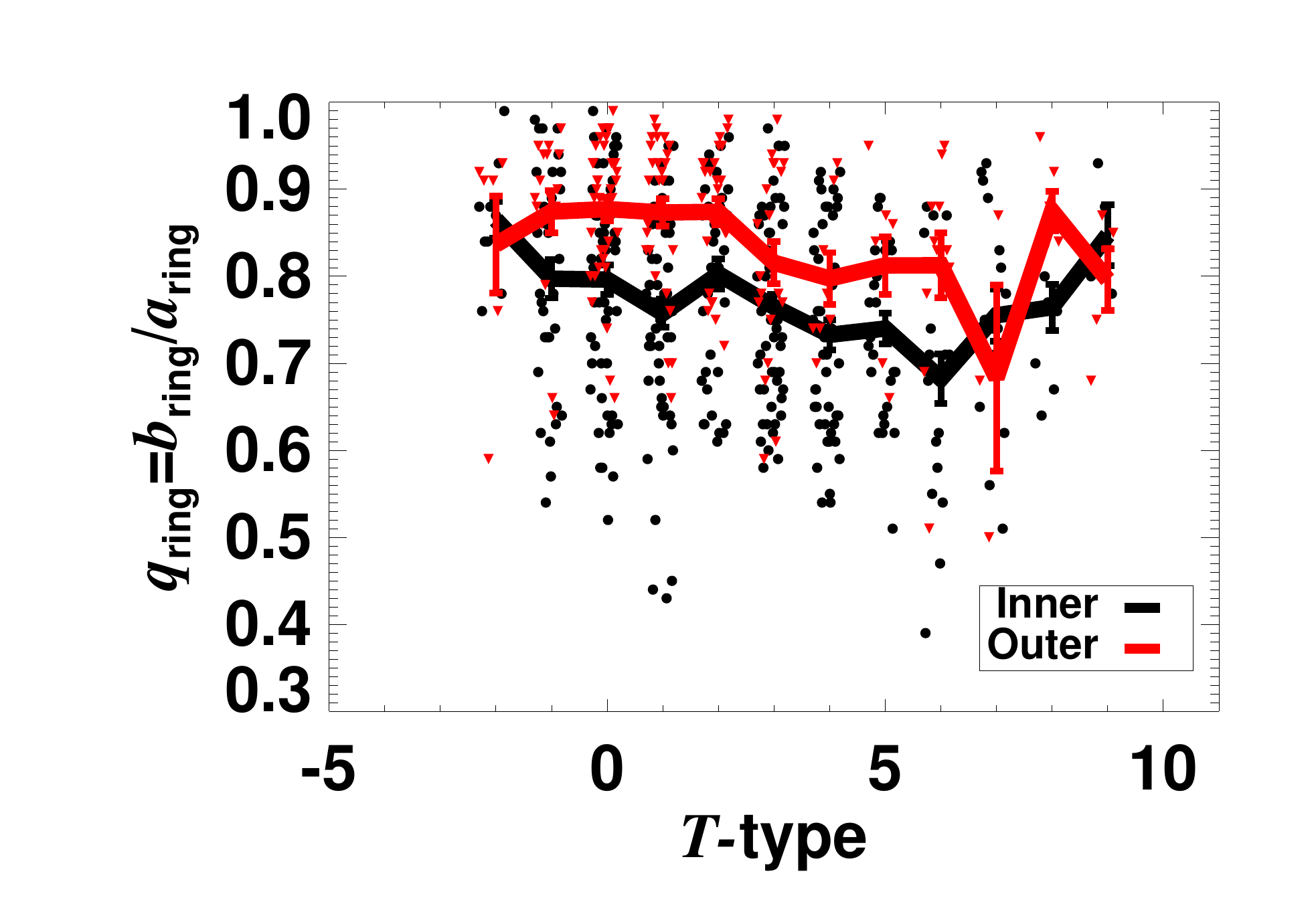}
\caption{
De-projected semi-major axis of inner and outer rings, in physical units (upper panel) and normalised to $R_{25.5}$ (middle panel), 
and intrinsic ellipticity (lower panel) as a function of the revised numerical Hubble stage. 
In the $x$-axis we have added small random offsets ($\lesssim 0.3$) to the $T$ values (integers) for the sake of avoiding point overlapping. 
The colour palette and symbol style separates inner and outer rings. The lines indicate the running mean. 
For each bin, the error bars indicate the standard deviation of the mean.
}
\label{inner_outer_ttype}
\end{figure}
%
%
\begin{figure}
\centering
\includegraphics[width=0.45\textwidth]{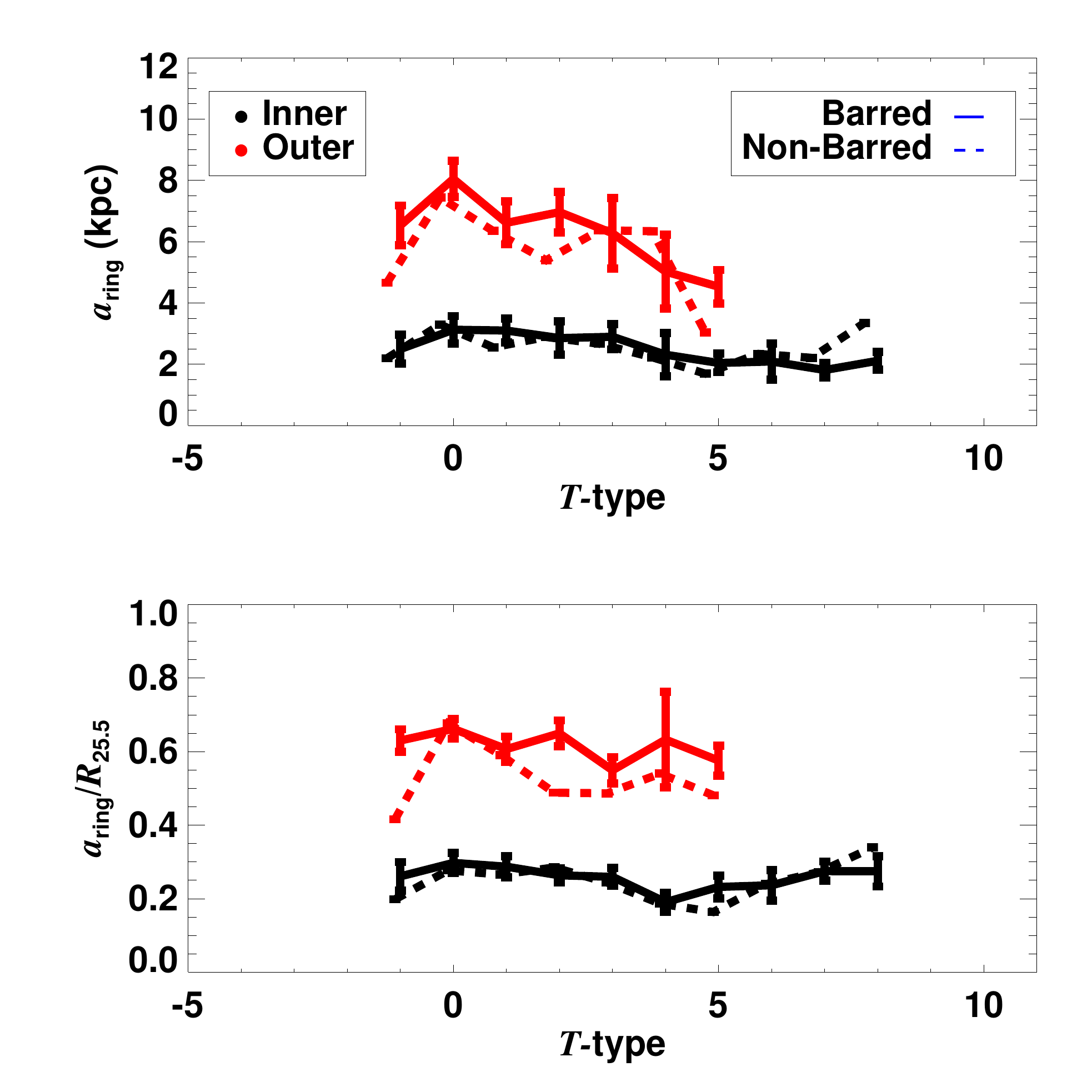}\\
\includegraphics[width=0.45\textwidth]{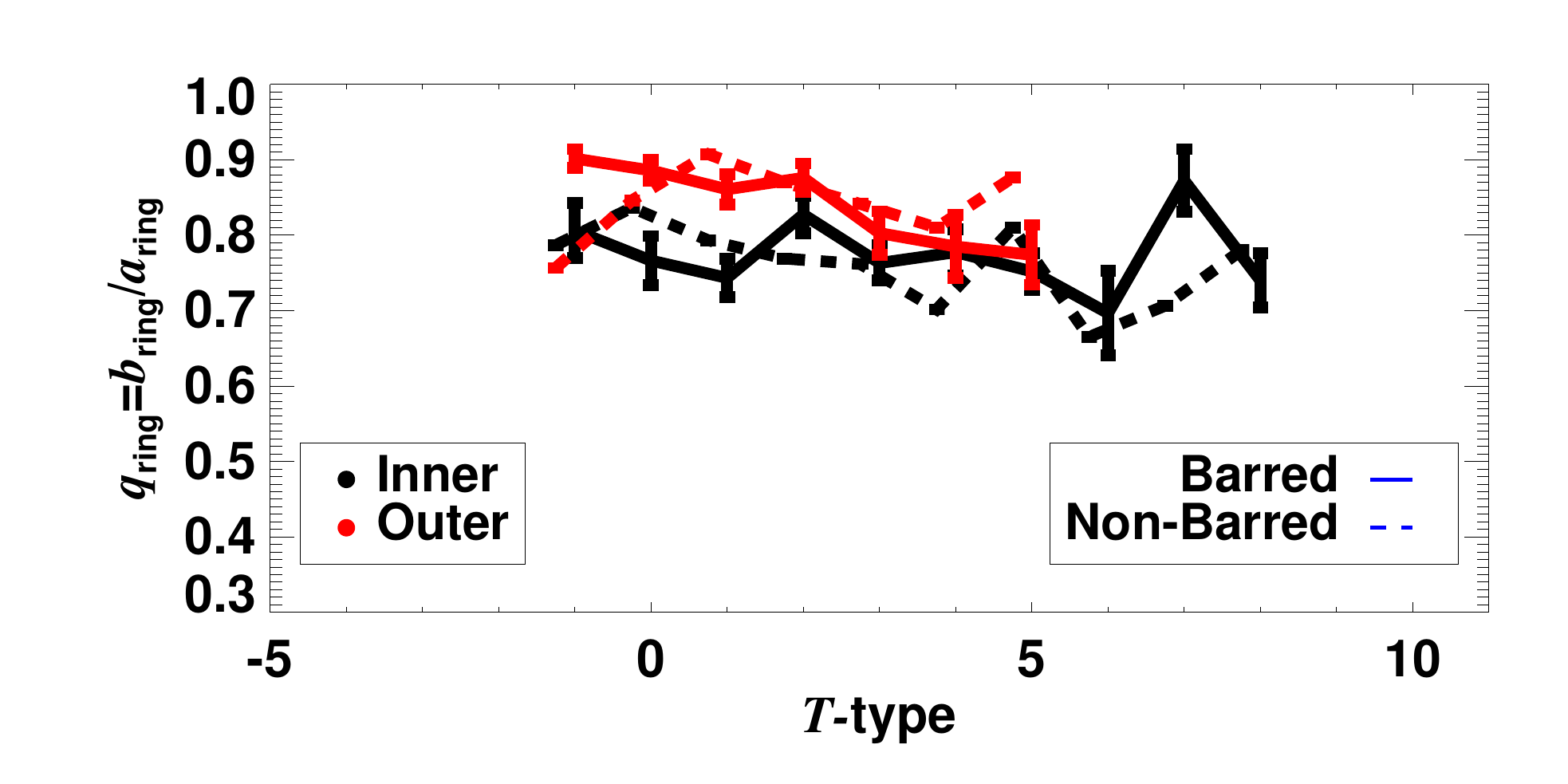}
\caption{
Same as in Fig.~\ref{inner_outer_ttype} but separating barred and non-barred galaxies, and showing only the running mean.
The standard deviation of the mean (error bars) is shown for the barred galaxies.
}
\label{inner_outer_ttype_barred}
\end{figure}
%
%
\begin{figure*}
\centering
\begin{tabular}{c c}
\includegraphics[width=0.99\textwidth]{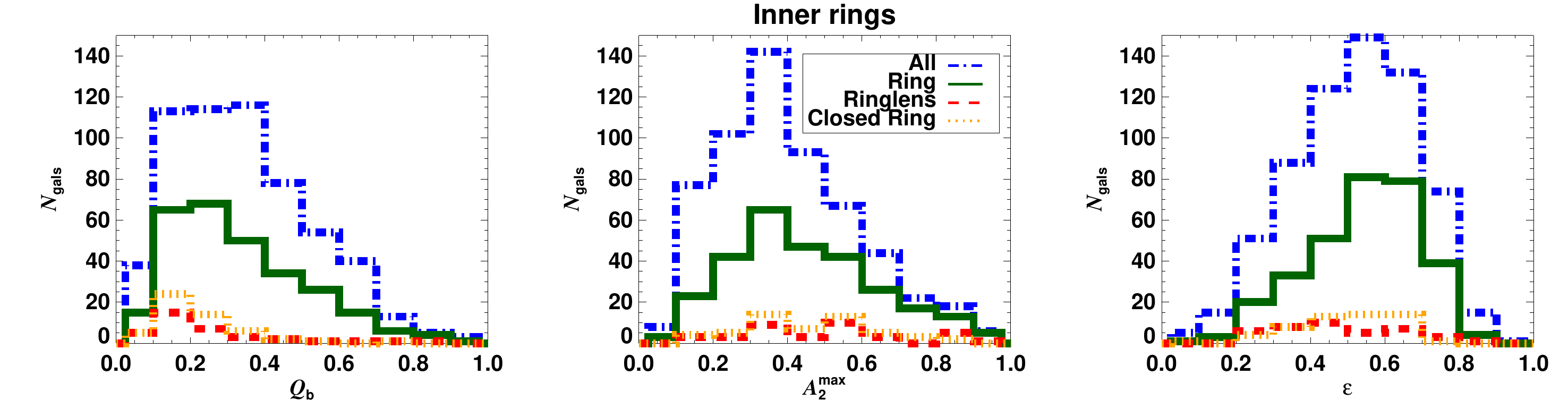}\\[5ex]
\includegraphics[width=0.99\textwidth]{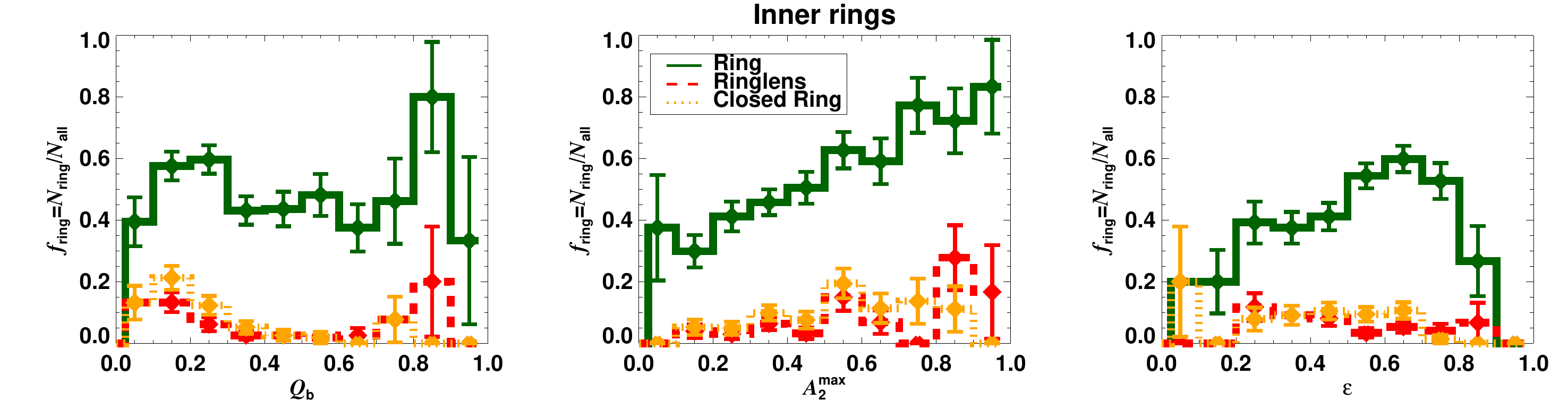}
\end{tabular}
\caption{
Upper panels: Histograms of the distributions of the different bar strength proxies for the barred galaxies in our sample 
and for those hosting inner rings in particular.
We also show separately closed rings ($r$ and $R$, i.e. excluding pseudo-rings) and ringlenses with different colours. 
Lower panels: Same as above but for the fraction of inner rings. 
The error bars correspond to the uncertainties estimated assuming a binomial distribution.
}
\label{inner_qb_a2_histo}
\end{figure*}
%
%
\begin{figure*}
\centering
\begin{tabular}{c c}
\includegraphics[width=0.99\textwidth]{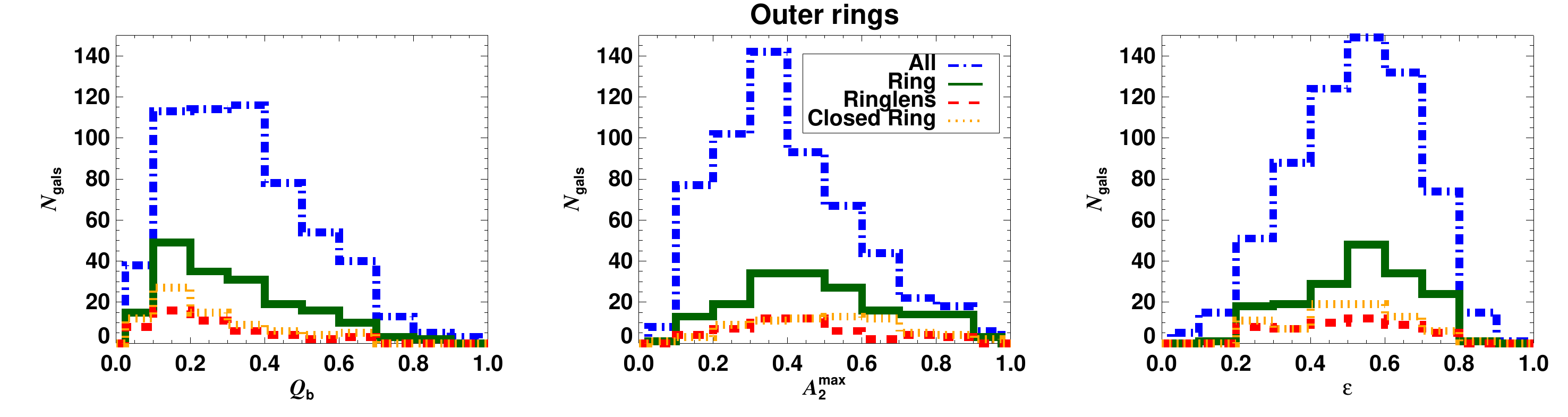}\\[5ex]
\includegraphics[width=0.99\textwidth]{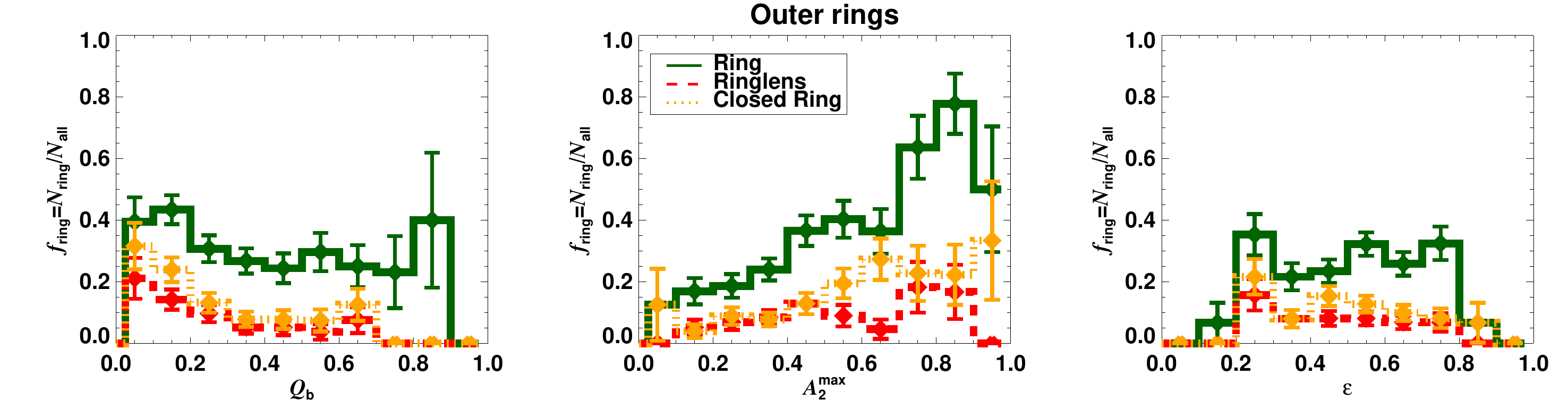}\\
\end{tabular}
\caption{
Same as in Fig.~\ref{inner_qb_a2_histo} but for barred galaxies hosting outer rings.
}
\label{outer_qb_a2_histo}
\end{figure*}
\begin{figure*}
\centering
\includegraphics[width=0.99\textwidth]{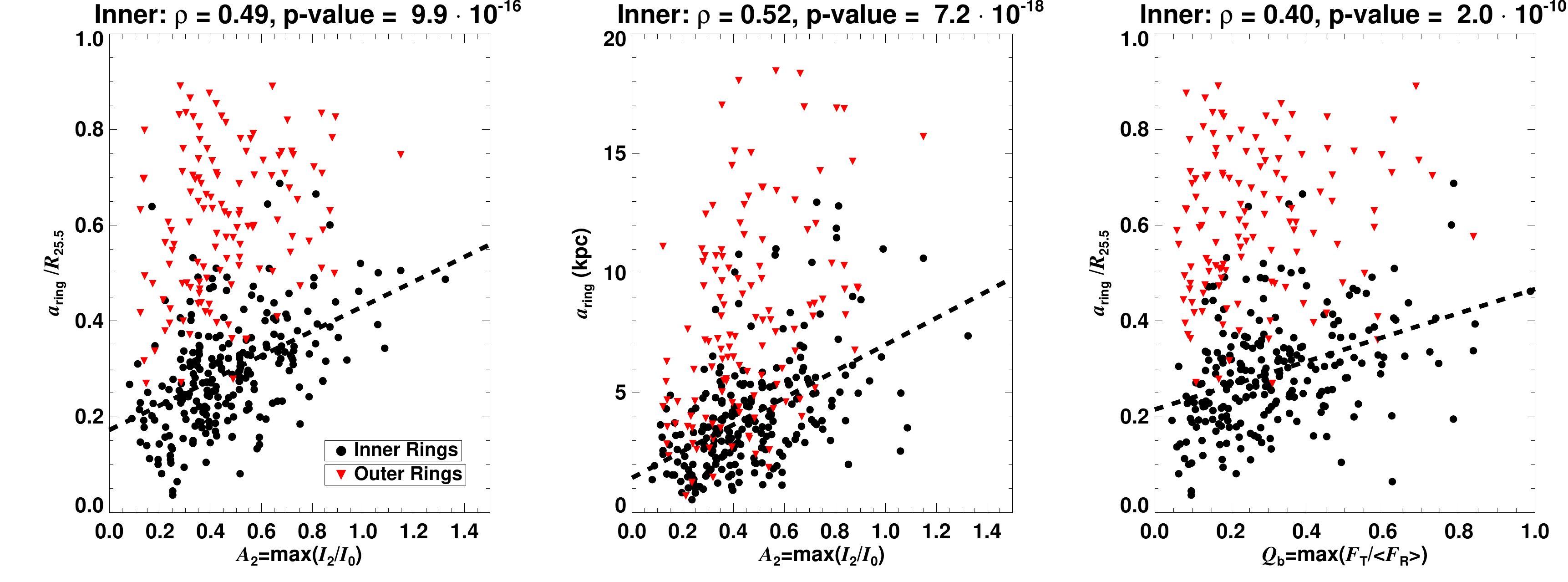}\\
\caption{
De-projected semi-major axis of rings, normalised to $R_{25.5}$ and in physical units, 
as a function of the bar strength, estimated from the $m=2$ bar Fourier amplitude (left and central panels). 
The disc-relative sizes of rings are also compared to the bar torque parameter (right panel). 
The colour palette is the same as in Fig.~\ref{inner_outer_ttype}, separating inner and outer rings. 
The Spearman's correlation coefficient and significance of the correlation for the inner rings is shown above the figures. 
The dashed black line corresponds to the linear fit to the cloud of points for the inner rings.
}
\label{inner_outer_qb_a2}
\end{figure*}
%
%
\begin{figure}
\centering
\includegraphics[width=0.4\textwidth]{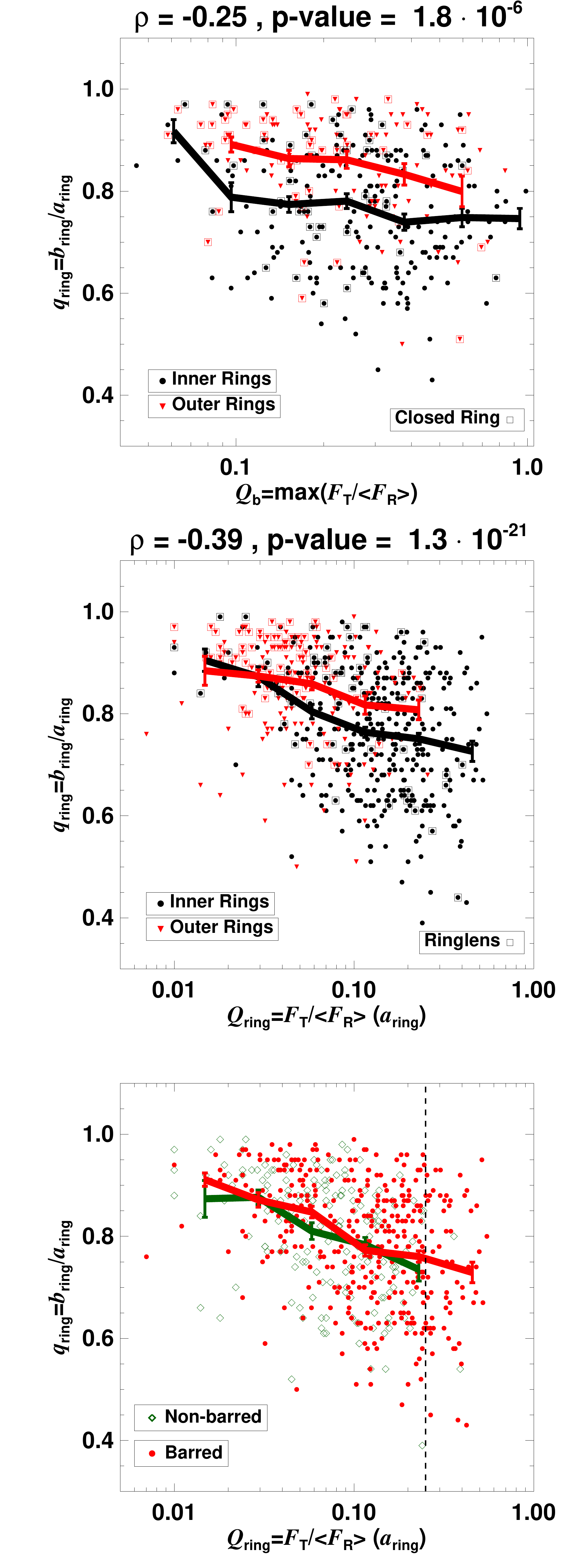}
\caption{
Upper panel: De-projected ring axis ratio versus bar torque parameter, separating inner and outer rings, and highlighting closed rings. 
Central panel: De-projected ring axis ratio versus tangential-to-radial forces ($Q_{\rm T}$) evaluated 
at the location of the ring semi-major axis. Ringlenses are displayed with different symbols. 
Lower panel: Same as above but separating barred and non-barred systems. The vertical line corresponds to $Q_{\rm ring}=0.25$. 
For the different subsets the solid lines correspond to the running mean, while the vertical lines indicate the standard deviation of the mean. 
The Spearman's correlation coefficient and significance are indicated on top of the uppermost panels.
}
\label{q_rings_vs_q_bar}
\end{figure}
%
%
\begin{figure}
\centering
\includegraphics[width=0.4\textwidth]{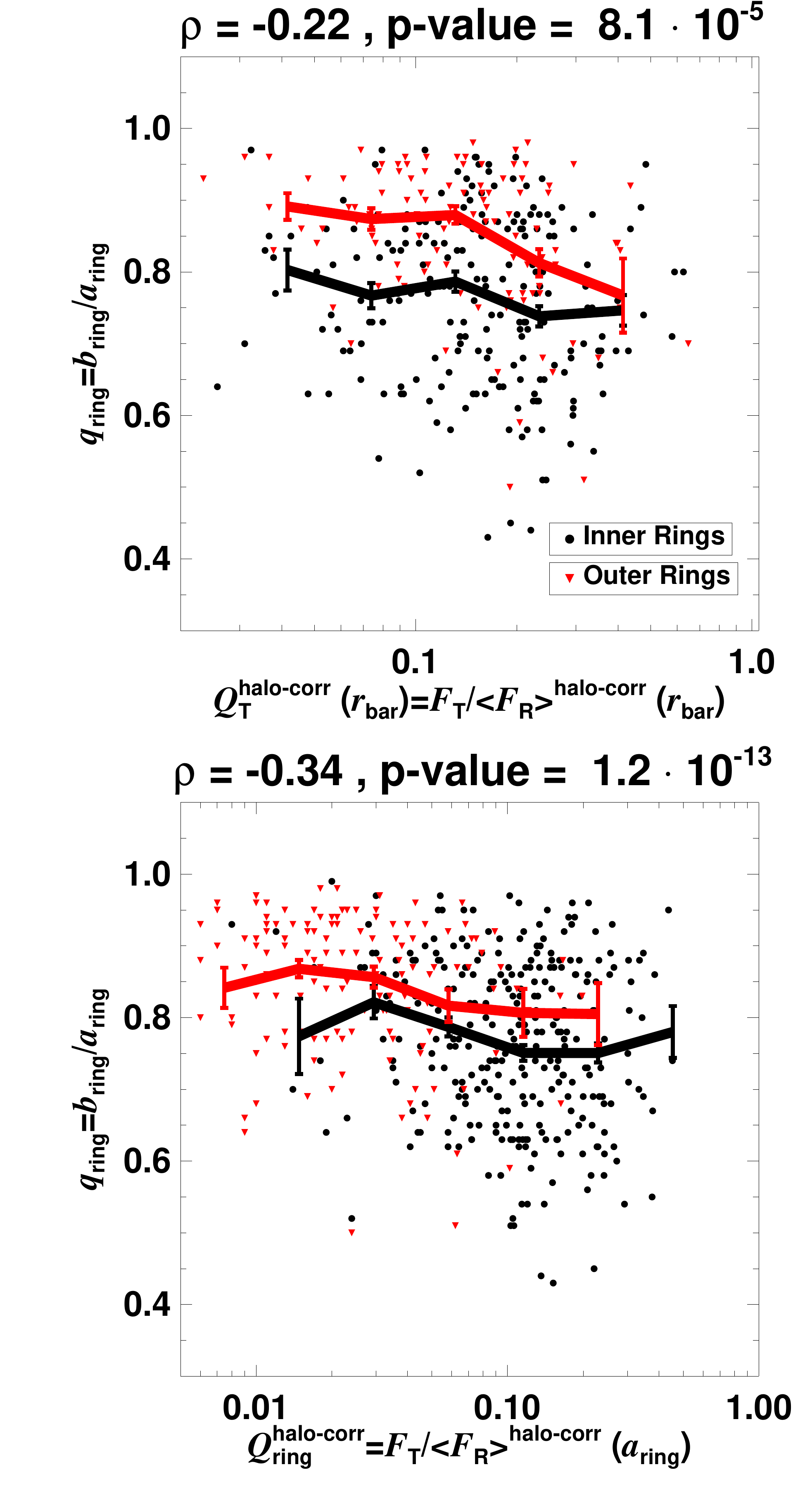}
\caption{
Same as in Fig.~\ref{q_rings_vs_q_bar} but correcting the force profiles for the halo dilution 
(see text) and evaluating $Q_{\rm T}^{\rm halo-corr}$ at the bar end and at the ring semi-major axis.
}
\label{q_rings_vs_q_bar_halo_corr}
\end{figure}
%
%
\section{Ring properties in the Hubble sequence}\label{intrinsic_rings_ttype}
%
%
Ring sizes in physical units in galaxies with $T<5$ are $\sim 30\%$ larger than in late-type galaxies with $T\ge 5$ 
(Fig.~\ref{inner_outer_ttype}, upper panel), but this is mainly a consequence of the former galaxies being more massive and bigger in general 
(see also the upper left panel in Fig.~\ref{inner_outer_mass}). 
Ring sizes in physical units decrease with increasing $T$ for the spirals, and more clearly for the outer features. 

When ring dimensions are normalised to the disc size (Fig.~\ref{inner_outer_ttype}, middle panel), a new picture arises. 
The distribution of inner ring sizes has a minimum at $T\approx4$. 
Their disc-relative sizes increase (decrease) with increasing $T$ when $T\ge 5$ ($T<5$), less clearly for outer features. 
Outer ring sizes (in kiloparsecs and normalised to the disc size) seem to decrease amongst S0s\footnote{
We note that the sampling of lenticulars in the S$^4$G is relatively poor.
}.

Considering their intrinsic ellipticity (Fig.~\ref{inner_outer_ttype}, lower panel), 
inner rings are found to be typically more elongated than outer rings, 
regardless of $T-$type, in agreement with \citet[][]{2014A&A...562A.121C}. 
On average, S0s and early-type spirals host somewhat more circular inner and outer rings than 
their late-type counterparts \citep[see also][]{2014A&A...562A.121C}, but the trend is very mild 
(at the end of the Hubble sequence, inner rings seem to be more circular for larger $T$).
Interestingly, when we separate barred and non-barred galaxies and the ring sizes 
and axis ratios are compared, we find roughly the same running means (Fig.~\ref{inner_outer_ttype_barred}). 
%
%
\section{Ring properties as a function of bar strength}\label{rings_bars}
%
%
\subsection{Ring fraction versus bar strength}
%
%
In order to investigate the role played by bars in ring formation, 
in Figs.~\ref{inner_qb_a2_histo} and~\ref{outer_qb_a2_histo} we show the distribution of the bar strength measurements for the galaxies 
with and without inner and outer rings, respectively. 
Ringlenses and closed rings (i.e. excluding pseudo-rings) are shown separately, and all bar strength proxies are taken into account. 
We also display the fraction of rings in bins of bar strength (lower panels).

The fraction of inner and outer rings increases with increasing $A_{2}^{\rm max}$. 
Bars with the largest $A_2$ values ($\gtrsim 0.7$) host inner rings in nearly $80\%$ of the cases, 
which is a factor of two larger than the frequency for the weakly barred galaxies. 
This trend is also seen in $\epsilon$, but it becomes rather flat for $Q_{\rm b}$ (with two local maxima for low and high values). 
In addition, galaxies with the largest bar Fourier amplitudes host outer rings in more than $60\,\%$ of the studied cases, 
which is about three times larger than in their weakly barred counterparts. 
This tendency is less clear when the ring fraction is studied as a function of $Q_{\rm b}$ or $\epsilon$. 
%
%
\subsection{Ring sizes versus bar strength}
%
%
In Fig.~\ref{inner_outer_qb_a2} we compare the sizes of rings 
to the strength of the stellar bars, measured from $Q_{\rm b}$ and $A_{2}^{\rm max}$. 
This test aims to shed more light on the bar-rings coupling.

There is a clear correlation between the inner ring SMA, normalised to the disc size ($R_{25.5}$), and the bar strength, as 
confirmed by the Spearman's correlation coefficient ($\rho$) and $p-$value\footnote{The Spearman's rank correlation assesses 
the existence of a monotonic relation between two variables: 
$\rho=(-1)+1$ implies perfect (anti)correlation, and small $p-$values ($<0.01$) indicate significant correlation.
}, shown above the plots. 
Nevertheless, the outer ring SMA is not (or very weakly) dependent on $Q_{\rm b}$ or $A_{2}^{\rm max}$. 
Likewise, $A_{2}^{\rm max}$ correlates with the inner ring size in physical units, 
but we checked that this correlation is weaker for $Q_{\rm b}$.
%
%
\subsection{Ring ellipticity versus bar strength}\label{ring_ellip_bar_force}
%
%
The next step is to analyse the link between the shape of the rings and the bar-induced perturbation strength.
In Fig.~\ref{q_rings_vs_q_bar} we test the dependence of the intrinsic shape of inner and outer rings on 
(i) the bar torque parameter $Q_{\rm b}$ (upper panel), and 
(ii) the tangential-to-radial forces evaluated at the de-projected SMA of the ring: 
$Q_{\rm ring}=F_{\rm T}/<F_{\rm R}>(a_{\rm ring})$ (central and lower panels). 

We find that, on average, the ring axis ratio $q_{\rm ring}$ decreases with increasing $Q_{\rm b}$ 
(more circular rings in weakly barred galaxies). 
The Spearman's correlation coefficient and significance are $\rho=-0.25$ and $p=1.8\cdot10^{-6}$, respectively, 
and thus the relation is fairly weak. When the correlation is studied separately, it gets weaker for both outer 
($\rho=-0.27$, $p=0.003$) and inner ($\rho=-0.16$, $p=0.017$) rings. 
The trend is the same when pseudo-rings are excluded, that is, when we only consider closed rings. 

A stronger relation is shown between the intrinsic axis ratio of (inner and outer) rings and 
the tangential-to-radial forces evaluated at the ring semi-major axis ($\rho=-0.39$ and $p=1.32\cdot10^{-21}$). 
The scatter increases with increasing $Q_{\rm ring}$. 
The same trend is found for ringlenses, which tend to be rounder than normal rings. 
Also, the tendency of $q_{\rm ring}$ versus $Q_{\rm ring}$ 
is exactly the same for barred and non-barred galaxies when $0.01 \lesssim Q_{\rm ring} \lesssim 0.25$. 
However, we note that only two of the 64 inner ringed galaxies with $Q_{\rm ring} \gtrsim 0.25$ are non-barred. 
%
%
\subsection{Ring ellipticity versus halo-corrected forcing}\label{rings_bars_halo}
%
%
At the ring radii the contribution of the dark matter halo to the overall radial force field might be significant, 
especially in the case of outer features. Thus, it is important to correct the $F_{\rm T}/<F_{\rm R}>$ ratio for halo dilution 
(Eq.~\ref{halo_dilution}) when we assess the relationship between the intrinsic ellipticity of rings ($q_{\rm ring}$) 
and the amplitude of non-axisymmetries. 
This is done in Fig.~\ref{q_rings_vs_q_bar_halo_corr}, using a smaller sample of 439 galaxies (barred and unbarred) 
with estimates of the rotation curve decomposition model. 

Halo-corrected tangential-to-radial forces evaluated at the bar end -- $Q_{\rm T}^{\rm halo-corr}(r_{\rm bar})$ -- 
and at the ring SMA  -- $Q_{\rm ring}^{\rm halo-corr}$ -- correlate weakly with $q_{\rm ring}$. 
The trend is less clear when outer ($\rho=-0.31$, $p=0.001$) and inner ($\rho=-0.13$, $p=0.06$) rings are studied separately.
The scatter in this relation is not noticeably reduced after the halo correction is performed.

Also, we note that the effect of halo dilution at the inner ring SMA is strongly dependent on the inner profile of dark matter density that we adopt 
(Sect.~\ref{forces}). 
Due to the uncertainty in $V_{\rm halo}$ \citep[e.g. core versus cusp, see][and references therein]{2008AJ....136.2648D} 
we also stress the uncertainty in the halo-corrected force profiles at $a_{\rm inner}$. 
On the contrary, the halo correction at $a_{\rm outer}$ is likely to be more reliable.
%
%
\section{Ratio of outer-to-inner ring sizes}\label{rings_ratio}
%
%
The ratio of the de-projected sizes of outer and inner rings ($a_{\rm outer}/a_{\rm inner}$) 
is an important constraint for numerical models (see discussion in Sect.~\ref{manifold}) that we probe observationally here. 
We use a sub-sample of 103 galaxies simultaneously hosting inner and outer rings, of which 82 are barred. 
The distribution of $a_{\rm outer}/a_{\rm inner}$ 
as a function of the total stellar mass of the host galaxy is fairly flat (Fig.~\ref{Ring_ratios_qb_a2_mass}). 

Under the assumption of a flat rotation curve, the linear treatment of resonances implies that 
$R_{\rm OLR}/R_{\rm CR}=1+\sqrt{2}/2$ and $R_{\rm UHR}/R_{\rm CR}=1-\sqrt{2}/4$ \citep[e.g.][]{1987gady.book.....B}, 
where $R_{\rm OLR}$, $R_{\rm UHR}$ , and $R_{\rm CR}$ refer to the location of the outer Lindblad resonance, 
the inner ultraharmonic resonance, and the corotation radii, respectively. 
This implies that $R_{\rm OLR}/R_{\rm UHR}\approx 2.64$, 
which is consistent with the mean values that we obtain for $a_{\rm outer}/a_{\rm inner}$ (Fig.~\ref{Ring_ratios_qb_a2_mass}). 
We note that $a_{\rm outer}/a_{\rm inner}$ tends to be slightly 
lower than 2.64 for all the data points with $M_{\ast}>10^{10.75}M_{\odot}$ (13 galaxies). 
This, and the fairly large scatter in the plot, could be due to differences in the mass distribution and 
shape of the rotation curves of each of the galaxies (e.g. declining rotation curves amongst the most massive galaxies). 
Alternatively, these could be partly explained by the fact that some outer rings might not form exactly at the OLR, 
but at the outer 1:4 resonance \citep[][]{2017MNRAS.470.3819B}.

In order to shed more light on the connection between ring and bar properties, 
in Fig.~\ref{Ring_ratios_qb_a2_bar_strength} we study $a_{\rm outer}/a_{\rm inner}$ 
as a function of $Q_{\rm b}$, $Q_{\rm T}(r_{\rm bar}),$ and $A_{2}^{\rm max}$. 
This allows us to test the expected correlation from the manifold theory \citep[e.g.][and references therein]{2009MNRAS.400.1706A} 
between $a_{\rm outer}/a_{\rm inner}$ and the bar-induced perturbation strength. 
A weak anti-correlation is found with $A_{2}^{\rm max}$ or $Q_{\rm T}$ evaluated at the bar end. 
Bars with the largest $A_2$ do not present large outer-to-inner ring ratios. 
Amongst weaker bars ($A_2 \lesssim 0.6$) the bar prominence does not seem to control the ratio of the ring semi-major axes. 
When we only consider the seven inner ringed galaxies in our sample 
hosting $R_{1}$ or $R_{1}R_{2}$ outer rings (we do not sample galaxies hosting rings of type $R_{2}$ exclusively), 
the trends are roughly the same. Overall, there is not a strong link between $a_{\rm outer}/a_{\rm inner}$ and bar strength. 
%
%
\section{Insights from unsupervised machine learning}\label{ml_analysis}
%
%
%
In this section, we apply unsupervised machine learning (ML) techniques to the S$^4$G. 
ML is well suited for our purposes for various reasons: 
i) Several factors govern the nature of rings and bars, which is hard to understand in terms of traditional scaling relations 
between pairs or trios of quantities \citep[e.g.][]{1977A&A....54..661T} or classical approaches for  statistical analysis; 
ii) rather than using numerical Hubble stages, it is preferable to take into account large sets of physical properties of galaxies 
to shed light on the formation and evolution of stellar substructures\footnote{
In fact, there is a substantial scatter in the relation between morphology and colour, 
stellar mass, and other global parameters of disc galaxies (see e.g. Fig.~\ref{morpho_type_SOM}).}; and 
iii) we can check whether visual classifications reflect different families of galaxies 
(e.g. ringed) based on global parameters of S$^4$G galaxies.
%
%
\subsection{Self-organising maps (SOMs)}
%
%
We use self-organising maps \citep[SOMs;][]{2001som..book.....K}, which are well suited to analyse "small" data sets (on the order of $\sim 10000$ records). 
In particular, we use the SOM Toolbox 2.0 for Matlab \citep[][]{2000Vesanto} and apply it to the S$^4$G data. 
SOMs resemble vector quantisation algorithms (e.g. $k-$means) 
by which one can identify representative prototypes that capture fundamental properties of the data.

SOMs are groups of neurons that are connected with their neighbours and organised on a low-dimensional grid. 
Each of these neurons is a $d-$dimensional prototype (or weight vector), where $d$ refers to the dimension of the input vectors. 
A pre-defined grid of neurons is trained iteratively to stretch towards the training sample, 
and thus represents the density of data. This is done by finding the so-called best matching units (BMUs). 
The BMU of a data vector corresponds to the map unit whose model vector best fits it. 
BMUs are determined from the minimum Euclidean distance. 
More specifically, a SOM self-organises by learning the position of the data cloud through cooperative learning: 
not only the most similar prototype vector is updated (competitive learning), 
but also its neighbours on the map are moved towards the data vector. 

We trained the SOMs using the default configuration (toolbox "SOM\_MAKE" routine) that applies the batch algorithm 
\citep[see][]{2000Vesanto}. We used an hexagonal lattice with a Gaussian relation between neighbours, 
and 16x10 neurons. We tested the consistency of our method and results against changes in the number of neurons, neighbour 
relations between neurons and radius, dimensions of the SOMs, training time, and using the sequencial algorithm. 
%
%
\begin{figure}
\centering
\includegraphics[width=0.49\textwidth]{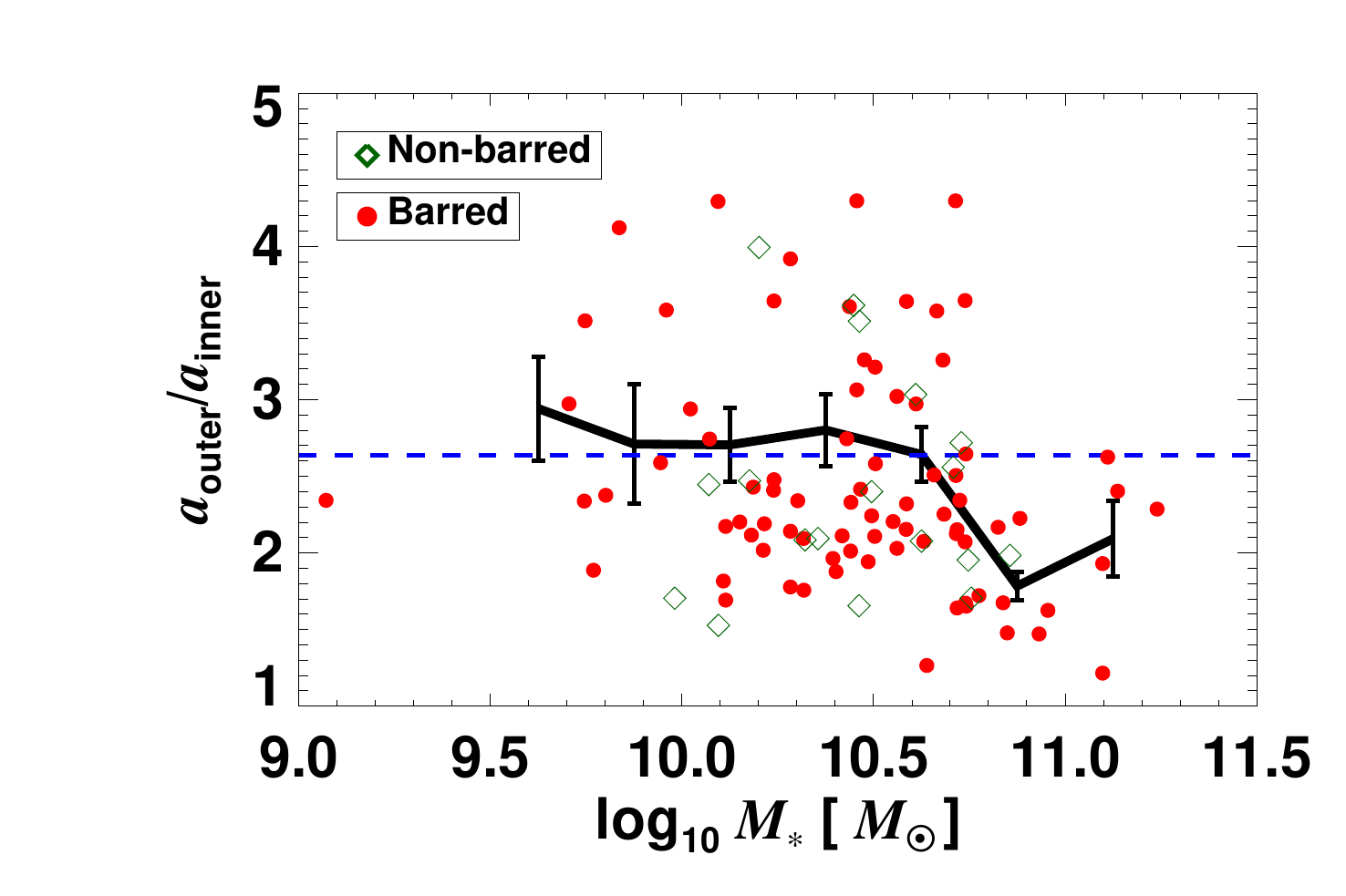}
\caption{
Ratio of the de-projected semi-major axes of outer and inner rings versus the total stellar mass of the host galaxy, for a sub-sample of 103 galaxies. 
The horizontal dashed blue line corresponds to the expected ratio for a galaxy with a flat rotation curve 
and a linear treatment of its resonances (see text). 
With a black line we show the running mean and standard deviation of the mean, in bins of 0.25 dex. 
Barred and non-barred galaxies are shown with different colours and symbols.
}
\label{Ring_ratios_qb_a2_mass}
\end{figure}
%
%
\begin{figure}
\centering
\includegraphics[width=0.49\textwidth]{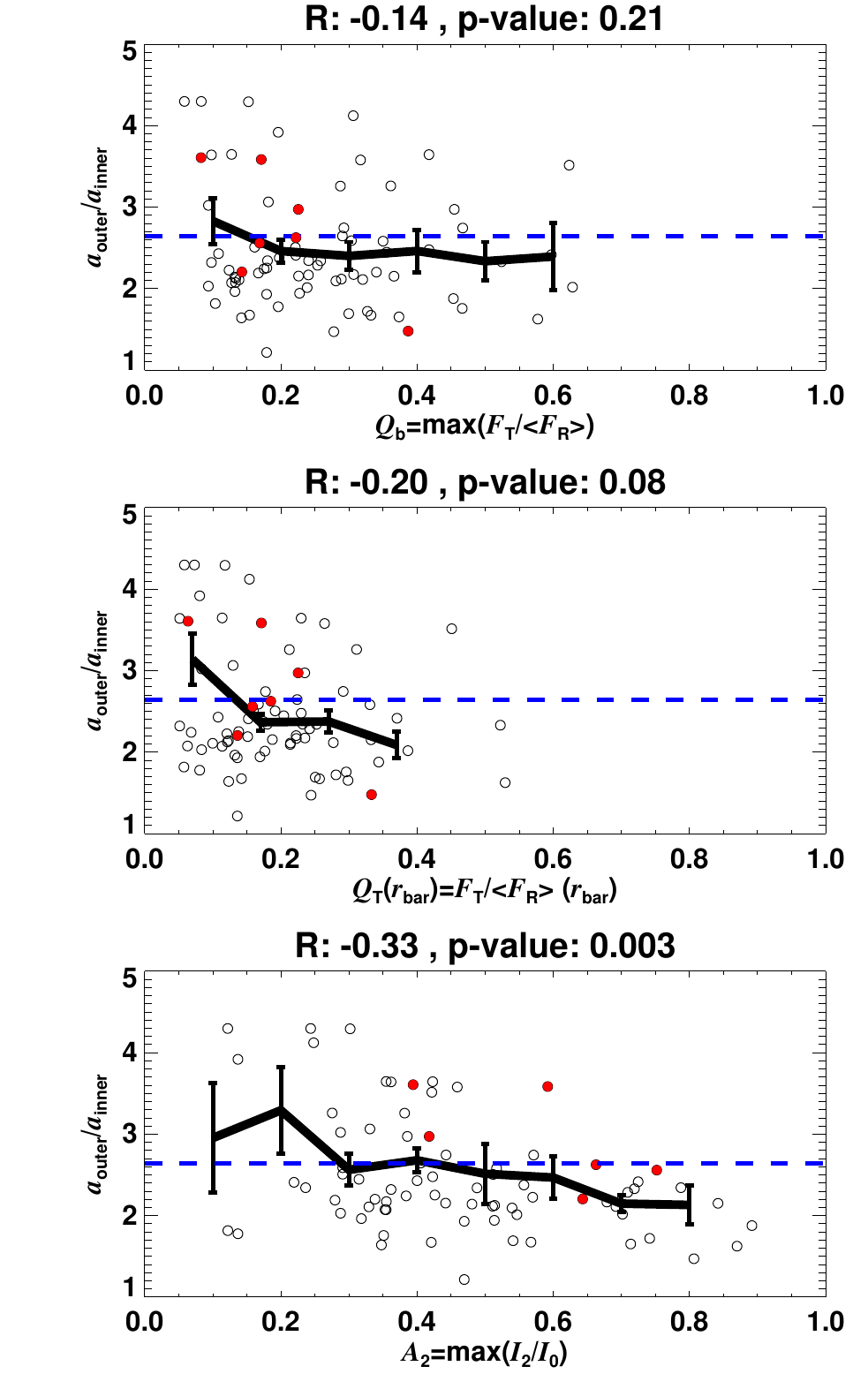}
\caption{
Ratio of the de-projected semi-major axes of outer and inner rings versus bar torque parameter (top panel), 
gravitational torques evaluated at the bar end (central panel), 
and the bar maximum m=2 Fourier amplitude (bottom panel), shown as in Fig.~\ref{Ring_ratios_qb_a2_mass}. 
The red solid circles correspond to the few galaxies in our sample 
hosting outer (pseudo-)rings of type $R_{1}$ or $R_{1}R_{2}$, according to \citet[][]{2015ApJS..217...32B}.
}
\label{Ring_ratios_qb_a2_bar_strength}
\end{figure}
%
%
\subsection{Measurements of fundamental parameters used for training self-organising maps}
%
%
We trained the SOM using the following 20 measurements for the 1320 galaxies in our sample
(in parentheses we indicate the percentage of the sample with available measurements)\footnote{We note 
that SOM allows the data matrix to contain unknown values that are excluded from the distance calculations.
}:

\begin{enumerate}
%
%
\item $M_{\ast}$ ($99.3\%$). Total stellar masses from \citet[][]{2015ApJS..219....3M}. 
%
%
\item $h_{\rm R}$ ($81.8\%$). Disc exponential scale length from \citet[][]{2015ApJS..219....4S} ("ok" flags).
%
%
\item $R_{25.5}$ ($99.3\%$). 25.5~mag~arcsec$^{-2}$ isophotal radius at $3.6\,\mu$m from \citet[][]{2015ApJS..219....3M}.
%
%
\item $M_{\rm halo}/M_{\ast}(<R_{\rm opt})$ ($97.0 \%$). 
First-order estimate of the halo-to-stellar mass ratio within the optical radius ($R_{\rm opt} \approx 3.2h_{\rm R}$) 
from \citet[][]{2016A&A...587A.160D,2016arXiv161101844D}\footnote{\citet[][]{2016A&A...587A.160D,2016arXiv161101844D} showed that 
the $M_{\rm halo}/M_{\ast}$ versus $M_{\ast}$ relation is in good agreement with the best-fit model at z$\approx$0 in $\Lambda$CDM 
cosmological simulations, assuming that the dark matter halo within the optical radius comprises a constant fraction ($\sim4\%$) of its total mass.}:
\begin{equation}
M_{\rm halo}/M_{\ast}(<R_{\rm opt})\approx 1.34 \cdot\bigg(\frac{(V_{\rm HI}^{\rm max})^{2}}{V_{3.6 \mu \rm m}^{2}(R_{\rm opt})}-1\bigg),
\label{halo-to-stellar-eq}
\end{equation}
where $V_{\rm HI}^{\rm max}$ refers to the inclination-corrected H{\sc\,i} velocity amplitude.
%
%
\item $M_{\rm dyn}$ ($100 \%$). Dynamical mass within the optical radius, calculated under the assumption of a spherical mass distribution:
\begin{equation}
M_{\rm dyn}(<R_{\rm opt})\approx \frac{(V_{\rm HI}^{\rm max})^{2} \cdot R_{\rm opt}}{G},
\label{m_dyn}
\end{equation}
where $G$ is the Newtonian constant of gravitation. 
%
%
\item $d_{\rm r}v_{\ast}(0)$ ($97.5 \%$). Inner gradient of stellar component of the rotation curve \citep[from][]{2016A&A...587A.160D}, 
used as a proxy of the central stellar mass concentration \citep[e.g.][]{2016MNRAS.458.1199E,2017ApJ...835..252S}. 
It was calculated from a polynomial fit to the inner part of $V_{\rm 3.6\mu m}$, following \citet[][]{2013MNRAS.433L..30L}, 
and taking the linear term as an estimate of the inner slope: $a_{1}={\rm lim}_{\rm r\rightarrow0}dv/dr=d_{\rm r}v_{\ast}(0)$. 
%
%
\item $M_{\rm HI}$ ($96.7 \%$). Atomic gas masses calculated using HyperLEDA data \citep[e.g.][]{2018MNRAS.474.5372E}:
\begin{equation}\label{fourierdec}
M_{\rm HI}=2.356 \cdot 10^5 \cdot D^2 \cdot 10^{0.4 \cdot (17.4-m21c)},
\end{equation}
where $m21c$ is the corrected 21-cm line flux in magnitude from LEDA, 
and $D$ is the distance adopted by \citet[][]{2015ApJS..219....3M}.
%
%
\item $M_{\rm HI}/M_{\ast}$ ($99.3 \%$). Cold gas fraction.
%
%
\item $\Sigma_{\rm HI}$ ($99.8 \%$). H{\sc\,i} gas surface mass density, estimated as
\begin{equation}
\Sigma_{\rm HI}=M_{\rm HI}/(\pi \cdot R_{25.5}^2).
\end{equation}
%
%
\item $M_{\rm BH}$ ($23.6 \%$). Mass of the supermassive black hole estimated from central velocity dispersion 
($\sigma_{\ast}$) taken from HyperLEDA, using the calibration from \citet[][]{2009ApJ...698..198G} 
(from Díaz-García et al. 2019, in prep).
%
%
\item $[{\rm FUV}]-[{\rm NUV}]$ ($83.0 \%$). 
Colour calculated from the \emph{GALEX} near-UV ($\lambda_{\rm eff}=1516\,\AA$) and 
far-UV ($\lambda_{\rm eff}=2267\,\AA$) magnitudes from the \emph{GALEX}/S$^4$G UV-IR catalogue \citep[][]{2018ApJS..234...18B}. 
It traces variations in recent star formation (<1 Gyr). 
%
%
\item $[{\rm FUV}]-[3.6]$ colour ($83.0 \%$) that traces the specific star formation (as does the next parameter).
%
%
\item $[{\rm NUV}]-[3.6]$ colour ($83.0 \%$).
%
%
\item SFR ($67.9 \%$). Total star formation rate (SFR), from \citet[][]{2015ApJS..219....5Q}. These were calculated from the global 
\emph{IRAS} photometry at 60 $\mu$m and 100 $\mu$m, following \citet[][]{2000A&A...354..836L}. 
%
%
\item sSFR ($67.9 \%$). Specific star formation rate: $sSFR=SFR/M_{\ast}$, where $SFR$ was derived from IRAS far-IR photometry. 
%
%
\item $\tau$ ($66.4 \%$). H{\sc\,i} gas depletion times, calculated as \citep[e.g.][]{2009ApJ...698.1437K}
\begin{equation}
\tau=\dfrac{2.3 \cdot M_{\rm HI}}{0.6 \cdot {\rm SFR}}.
\end{equation}
%
%
\item $Q$ ($72.4 \%$). Dahari parameter \citep[][]{1984AJ.....89..966D}, 
estimated as the tidal interaction strength between galaxies \citep[see also][]{2007A&A...472..121V,2019arXiv190309384W}. 
Values are taken from \citet[][]{2014MNRAS.441.1992L}, who calculated $Q$ 
from galaxies with recession velocities of $\pm 1000$ km/s in an area centred on the primary system.
%
%
\item $\Sigma_{3}^{A}$ ($72.4 \%$). Projected surface density to the third nearest neighbour galaxy from \citet[][]{2014MNRAS.441.1992L}.
%
%
\item Mean pitch angle of the spiral pattern ($29.6 \%$). Calculated as the mean of the absolute value of the pitch angle of the different spiral segments 
\citep[from Díaz-García et al. 2019, in prep, and][]{2015A&A...582A..86H}. 
%
%
\item Weighted mean pitch angle of the spiral pattern ($29.6 \%$). Average of the absolute value of the pitch angle measurements, 
weighted by the relative length of the spiral segments (from Díaz-García et al. 2019, in prep).
%
%
\end{enumerate}
%
%
We need to normalise these variables to avoid any of them having an overwhelming influence on the training. 
We use a continuous histogram equalisation \citep["histC", see Sect~4.2.4 in][]{{2000Vesanto}}, 
which linearly scales the pre-ordered values (ranking) in bins of almost equal size, so that they eventually lie within $[0,1]$
\footnote{The components defined in log scale (11-12-13-17-18) were linearised 
(e.g. ratio of FUV and NUV fluxes instead of colours derived from magnitudes) 
before normalisation and initialisation of the SOM.
}. 
Other normalisation methods were also tested, yielding similar results. 
We confirmed that reducing the number of measurements used as an input for SOM hardly changes the results, 
especially when such measurements are similar or redundant (e.g. parameters 2-3, or 12-13, or 19-20) and 
could potentially induce an over-weighting of certain properties (but we note that using several proxies of the same property increases accuracy).
%
%
\subsection{Analysis of self-organising maps}\label{SOM_analysis}
%
%
%
\begin{figure*}
\centering
\begin{tabular}{c c c c c c c c}
\includegraphics[width=0.255\textwidth]{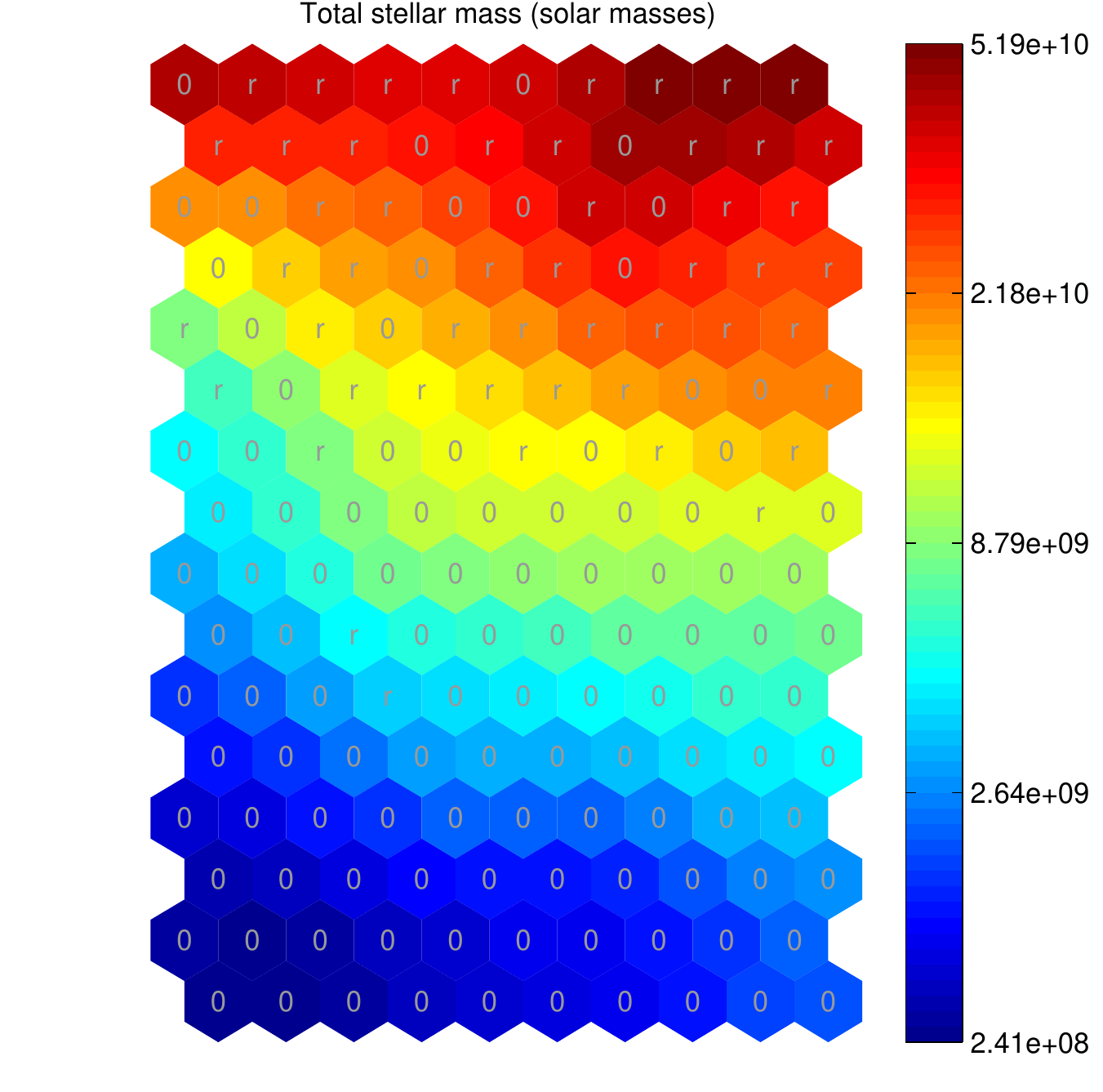}
\includegraphics[width=0.245\textwidth]{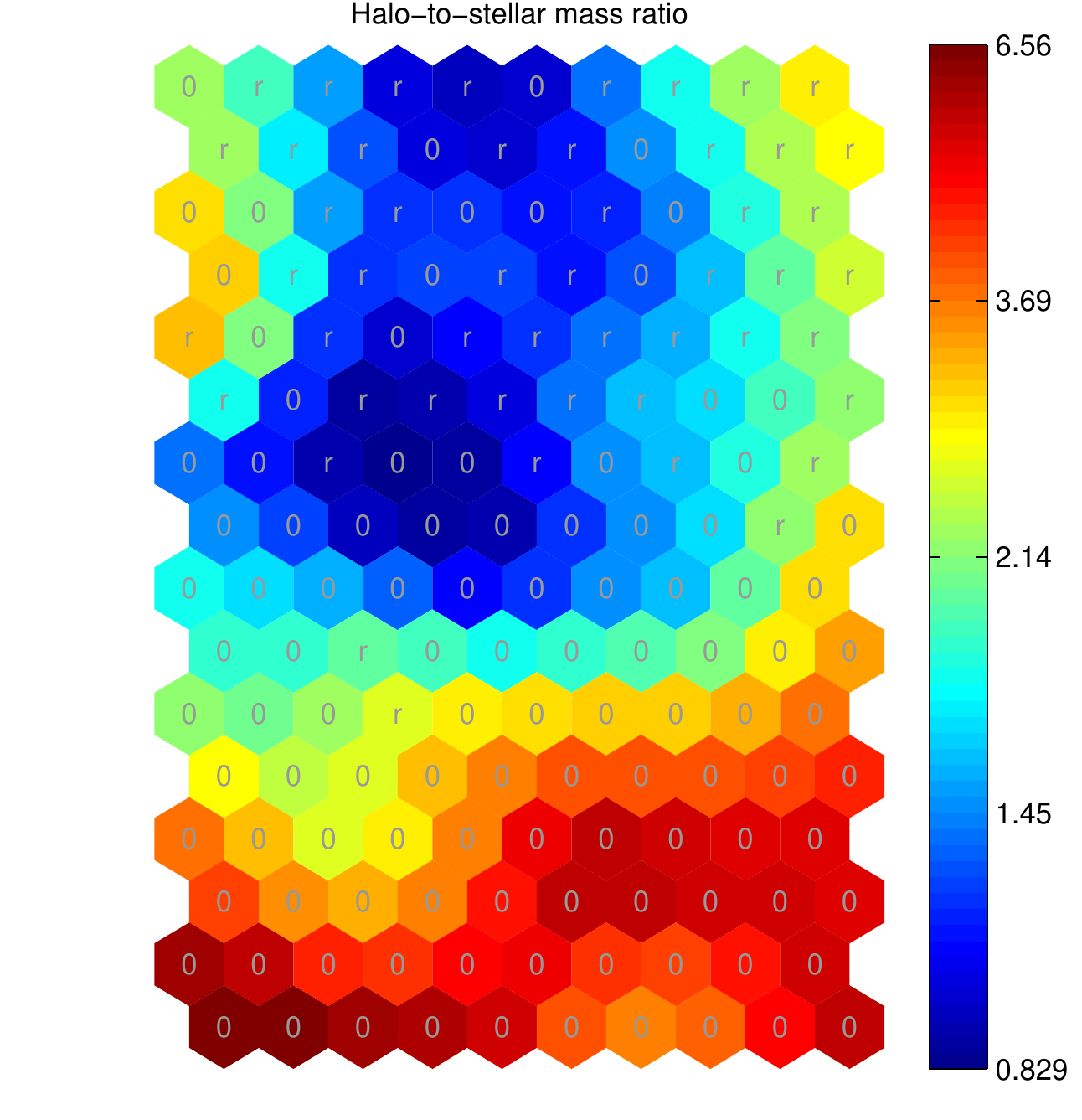}
\includegraphics[width=0.245\textwidth]{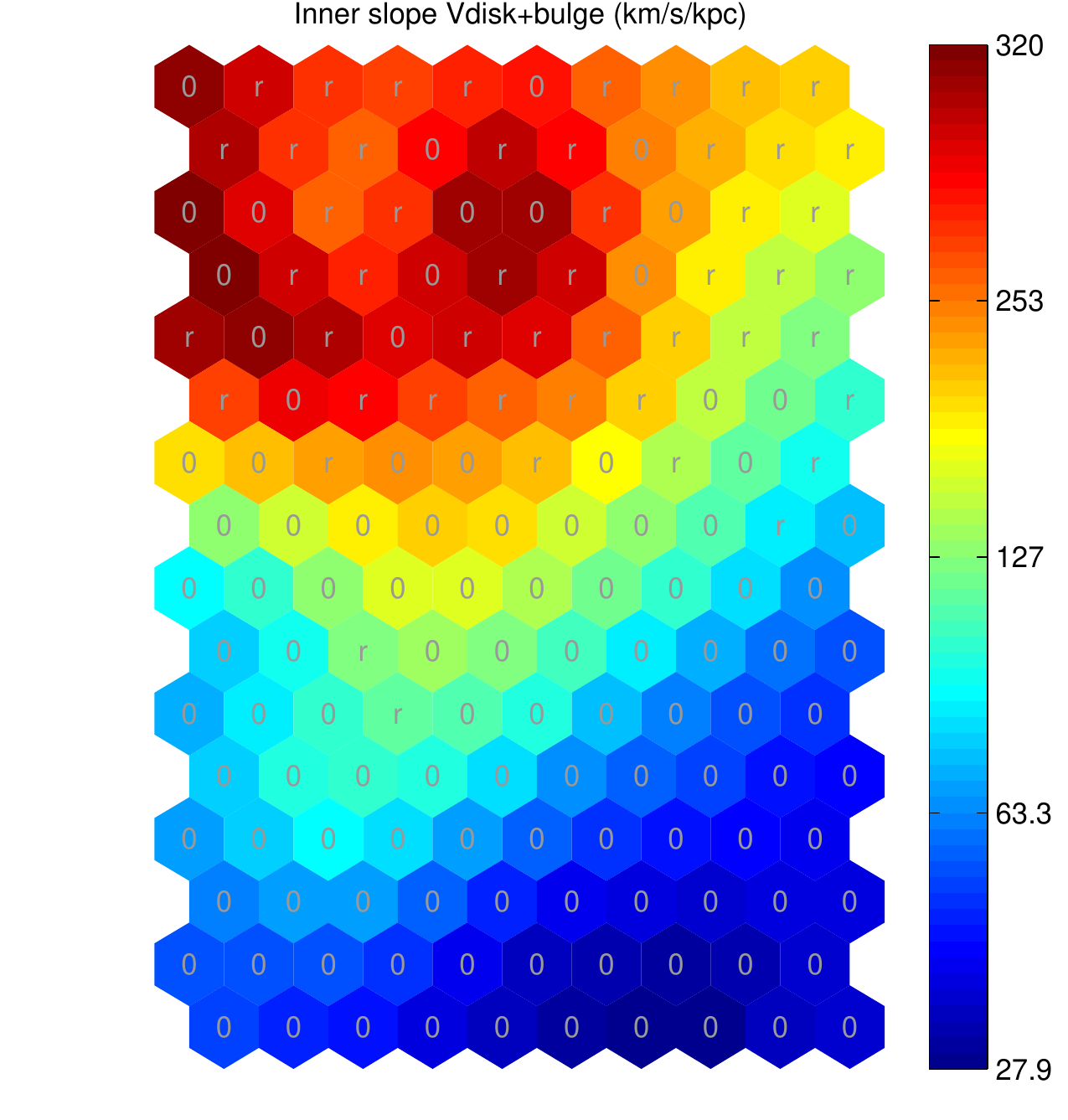}
\includegraphics[width=0.255\textwidth]{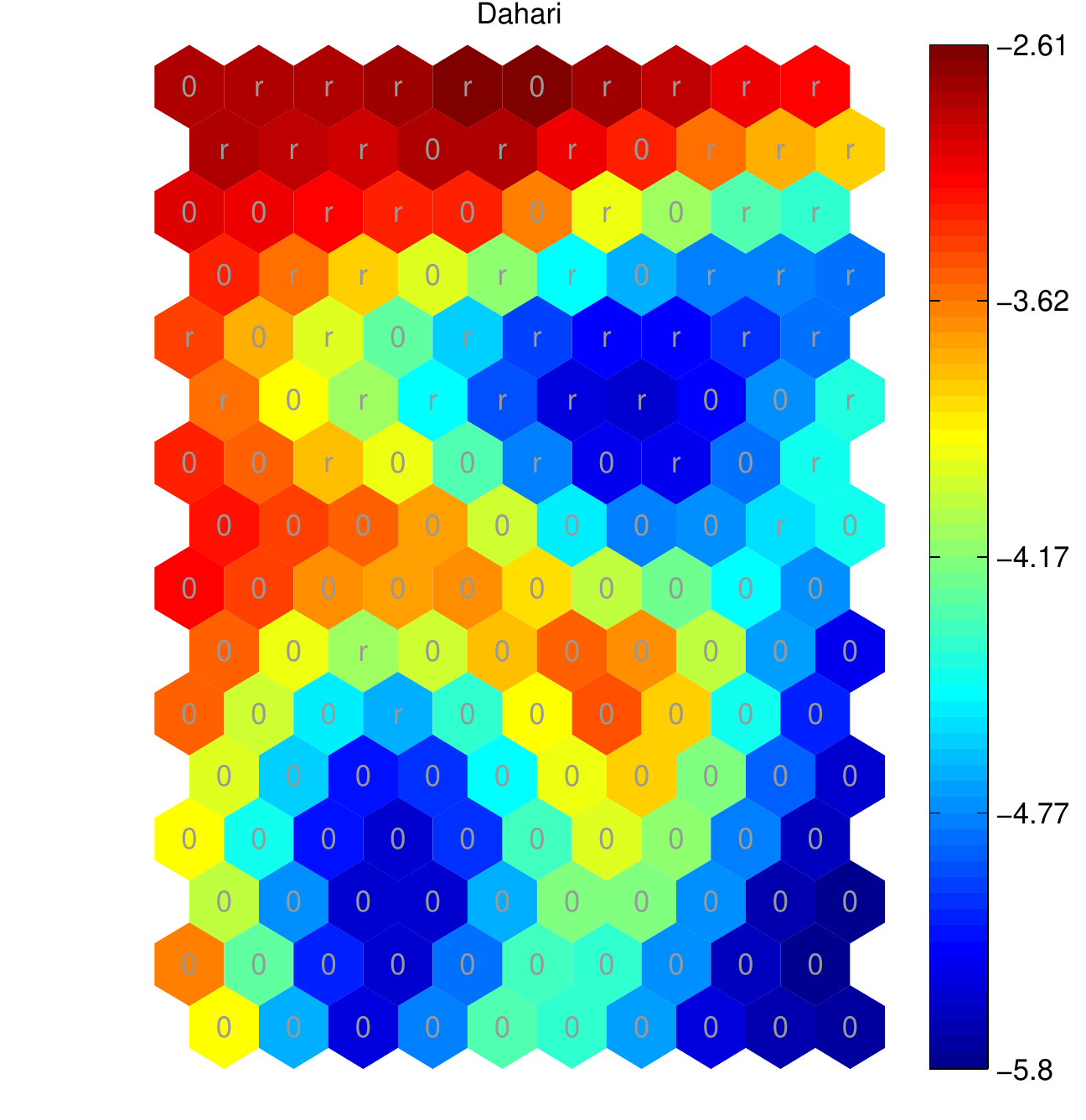}\\
\includegraphics[width=0.25\textwidth]{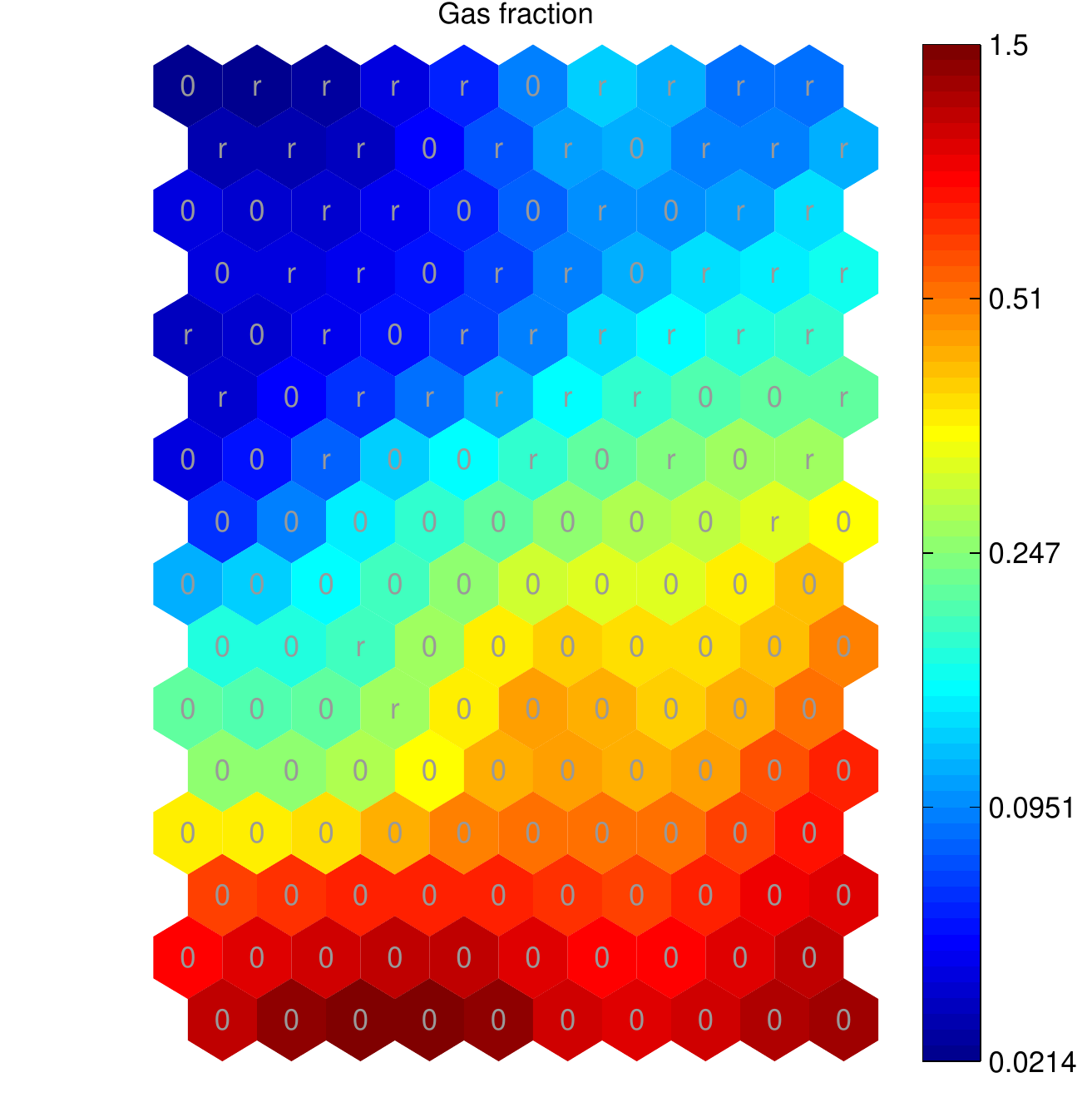}
\includegraphics[width=0.245\textwidth]{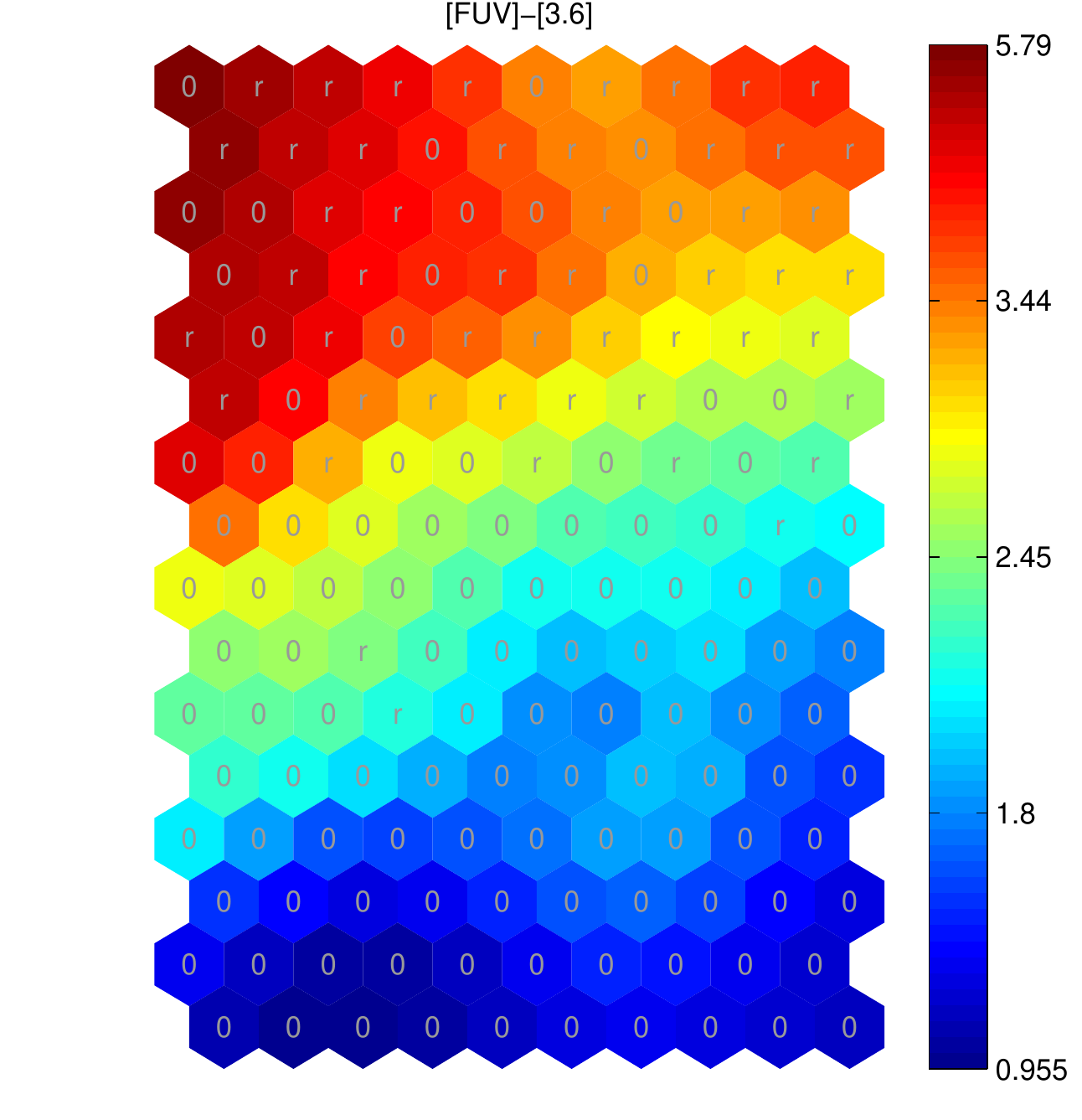}
\includegraphics[width=0.245\textwidth]{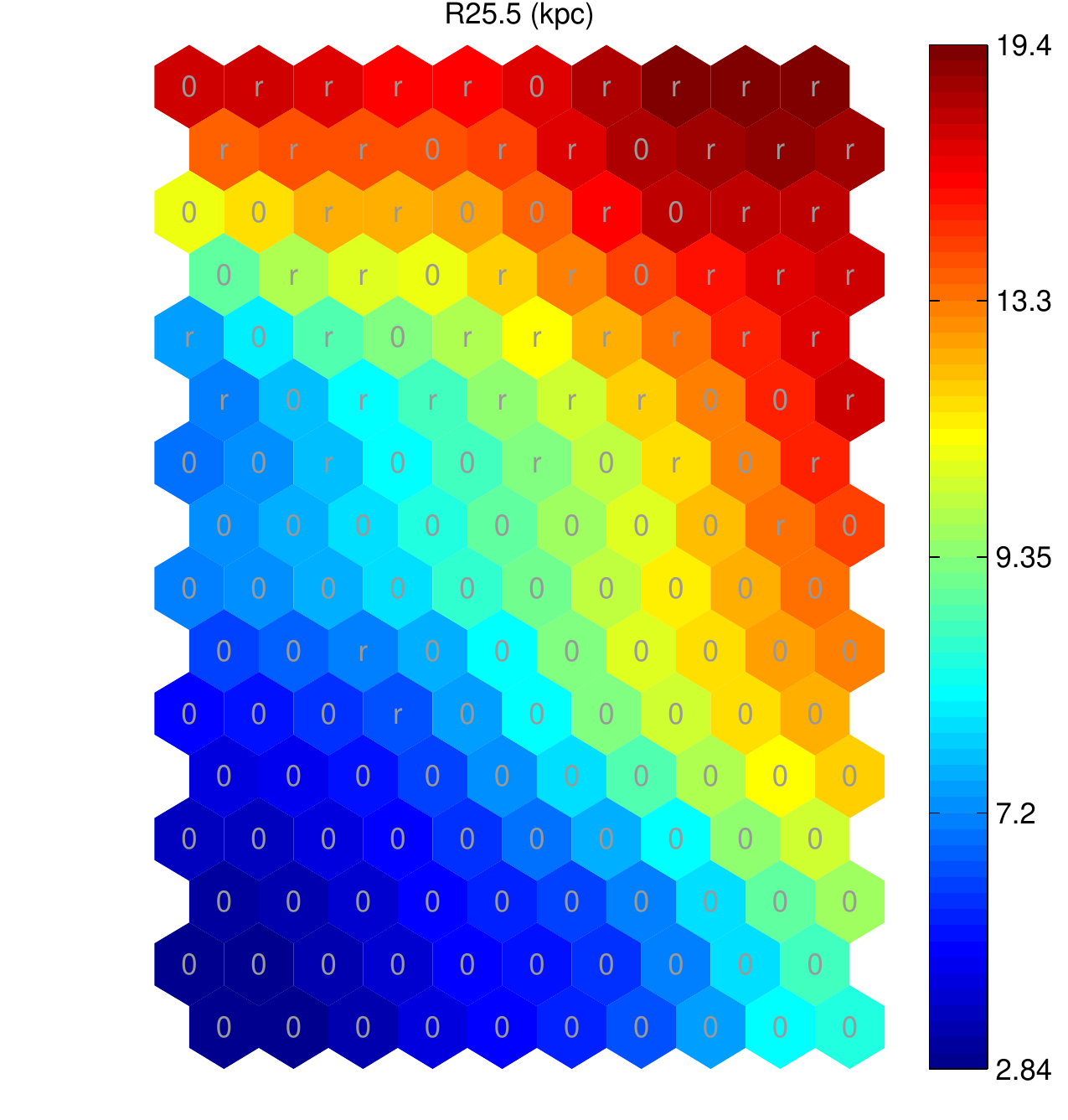}
\includegraphics[width=0.255\textwidth]{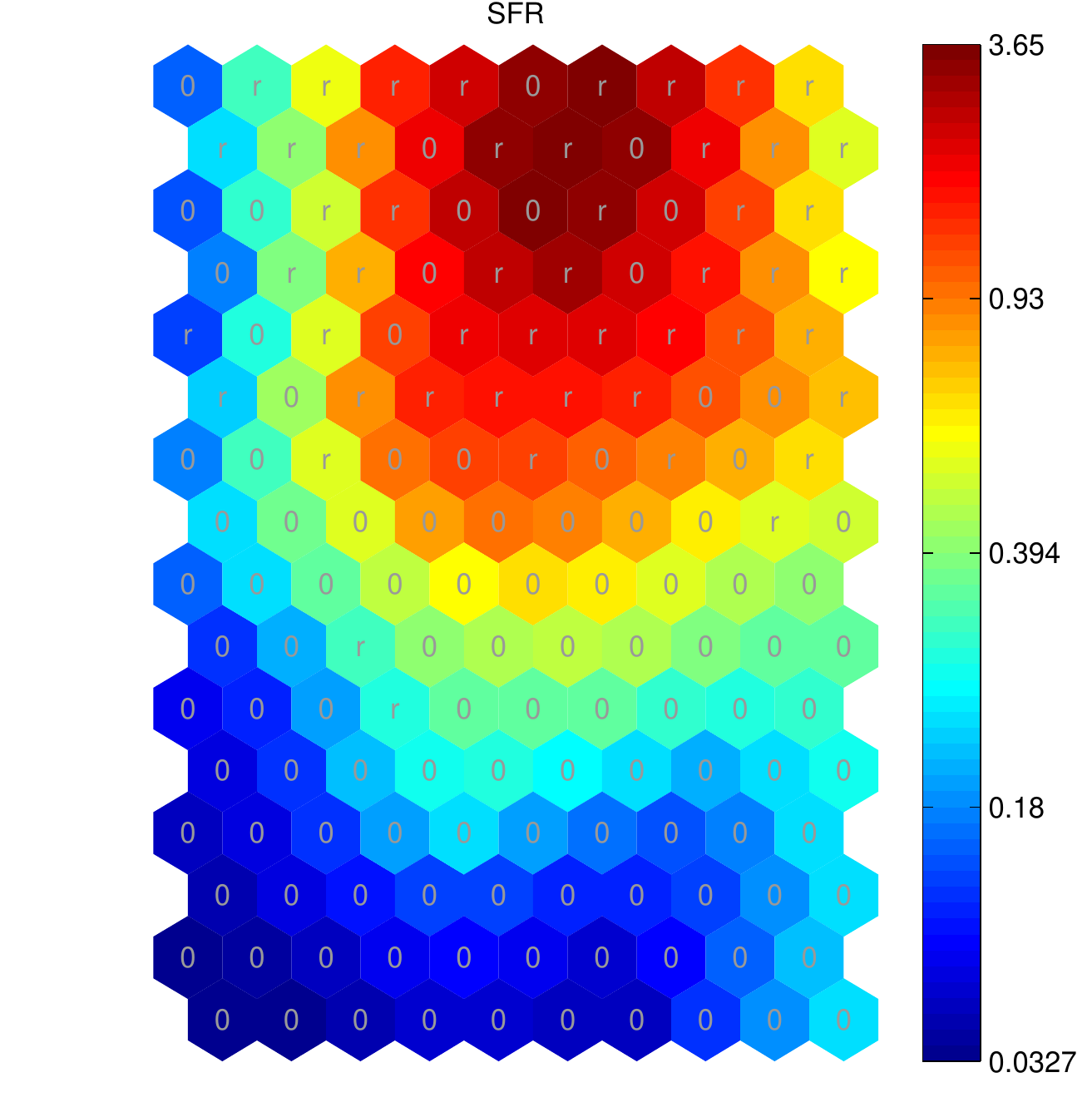}
\end{tabular}
\caption{
Component planes after training a self-organising map \citep[][]{2001som..book.....K} using the Toolbox package \citep[][]{2000Vesanto} 
(see text for details on the method and the data set used for the training). 
A sample of 1320 non-highly inclined ($i<65^{\circ}$) disc galaxies, with and without inner rings, has been used. 
Each hexagon corresponds to a neuron of the SOM, and the colour palette indicates the value of the prototype associated to a given measurement 
(see the vertical bar); the values of the different components have been de-normalised to the original range. 
The component names are indicated in the titles of the subplots. 
We show, from left to right and top to bottom, the total stellar mass, halo-to-stellar mass ratio within the optical radius, 
inner slope of the disc+bulge contribution to the circular velocity, 
Dahari parameter, cold gas fraction, [FUV]-[3.6] colour, disc size estimated from $R_{25.5}$, and star formation rate. 
We overplot the labels of each map unit with a letter that indicates the winning class: inner ringed ("r") or not ("0"). 
This is calculated by counting the times that this neuron acts as a best matching unit of data vectors associated to inner 
ringed or non-ringed galaxies.
}
\label{ml_properties_inner}
\end{figure*}
%
%
\begin{figure*}
\centering
\begin{tabular}{c c c c}
\includegraphics[width=0.25\textwidth]{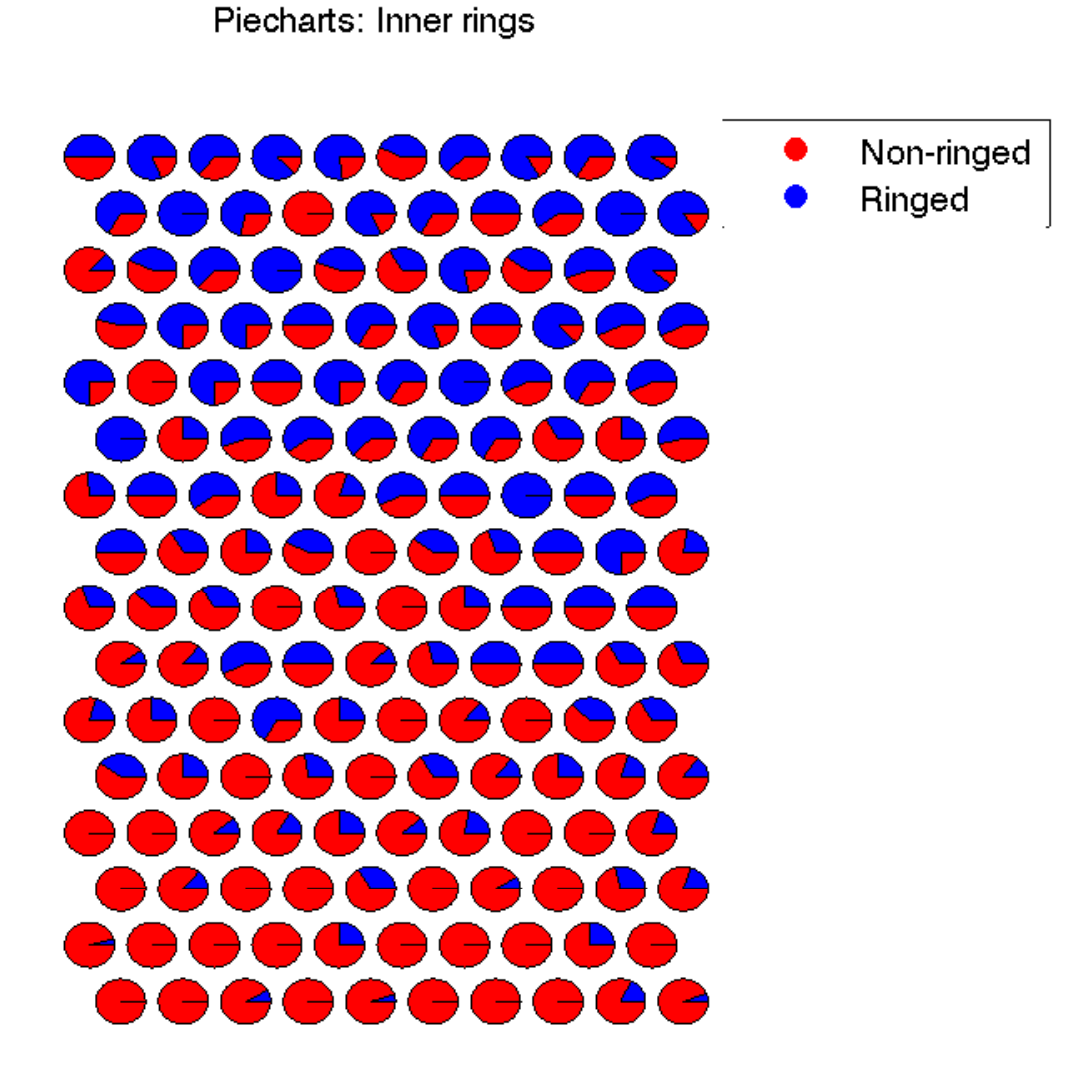}
\includegraphics[width=0.25\textwidth]{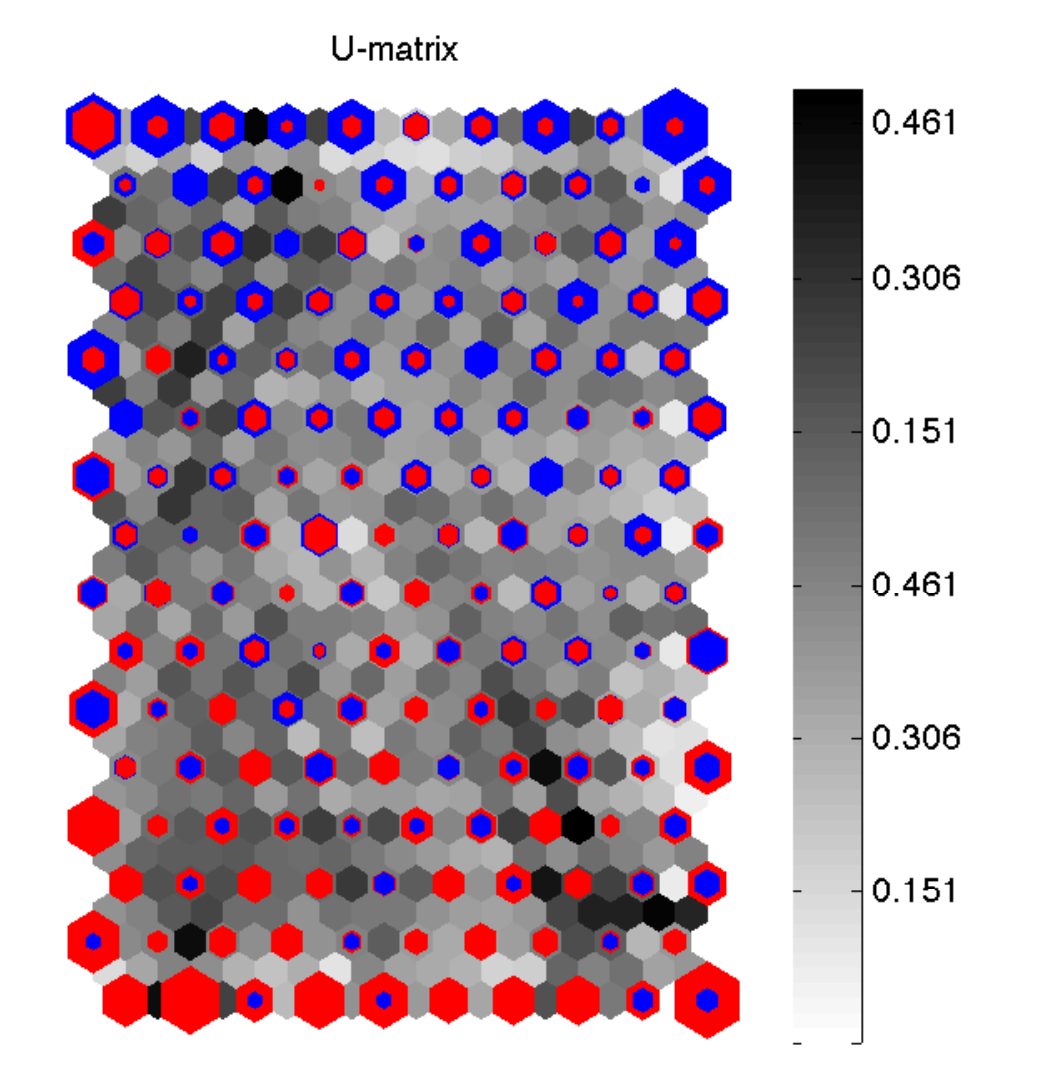}
\includegraphics[width=0.25\textwidth]{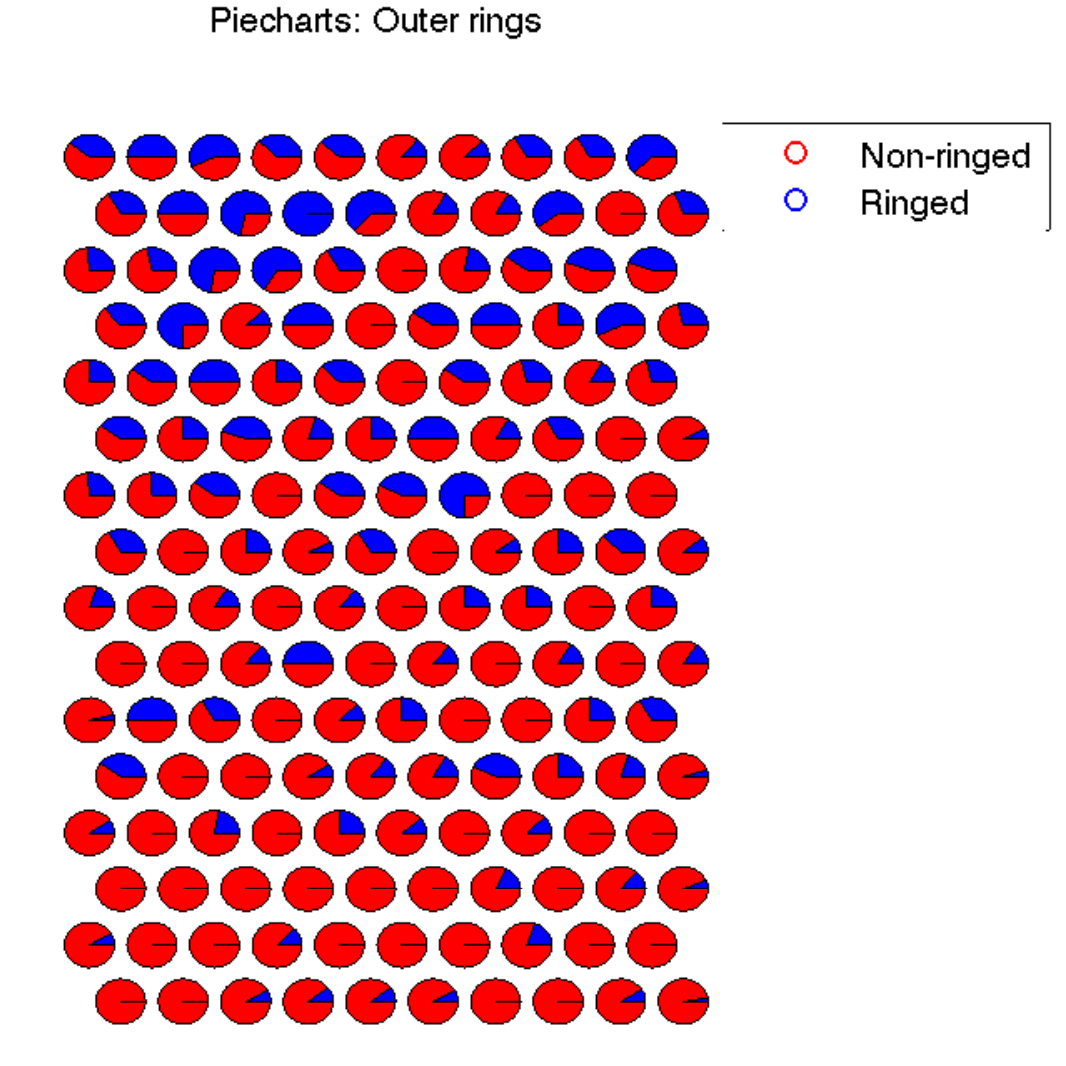}
\includegraphics[width=0.25\textwidth]{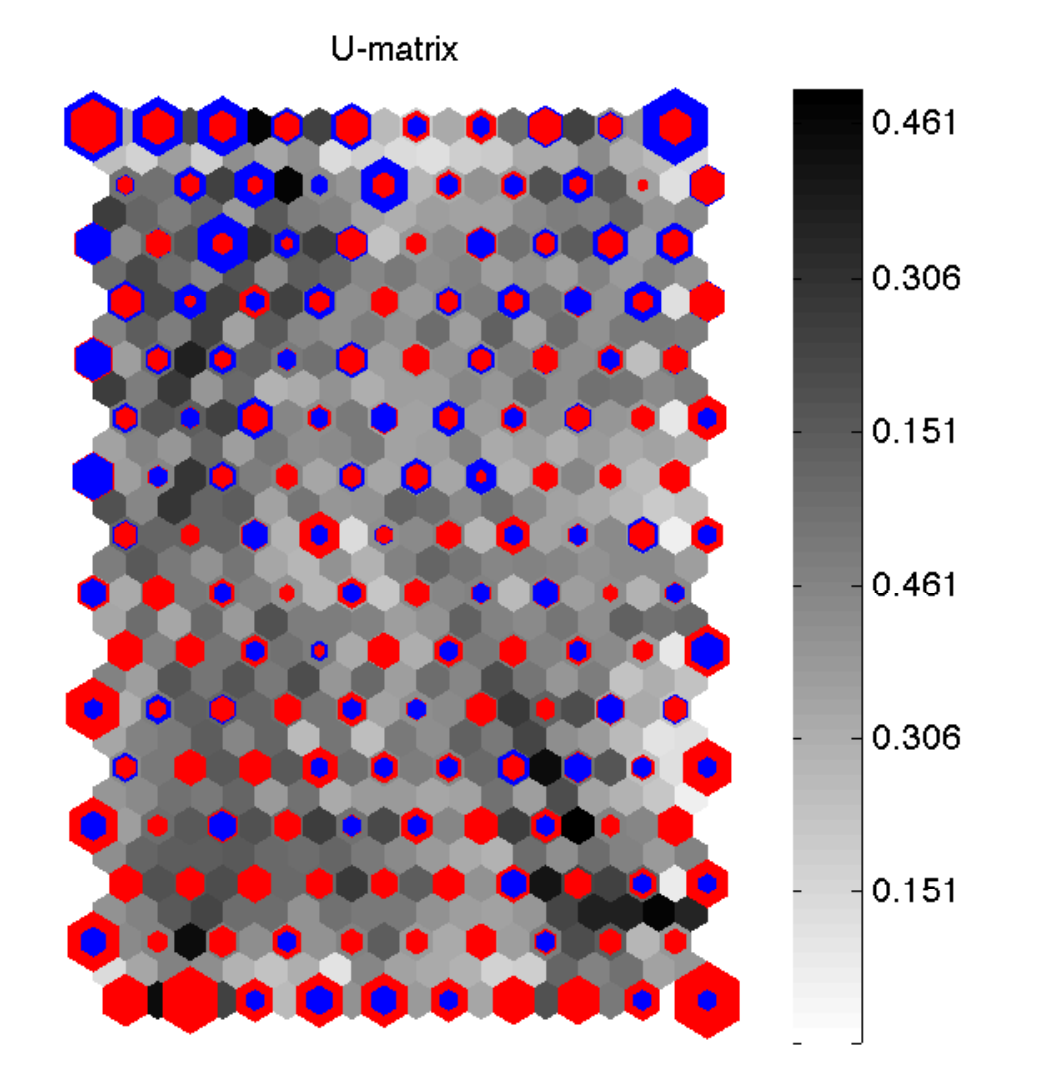}
\end{tabular}
\caption{
Pie charts showing the hit histograms corresponding to the trained SOMs presented in Fig.~\ref{ml_properties_inner} (inner rings, left) 
and Fig.~\ref{ml_properties_outer} (outer rings, right). 
These are calculated from the number of times that a neuron acts as a best matching unit (BMU) of a given data sample of certain class, 
namely ringed (blue) or non-ringed (red) (see the text for further details). 
We also show the $U-$matrix (second and fourth panels), which displays distances between neighbouring map units, shown in grey, 
with the hit histograms overplotted (coloured areas are proportional to the number of times that the neuron has acted as a BMU). 
}
\label{pie_charts}
\end{figure*}
%
%
In Fig.~\ref{ml_properties_inner} we show eight of the plane components used as input for SOM, which appear ordered after the training. 
The SOM has ordered in a way that the most massive, centrally concentrated, reddest, 
and gas-poor prototypes appear in the upper part. 
We easily detect various correlations between the different components: 
for instance, between total stellar mass and halo-to-stellar-mass ratio \citep[e.g.][]{2016A&A...587A.160D,2016arXiv161101844D}, 
gas fraction, [FUV]-[3.6] colour, or even weakly with the Dahari parameter.

Overlaid in Fig.~\ref{ml_properties_inner} are the winning classes per neuron: inner ringed ("r") or non-inner ringed ("0") galaxies. 
In Fig.~\ref{ml_properties_outer} in Appendix~\ref{app_ml} we show the same trends for the outer rings. 
In this phase, called post-labelling, we assign to each prototype the majority class of the data that it wins. 
These are determined from the hit histograms, 
which are created by increasing the counter of a map unit every time it acts as a BMU of a given data sample. 
In Fig.~\ref{pie_charts} we visualise hits with the aid of pie charts for each of the SOMs (first and third panels). 
They indicate the percentage of classes that win in terms of hits in each map unit. 
We also show the unified distance matrix or $U-$matrix (grey scale in second and fourth panels). 
It displays distances between neighbouring map units (prototypes), and therefore it has more elements than the component planes. 
The $U-$matrix helps to detect the cluster structures of the map, 
which should appear when an area of low uniform values is surrounded by high values (borders). 

There are quite a few neurons with winning class "r" in the SOM ($\sim 30 \%$), 
but they tend to be in the top-half, which mainly traces early-type galaxies. 
Ringed and non-ringed galaxies are randomly distributed within the SOM beyond a certain mass or colour, 
although the dominant class in the uppermost lines of neurons (most massive/reddest galaxies) is "r".

In the lower-half of the SOM, which corresponds to the prototypes associated to the faintest, more dark matter dominated, 
gas-rich, bluest, bulgeless galaxies (see Fig.~\ref{ml_properties_inner}), there are no neurons probing inner ringed galaxies. 
The same conclusion applies to outer rings, which are less frequent and appear only in the uppermost neurons of the SOM 
(see Fig.~\ref{ml_properties_outer}), corresponding to the most massive galaxies (e.g. S0s). 
No correlation is found between the presence of rings and the parameters probing the environment. 

We checked that the results are practically the same when the post-labelling is done for barred and non-barred galaxies 
(Fig.~\ref{morpho_type_SOM} in Appendix~\ref{ttype_ml}): neurons hitting barred galaxies as BMUs are not clustered within the SOM. 
We conclude that ringed or barred galaxies cannot be univocally distinguished using global S$^4$G parameters and a SOM clustering algorithm. 
%
%
\section{Discussion: Interplay between bars and rings}\label{Discussion}
%
%
We now discuss the formation and evolution of inner and outer rings and their hypothetical coupling with stellar bars 
-- complementing previous studies by \citet[][]{2014A&A...562A.121C} and \citet[][]{2015A&A...582A..86H} -- 
based on the analysis of the 3.6~$\mu$m images from the S$^4$G survey \citep[][]{2010PASP..122.1397S}, 
as presented in previous sections.
%
%
\subsection{Bars and the resonant origin of rings}\label{bars_spirals}
%
%
Bars, which are believed to be the dominant element in ring formation \citep[e.g.][]{1993RPPh...56..173S}, 
form spontaneously in $N-$body simulations from massive and cold galactic discs  
\citep[e.g.][]{1970ApJ...161..903M,1971ApJ...168..343H,1977ARA&A..15..437T,1982MNRAS.199.1069E,1980A&A....89..296S,1981seng.proc..111T,1984PhR...114..321A}. 
The so-called bar instability is believed to be the mechanism for the formation of the bar of the Milky Way at $z\approx1$ 
\citep[e.g.][]{2016ASSL..418..233S} \citep[but see][]{2000ASPC..197....3S}. 

Inner and outer rings most likely formed through secular evolution via the rearrangement of cold gas by 
non-axisymmetric structures \citep[e.g.][]{1996FCPh...17...95B}, 
more specifically, from the gas piled up near the 
1/4 ultraharmonic resonance \citep[e.g.][]{1984MNRAS.209...93S}, 
and near the outer Lindblad resonance\footnote{Recently, 
\citet[][]{2017MNRAS.470.3819B} identified dark gaps in barred and oval ringed galaxies and linked them to the loci of corotation. 
He concluded that $R_{1}$ and $R_{1}^{\prime}$ outer rings and pseudo-rings are likely associated with the 4:1 resonance, 
not the outer Lindblad resonance, 
while the gap method links $R_{2}^{\prime}$ outer pseudo-rings to the OLR.
}
\citep[e.g.][]{1982A&A...107..101A}, followed by episodes of star formation. 

The distribution of the sizes of inner and outer rings shown by \citet[][and references therein]{2015A&A...582A..86H} 
is consistent with their resonant origin: they tend to lie in the expected loci of disc resonances. 
We also show (Sect.~\ref{rings_ratio}) that the ratio of outer-to-inner 
ring SMA is close to the expected value in galaxies with flat rotation curves, 
when $M_{\ast} \in [10^{9.5},10^{10.75}] \cdot M_{\odot}$, under a linear treatment of their resonances. 
Nonetheless, \citet[][]{2015A&A...582A..86H} also found in their distribution 
larger rings (relative to the bar extent) than expected from a resonant origin, 
and we show that the distribution of $a_{\rm outer}/a_{\rm inner}$ is very scattered and drops in the high-$M_{\ast}$ end. 
Thus, other mechanisms might also be responsible for ring formation. 
In fact, a few outer features might be polar rings \citep[e.g.][]{1976ApJ...209..382L,1983AJ.....88..909S} 
or accretion features that have not been recognised as such: 
\citet[][]{2015ApJS..217...32B} mentioned the case of NGC$\,$4772 in the S$^{4}$G database.
%
%
\subsection{Higher ring frequency in barred galaxies than in non-barred ones}\label{ring_frac_barred_non_barred}

In Sect.~\ref{intrinsic_rings_mass} we showed that the fraction of ringed galaxies increases with increasing $M_{\ast}$ and 
peaks for total stellar masses of order $10^{10.5}-10^{11}M_{\odot}$. 
This is the mass range with the lowest values of $M_{\rm halo}/M_{\ast}(<R_{\rm opt})$, 
that is, galaxies that are not dark-matter dominated, and the characteristic stellar masses 
at which the bar fraction has its maximum in the S$^4$G \citep[][]{2016A&A...587A.160D}.

We also showed that the fraction of outer (inner) rings in 
barred galaxies is larger than in non-barred ones by a factor of $1.88 \pm 0.25$ ($1.41 \pm 0.12$). 
These numbers drop by $10-15 \%$ if a more complete and smaller volume-limited sample (distances $< 30 $ Mpc) is used 
\citep[Appendix.~\ref{app_frac}; see also Section 2.2 in][]{2018MNRAS.474.5372E}.
Our results are consistent (within the uncertainties) with the analysis in \citet[][]{2014A&A...562A.121C}, 
who used a slightly more conservative inclination cut-off ($60^{\circ}$ instead of the $65^{\circ}$ limit used in this work). 
In addition, we find that the differences increase with increasing stellar masses, especially in the case of outer features, 
which is an interesting observational constraint for numerical models. 
Overall, the larger fraction of rings amongst barred galaxies supports the role of bars in ring formation.
%
%
\subsection{Rings in non-barred galaxies}\label{rings_nonbarred}
%
%
Approximately $1/3$ ($1/4$) of the disc galaxies in the S$^4$G hosting inner (outer) rings are not barred (Sect.~\ref{intrinsic_rings_mass}). 
In addition, in Sect.~\ref{intrinsic_rings_ttype} we showed that the distributions in the Hubble sequence of 
ring intrinsic ellipticities and sizes for non-barred galaxies are nearly the same as for barred ones. 
\citet[][]{2015A&A...582A..86H} came to the same conclusions after analysing ring sizes as a function of $M_{\ast}$ and using 
the disc scale length for the normalisation. 

The existence of ringed galaxies lacking bars, and the similar distribution of ring sizes and ellipticities for barred and non-barred galaxies, 
points towards an unknown mechanism that leads some bars to dissolve into axisymmetric substructures 
\citep[e.g.][]{1979ApJ...227..714K,1991Natur.352..411R,2005MNRAS.364L..18B}.
Nevertheless, bars in simulation models are very difficult to destroy \citep[e.g.][]{2003MNRAS.341.1179A,2004ApJ...604..614S,2010ApJ...719.1470V,2013MNRAS.429.1949A}, 
unless the disc is extremely gas rich \citep[e.g.][]{2002A&A...392...83B}. 
In addition, the existence of non-barred galaxies is in itself not well understood \citep[but see][]{2018ApJ...858...24S}. 
On the other hand, ovals might be sometimes overlooked in galaxy classifications, 
but a massive oval could be just as important as a conventional bar on galactic dynamics and ring formation.

Alternatively, rings in non-barred galaxies could be created by long-lived spiral modes \citep[see e.g.][]{2000A&A...362..465R} 
or by non-axisymmetries set up by present (or past) interactions. 
As discussed in \citet[][]{2014A&A...562A.121C}, the lack of outer rings in non-barred galaxies could be due to the 
effect of galaxy interactions destroying them \citep[e.g.][]{1992A&A...257...17E,2003MNRAS.341..343B}. 
Otherwise, it could be due to the fact that the gas and stars pushed outwards by weak ovals takes longer to form outer features.
%
%
\subsection{Possible dependence of ring fraction on bar strength}
%
%
We have found that the fractions of inner and outer rings 
increase with increasing relative bar surface density amplitude ($A_{2}^{\rm max}$) (Sect.~\ref{rings_bars}). 
This could be a hint of the role of bars driving the formation of rings: 
stronger bars are more efficient at collecting matter into the 1/4 ultraharmonic and outer Lindblad resonances.

Alternatively, it could arise from the simultaneous dependence on $M_{\ast}$ of $f_{\rm ring}$ (Sect.~\ref{intrinsic_rings_mass}) 
and $A_{2}^{\rm max}$ \citep[][]{2016A&A...596A..84D,2016A&A...587A.160D}, and thus the correlation 
would not necessarily imply causation\footnote{It is possible that the $f_{\rm ring}$-$M_{\ast}$ relationship 
is more fundamental than any correlation between $f_{\rm ring}$ and bar strength. 
The latter should be addressed with even larger galaxy samples at different $z$ that allow for taking $M_{\ast}$-bins 
and probing secondary dependences without compromising the statistical significance.
}. 
On the other hand, the histograms of inner ringed galaxies peak for slightly lower $Q_{\rm b}$ values as compared to the control sample. 
This is probably a consequence of $Q_{\rm b}$ being sensitive to the 
central concentration \citep[bulge dilution, e.g.][]{2001A&A...375..761B,2002MNRAS.337.1118L} 
and to the total stellar mass of the galaxy \citep[e.g.][]{2016A&A...596A..84D,2016A&A...587A.160D}\footnote{
The slightly different trends of $f_{\rm ring}$ versus $Q_{\rm b}$ and $A_{2}^{\rm max}$ 
are likely to be due to the bimodality reported by \citet[][see their Fig.~17]{2016A&A...587A.160D} when 
these bar strength measurements were compared.
}.

Interestingly, \citet[][]{2015A&A...582A..86H} showed that inner rings are preferentially found 
in $\rm S\underline{A}B$ and non-barred galaxies, whereas outer rings occur independently of the family of the galaxy 
\citep[see also][]{1996FCPh...17...95B,2011MNRAS.418.1452L,2014A&A...562A.121C}. 
Likewise, outer resonant rings are mainly seen in SAB galaxies in the analysis by \citet[][]{2017MNRAS.471.4027B}, 
who made the classifications of 3962 ringed galaxies from the Galaxy Zoo 2 \citep[][]{2013MNRAS.435.2835W}. 
These trends would imply that weak bars are more efficient at resonance ring formation, 
which is rather counter-intuitive, or that bars tend to dissolve after the rings are assembled \citep[][]{1979ApJ...227..714K}, 
which is challenging numerical models. 
These trends are not confirmed when the bar strength is used instead of the galaxy family (Sect.~\ref{rings_bars}), 
even though we find a local maximum of $f_{\rm ring}$ for the lowest $Q_{\rm b}$ values (clearer for closed rings and ringlenses). 
We conclude that $f_{\rm ring}$ does not peak for weakly barred galaxies when quantitative measurements of the bar strength are used. 
%
%
\subsection{How non-axisymmetries control the dimensions of rings}
\subsubsection{Dependence of ring size on bar strength}
%
%
It is known that the strength and the length of the bars in disc galaxies are correlated 
\citep[e.g.][]{2011MNRAS.415.3308G,2016A&A...587A.160D}, more tightly amongst early- and intermediate-type spirals (S0/a-Sc). 
The relation is very clear for $A_2^{\rm max}$ \citep[e.g.][]{2007ApJ...670L..97E,2009MNRAS.397.1756D,2016A&A...587A.160D}, 
which is the most common proxy for bar strength used in $N$-body numerical models. 
This correlation agrees with the evolutionary track followed by bars in these simulations \citep[e.g.][]{2003MNRAS.341.1179A,2006ApJ...637..214M}, 
in which stellar bars become longer, narrower, and stronger in time as they trap particles from the disc and lose angular momentum. 
In Sect.~\ref{rings_bars} we found that the absolute and relative sizes of inner rings are correlated with bar strength. 
Given the coupling of the sizes of bars and inner rings \citep[e.g.][]{2014A&A...562A.121C}, 
this correlation could be explained by their concurrent growth as the galaxy evolves secularly. 
While the bar traps particles, it loses angular momentum and the pattern speed decreases, 
causing the 1/4 ultraharmonic resonance radius to move outwards in the disc.
However, we note that the decrease of the bar pattern speed has not been confirmed 
with observations \citep[e.g.][]{2005ApJ...631L.129R,2008MNRAS.388.1803R,2012A&A...540A.103P,2015A&A...576A.102A}, 
but see \citet[][]{2017ApJ...835..279F}.

\citet[][]{2016A&A...587A.160D} and \citet[][]{2016MNRAS.458.1199E} confirmed that the 
bars with the largest sizes and $A_2^{\rm max}$ values are typically hosted by galaxies with larger inner velocity gradients. 
These are indeed systems that evolve more quickly, given their larger central densities \citep[e.g.][]{2007ApJ...670L..97E}. 
This might explain why early-type spirals have larger rings (Sect.~\ref{intrinsic_rings_ttype} in this work) and bars 
\citep[e.g.][]{2005ApJ...626L..81E,2007MNRAS.381..401L,2016A&A...587A.160D,2017ApJ...835..279F} 
than intermediate type spirals, which are the candidates to evolve secularly \citep[][]{2013seg..book....1K}.

On the contrary, outer ring sizes are not dependent on bar strength. 
This could be due to the fact that outer resonance features might have decoupled after they formed and the bar potential changed. 
It could also simply mean that their formation is unrelated to the mixing caused by bars.
%
%
\subsubsection{Dependence of ring ellipticity on bar strength}\label{ellip_st}
%
%
Inner rings are expected to be more elongated in strongly barred galaxies than in their non-barred and weakly barred counterparts 
\citep[e.g.][]{1984MNRAS.209...93S,1999AJ....117..792S}. 
Recently, \citet[][]{2014A&A...562A.121C} found that indeed inner and outer rings are, on average, more elliptical in SB than in SA galaxies. 

\citet[][]{2010AJ....139.2465G} used a sample of 44 galaxies (26 non-barred or weakly barred, and 18 strongly barred) 
to show that the de-projected axis ratios of inner rings correlate with the tangential-to-radial forcing ($Q_{\rm T}$), 
locally estimated at the ring SMA, regardless of whether the host galaxy is barred or not. 
This would indicate that non-axisymmetries -- including bars, ovals, and also spiral arms -- are responsible for the shaping of the rings. 
Previously, \citet[][]{2002ASPC..275..185B} and \citet[][]{2007dvag.book.....B} had reported that 
galaxies presenting similar values of $Q_{\rm b}$ can show different values of $q_{\rm ring}$. 
Highly elliptical rings were mainly found for strong bars.

In this work we have taken a step forward by using a factor of approximately ten bigger sample and 
by adding ringlenses to the analysis. In Sect.~\ref{rings_bars} we confirmed the result by \citet{2010AJ....139.2465G}: 
we found a stronger correlation between the intrinsic axis ratio of rings and $Q_{\rm T}$ evaluated at the ring's SMA 
than between $q_{\rm ring}$ and $Q_{\rm b}$. However, the trends present a large scatter, 
which can be understood as a consequence of a decoupled evolution of the bar potential since ring formation. 

Interestingly, we found that the distribution of $q_{\rm ring}$ as a function of $Q_{\rm ring}$ is the same for barred and non-barred galaxies: 
this reinforces the idea that the ring shapes are, to a certain extent, controlled by 
non-asymmetries that dominate the Fourier amplitude spectrum at the ring radius, 
rather than by exclusively the bar-induced perturbation strength. 
However, we also report that inner ringed galaxies with the largest $Q_{\rm ring}$ values 
are systematically barred (only $3\%$ are non-barred): 
this indicates that large gravitational torques at $a_{\rm inner}$ can only be induced by stellar bars. 
In these galaxies in particular, rings of different ellipticities are detected regardless of the large gravitational torques.

Curiously enough, we showed that for a given $Q_{\rm ring}$ bin, ringlenses tend to be rounder than normal rings.
This may be linked to the dissolution of rings into ringlenses \citep[e.g.][]{2013A&A...555L...4C}\footnote{
Under the assumption of a resonant origin, 
\citet[][]{2013A&A...555L...4C} estimated 200 Myr to be the lower bound for the dissolution of rings. 
}. 
Ringlenses are an intermediate type between rings and lenses \citep[e.g.][]{2011MNRAS.418.1452L}.

When the estimated radial forces exerted by the dark matter halo are included in the calculations (Sect.~\ref{rings_bars_halo}), 
the trend between the ring shape and the non-axisymmetric torques ($Q_{\rm T}^{\rm halo-corr}$) becomes somewhat weaker. 
Before strong final conclusions can be drawn, this needs to be further tested with proper estimates of the dark 
matter halo profiles from decompositions of observed rotation curves for a large sample of galaxies\footnote{
For a sub-sample of galaxies, 
\citet[][]{2016A&A...587A.160D} showed that the resulting rotation curve decomposition models more or less match 
observed rotation curves in the literature \citep[from][]{2006MNRAS.367..469D,2008MNRAS.385..553D,2008AJ....136.2648D}. 
Our models indicate that only $\sim 10\%$ of the discs in our sample are maximal according to 
the criterion by \citet[][]{1997ApJ...483..103S}, and the effect of the halo dilution on the 
bar forcing is larger for fainter systems ($T\gtrsim 5$).
}, 
which is beyond the scope of this paper. 

On the other hand, if dark matter does not exist, the correction of radial forces for the halo dilution is unjustified. 
In general, secular evolution is poorly understood in the framework of the modified Newtonian dynamics \citep[MOND;][]{1983ApJ...270..365M}. 
It is interesting that simulation models of disc galaxies in a MONDian framework by \citet[][]{2008A&A...483..719T}\footnote{MONDian bars 
form very early (it takes $\sim 1$ Gyr to form a bar with $A_2=0.25$), 
and then weaken in the late stages of the run \citep[][]{2016ASSL..418..413C}. Unlike when dark matter halos are present, 
the pattern speed of these bars is not found to decline, in agreement with observational work in the literature 
\citep[e.g.][]{2005ApJ...631L.129R,2015A&A...576A.102A}.
} 
show that gas is more efficiently redistributed by bars as compared to $\Lambda$CDM \citep[see also][]{2016ASSL..418..413C}: 
this could favour ring formation.

We conclude that the amplitude of non-axisymmetries might indeed have an effect controlling the ring shape. 
However, this effect is not as strong as expected from simulations.
%
%
\subsubsection{Predictions from the manifold theory}\label{manifold}
%
%
The manifold theory \citep[e.g.][]{2006A&A...453...39R,2006MNRAS.369L..56P,2006MNRAS.373..280V,2007A&A...472...63R,2009MNRAS.394...67A} 
makes specific predictions for the intrinsic shape of rings ($R_{1}$, $R_{2}$, $R_{1}R_{2}$, and also pseudo-rings) and spirals, 
and their dependence on bar structural properties. 
\citet[][]{2009MNRAS.400.1706A} showed a relation in their models between the bar-induced perturbation strength in the corotation region, 
measured by evaluating the tangential-to-radial forces at the $L_1$ point, and the axial ratio of the outer 
$R_{1}$ and $R_{1}^{\prime}$ rings (see their Fig.~7). 
The same trend is expected for inner rings (see their Fig.~9). 
They also found that the outer ring axial ratio correlates with $Q_{\rm b}$. 

Qualitatively, the expectations from the manifold theory are in agreement with the trends discussed in the previous Sect.~\ref{ellip_st}, 
but the scatter is larger for the observations. 
Also, we note that the conclusions on outer rings by \citet[][]{2009MNRAS.400.1706A} are based on the $R_{1}$ type, 
while we  use all types of rings and pseudo-rings in our analysis. 

\citet[][]{2009MNRAS.400.1706A} also reported a positive trend between the ratio of outer-to-inner ring SMAs and 
the non-axisymmetric perturbation strength (see their Fig.~10). 
In Sect.~\ref{rings_ratio} we also assessed this relation (Fig.~\ref{Ring_ratios_qb_a2_bar_strength}), but could not confirm this prediction. 
In fact, we found a weak trend of decreasing outer-to-inner ring SMA ratio with increasing bar strength, 
which is mainly driven by the lower $a_{\rm outer}/a_{\rm inner}$ 
values found for the most massive galaxies. 
We confirmed that the same trend was maintained when using $A_2$, or when evaluating $Q_{\rm T}$ at the bar end (which is closer to the $L_1$ point).

The fraction of $R_{1}$ or $R_{1}R_{2}$ rings in our sample is not large enough to perform a statistically significant analysis 
(the likely reason for the low number of such features is the bias towards extreme late-type galaxies in the S$^4$G sample). 
However, for direct comparison with the models discussed in this section, we checked that the tendencies are the same when only these types of rings are used. 

We conclude that the manifold origin of most extragalactic rings in the S$^4$G cannot be confirmed based on the ratio outer-to-inner ring SMAs. 
To further test the manifold theory, 
we will probe the expected dependence between spiral pitch angles and bar strength \citep[][]{2009MNRAS.400.1706A} in future papers. 
Recent observational work by \citet[][]{2019MNRAS.482.5362F} does not confirm this expectation.
%
%
\subsection{Large inner rings in late-type galaxies} \label{bars_spirals}
%
%
In Sect.~\ref{intrinsic_rings_ttype} we showed that inner ring size normalised by $R_{25.5}$ increases 
with increasing T-type for $T>4$ (Sbc) on average\footnote{
We checked that this trend holds true when the ring sizes are normalised by the disc scale length ($h_{\rm R}$) 
measured by \citet[][]{2015ApJS..219....4S} \citep[see also Fig.~10 in][]{2015A&A...582A..86H}. 
Thus, this is not an artificial effect caused by the lower central extrapolated surface brightness of discs in late-type galaxies 
\citep[see e.g. Fig.~5 in][]{2016A&A...596A..84D}.
}. 
Also, bars in late-type faint galaxies ($T>5$ or $M_{\ast} \lesssim 10^{10}M_{\odot}$) 
have been found to be unexpectedly long relative to their 
underlying discs \citep[][]{2016A&A...587A.160D,2018MNRAS.474.5372E,2019MNRAS.482.5362F}. 

In addition, fainter galaxies have been found to host larger inner rings and ringlenses relative to the bar, as shown by \citet[][]{2015A&A...582A..86H}. 
They pointed out that this result is consistent with the 2-D sticky particle simulations of the weakly barred galaxy IC$\,$4214 
by \citet[][]{1999AJ....117..792S}, who showed that inner rings increased in size by $\sim 10\%$ when the dark matter halo component was enhanced. 

In fact, \citet[][]{2016A&A...587A.160D,2016arXiv161101844D} showed a remarkably higher dark halo-to-stellar mass ratio for the faint galaxies 
with $T\gtrsim 5$, which probably causes them to have distinct disc stability properties. 
However, we checked and confirmed that the disc-relative sizes of inner and outer rings are not controlled 
by $M_{\rm halo}/M_{\ast}$ (or $M_{\ast}$) (see Appendix.~\ref{app_inner}). 
Further investigation is therefore needed to understand the physics governing the observational trends for rings in late-type faint galaxies.
%
%
\subsection{No striking differences in disc global properties for ringed and/or barred galaxies}
%
%
We have applied self-organising maps in its most natural form, that is, for data analysis and visualisation (Sect.~\ref{ml_analysis}; we are not doing modelling or automated classification of galaxies). 
This is motivated by the difficulty of unravelling the driving mechanisms for the secular evolution of disc galaxies when 
multiple parameters come into play. We have trained a SOM with as many internal and external fundamental parameters as possible from the S$^4$G, 
and looked for clusters associated to known galaxy classes. Post-labelling of winning classes was done based on the presence of rings. 

SOM confirms that ringless galaxies are the dominant class for the neurons representing faint, blue, gas-rich, star-forming, 
and dark-matter-dominated objects. This is not surprising, since these units are mainly probing late-type spirals, irregular, 
and Magellanic galaxies that are known to lack rings and ringlenses \citep[e.g.][]{2014A&A...562A.121C,2015A&A...582A..86H}. 

However, for the redder or more massive systems (early-type galaxies), 
none of the used measurements unequivocally determines whether a galaxy should host an inner or an outer ring, 
including the gas fraction, [UV]-[near-IR] colour, or star formation rate. Regarding the latter parameter, 
many rings present recent star formation and host young stars \citep[e.g.][]{1993AJ....105.1344B,1995ApJ...454..623K,2005A&A...429..141K}, 
but rings lacking star formation activity have also been found \citep[e.g.][]{1991ApJ...370..130B,2013A&A...555L...4C}. 

Likewise, the parameters used for the SOM training, including the environment, do not seem to be the main driver for the formation of stellar bars. 
We checked the post-labelling of the SOM for barred and non-barred galaxies, 
finding no clear clustering. This is in agreement with recent work in the literature reporting no strong dependence of the frequency of 
bars on the properties of the discs that harbour them \citep[e.g.][]{2016A&A...587A.160D,2018MNRAS.474.5372E}. 

It is intriguing that the galaxies hosting rings or bars do not present remarkable differences (with respect to non-ringed or non-barred galaxies) 
in their global internal or external properties for $T\lesssim 5$\footnote{Discs in barred galaxies have been found to have, on average, 
larger scale lengths and fainter extrapolated surface 
brightness than their non-barred counterparts \citep[][]{2013MNRAS.432L..56S,2016A&A...596A..84D}. 
This provides possible observational evidence for the role played by bars in the radial spread of the disc 
\citep[e.g.][]{2002MNRAS.330...35A,2011A&A...527A.147M,2012MNRAS.426L..46A}, 
rather than constraints on the kind of galaxies that are prone for developing a bar.}. 
This is in conflict with the clear predictions made with the aid of $N-$body simulations (even in a cosmological framework), 
whose outputs resemble S0-Sc galaxies. According to numerical models, secular evolution (and hence resonance ring formation) 
is essentially determined by the prominence and properties\footnote{For instance, 
bar-induced formation of rings requires an efficient transfer of angular momentum, 
which can be impeded in dynamically hot discs and haloes \citep[e.g.][and references therein]{2012ApJ...758..136S}.} 
of bulges, discs, dark matter haloes, or the fraction of gas, 
as well as by the dynamical interaction between these components \citep[e.g.][and references therein]{2013seg..book..305A} or 
by the environment \citep[e.g.][]{2019MNRAS.483.2721P}. 

Naturally, the tension between observations and simulations can be a problem arising from data contamination and lack of precision on our measurements. 
On the other hand, our analysis does not include other important parameters for bar and/or ring formation 
such as the triaxiality and inner profile of the dark matter halo, the availability and flow rate of extragalactic gas, 
the radius of the corotation resonance relative to the disc radius, or the age of the pattern, 
which are hard to measure reliably for large samples. Last but not least, SOM is not necessarily the only or the optimal ML algorithm for our purposes: 
we encourage a similar analysis to be done elsewhere by applying supervised machine learning algorithms (e.g. random forest) to S$^4$G data.

We have shown that SOMs are a powerful tool for the analysis of the S$^4$G survey 
and the visualisation of the typical disc properties that characterise ringed galaxies. 
In general, we do not find well-defined clusters that indicate the presence of different families of galaxies 
based on the 20 parameters used for the training. 
We conclude that there is a lack of definite observational constraints on the types of discs that are prone to ring and bar formation.
%
%
\section{Summary}\label{summarysection}
%
%
We used a sample of the 1320 not highly inclined disc galaxies ($i<65^{\circ}$) 
in the \emph{Spitzer} Survey of Stellar Structure in Galaxies \citep[S$^4$G;][]{2010PASP..122.1397S} 
to analyse the frequency of rings as a function of total stellar mass in barred and non-barred galaxies in the local Universe. 
We used the morphological classifications of galactic stellar substructures by \citet[][]{2015ApJS..217...32B}. 

We characterised the dimensions of rings (sizes and axial ratios) for $571$ galaxies using measurements from \citet[][]{2015A&A...582A..86H}, 
complementing previous analyses in the literature \citep[e.g.][]{2011MNRAS.418.1452L,2014A&A...562A.121C}. 
We studied the dependence of the properties and fraction of rings on the bar-induced perturbation strength. 
For the latter, we used measurements of gravitational tangential-to-radial torques, 
bar ellipticities, and $m=2$ Fourier density amplitudes from \citet[][]{2016A&A...587A.160D}.

Part of the statistical analysis is done with the aid of unsupervised machine learning techniques: 
we trained self-organising maps \citep[SOMs;][]{2001som..book.....K} by using the \emph{Matlab} SOM Toolbox \citep[][]{2000Vesanto}. 
Prototypes are fitted on measurements of 20 global galaxy parameters in the S$^4$G. 
This includes the total stellar mass, gas fraction, SFR, bulge prominence, [UV]-[near-IR] colours, dark matter content, 
or torques and density of nearby galaxies. 
We checked the post-labelling of the trained neural network to see if different families of galaxies are clustered.

The main results of this paper are the following:
%
%
\begin{itemize}
\item The fraction of rings increases with increasing total stellar mass (Fig.~\ref{ring_frec_vs_mass} and Tables~\ref{inner_ring_type}~and~\ref{outer_ring_type}). 
It peaks for $M_{\ast} \approx 10^{10.5}-10^{11}M_{\odot}$. 
The fraction of inner (outer) rings in barred galaxies is $1.41 \pm 0.12$ ($1.88 \pm 0.25$) times 
larger than in their non-barred counterparts. 
The differences are bigger for larger $M_{\ast}$, especially for outer features. 
These numbers go down by $10-15 \%$ if a more complete (but $\sim 50\%$ smaller) volume-limited (distances $< 30 $ Mpc, instead of $< 40 $ Mpc) 
sample is used (Fig.~\ref{ring_frec_vs_mass_volume} and Tables~\ref{inner_ring_type_volume_limited}~and~\ref{outer_ring_type_volume_limited}). 
\item We confirm previous results showing that ring sizes, relative to the disc, are larger amongst early-type spirals, 
compared to intermediate type spirals (Fig.~\ref{inner_outer_ttype}). Inner ring sizes increase with increasing Hubble stage for $T>4$.
\item The mean distributions of inner and outer ring sizes, and their de-projected axial ratios, 
are the same for barred and non-barred galaxies (Fig.~\ref{inner_outer_ttype_barred}). 
In addition, $\sim 1/3$ ($\sim 1/4$) of the disc galaxies in the S$^4$G hosting inner (outer) rings are not barred. 
These facts either rule out a strong interplay between rings and bars, 
or support a scenario in which bars dissolve \citep[e.g.][]{1979ApJ...227..714K} 
after ring formation. The latter is, however, not supported by the majority of recent simulations 
\citep[e.g.][and references therein]{2013MNRAS.429.1949A}.
\item The sizes of inner rings are correlated with bar strength (Fig.~\ref{inner_outer_qb_a2}). 
We interpret this as a consequence of the concurrent growth of bars (whose strength and length increase in time) 
and inner rings, as the 1/4 ultraharmonic resonance moves outwards in the disc while the pattern speed decreases. 
\item Outer ring sizes do not correlate with bar strength (Fig.~\ref{inner_outer_qb_a2}), 
which could be due to the larger timescales required for their formation from the gas redistributed by bars, 
whose potential might have changed in time.
\item Amongst barred galaxies, the fraction of inner and outer rings increases with the bar density amplitudes, 
possibly supporting the role of bars in ring formation (Figs.~\ref{inner_qb_a2_histo} and~\ref{outer_qb_a2_histo}). 
However, this might not imply bar-ring causality, {but rather arises} from a more fundamental dependence 
of both ring frequency and bar strength on the stellar mass of the host galaxy. 
In addition, this relation is not so clear with other proxies of bar strength. 
\item The fraction of inner rings does not peak for weakly barred galaxies 
when quantitative measurements of the bar strength are used (Fig.~\ref{inner_qb_a2_histo}), 
questioning previous results in the literature that were based on the galaxy family 
\citep[][]{2011MNRAS.418.1452L,2014A&A...562A.121C,2015A&A...582A..86H,2017MNRAS.471.4027B}.
\item The presence of inner and outer rings and/or stellar bars amongst early-type galaxies is not 
unambiguously determined by any of the parameters used to train the SOM 
(Figs.~\ref{ml_properties_inner},~\ref{pie_charts},~\ref{ml_properties_outer}, and~\ref{morpho_type_SOM}). 
We confirm, using SOM, that rings are rare in blue, faint, gas-rich, star-forming, dark-matter-dominated, and bulgeless galaxies.
\item On average, the intrinsic ellipticity of rings increases with increasing bar strength ($Q_{\rm b}$) (Fig.~\ref{q_rings_vs_q_bar}). 
However, the trend is weaker than expected from numerical models \citep[see e.g.][]{1993RPPh...56..173S}: 
this could be due to the different timescales that characterise the dynamical evolution of rings and bars. 
\item A stronger correlation is found between the ring axis ratio and the tangential-to-radial forces evaluated at the ring SMA, 
for both barred and non-barred galaxies, in agreement with \citet[][]{2010AJ....139.2465G} (our sample is approximately ten times bigger). 
The latter becomes somewhat weaker when the non-axisymmetric torques are corrected for the halo contribution to the radial force field 
(Fig.~\ref{q_rings_vs_q_bar_halo_corr}).
\item We do not find a positive correlation between the ratio of outer-to-inner ring semi-major axes 
and the strength of non-axisymmetries (Fig.~\ref{Ring_ratios_qb_a2_bar_strength}), 
as expected from a manifold origin of rings \citep[][]{2009MNRAS.400.1706A}. 
The distribution of $a_{\rm outer}/a_{\rm inner}$ peaks at the expected value for a resonant origin of rings in 
galaxies with flat rotation curves (Fig.~\ref{Ring_ratios_qb_a2_mass}).
\end{itemize}
%
%
%
%
\begin{acknowledgements}
We acknowledge financial support from the European Union's Horizon 2020 research and innovation programme under 
Marie Sk$\l$odowska-Curie grant agreement No 721463 to the SUNDIAL ITN network, 
and from the Spanish Ministry of Economy and Competitiveness (MINECO) under grant number AYA2016-76219-P. 
S.D.G. acknowledges the financial support from the visitor and mobility program of the Finnish Centre for Astronomy with ESO (FINCA), 
funded by the Academy of Finland grant nr 306531. 
J.H.K. acknowledges support from the Fundación BBVA under its 2017 programme of assistance to scientific research groups, 
for the project "Using machine-learning techniques to drag galaxies from the noise in deep imaging", 
and from the Leverhulme Trust through the award of a Visiting Professorship at LJMU. 
H.S. acknowledges financial support from the Academy of Finland (grant no: 297738). 
We thank the S$^{4}$G team for their work on the different pipelines and the data release. 
We thank Sébastien Comerón and Martín Herrera-Endoqui for valuable comments on the manuscript, 
and Aleke Nolte for useful discussions. 
We thank Alexandre Bouquin for providing us with the tabulated total integrated \emph{GALEX} FUV and NUV magnitudes used in this work. 
We thank the organisers of the SUNDIAL Training School "Applications of Computer Science Techniques in Galaxy Science" (February 2018, Groningen) 
for encouraging, and illustrating with examples, the application of SOM to S$^4$G data. 
We thank the anonymous referee for excellent comments that improved this paper.
{\it Facilities}: \emph{Spitzer} (IRAC).
%
%
%
%
%
%
%
%
\end{acknowledgements}
\bibliographystyle{aa}
\bibliography{bibliography}
\clearpage
%
%
%
\onecolumn
\begin{appendix}
\section{Ring fraction for a volume-limited sub-sample improved in completeness}\label{app_frac}
%
%
In this paper we have used an inclination-, magnitude-, and diameter-limited sample of 1320 galaxies (Sect.~\ref{samplesample}), 
which is not strictly complete in any quantitative sense. 
Here, we analyse again the fraction of inner and outer rings (as done in Sect.~\ref{intrinsic_rings_mass}) in a 
volume-limited sub-sample of 720 not highly inclined galaxies (i.e. $\sim 50\%$ smaller size), which is improved in completeness. 
This is done by imposing the more restrictive distance (d $<30$ Mpc) and stellar-mass ($M_{\ast}>10^{9}$ $M_{\odot}$) 
limits advocated by \citet[][]{2018MNRAS.474.5372E} (see his Fig.~1). 
By means of this volume cut, we also make sure that we do not miss any inner rings due to their smaller angular sizes amongst the most distant galaxies.

\begin{figure*}
\centering
\includegraphics[width=0.485\textwidth]{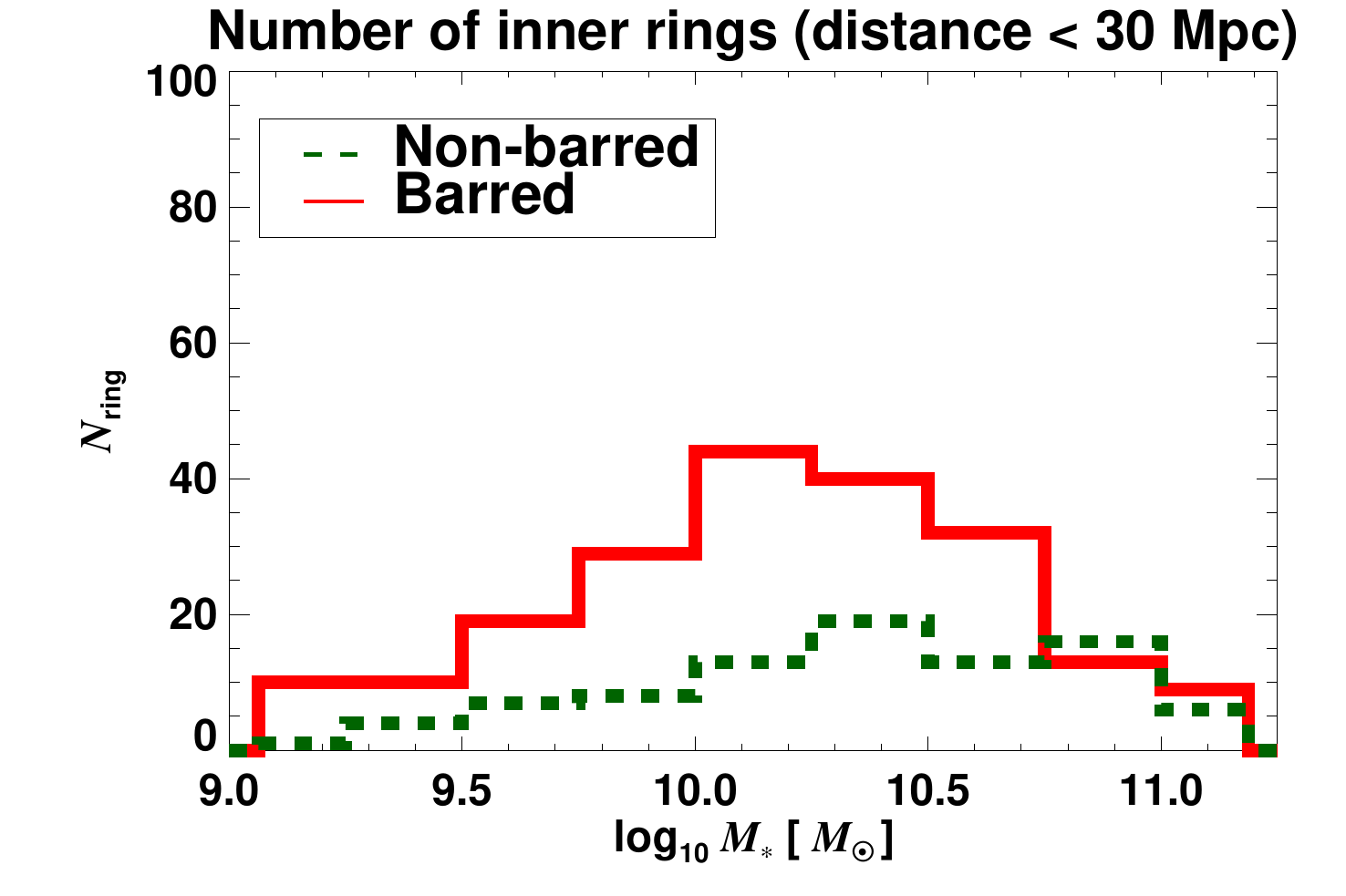}
\includegraphics[width=0.485\textwidth]{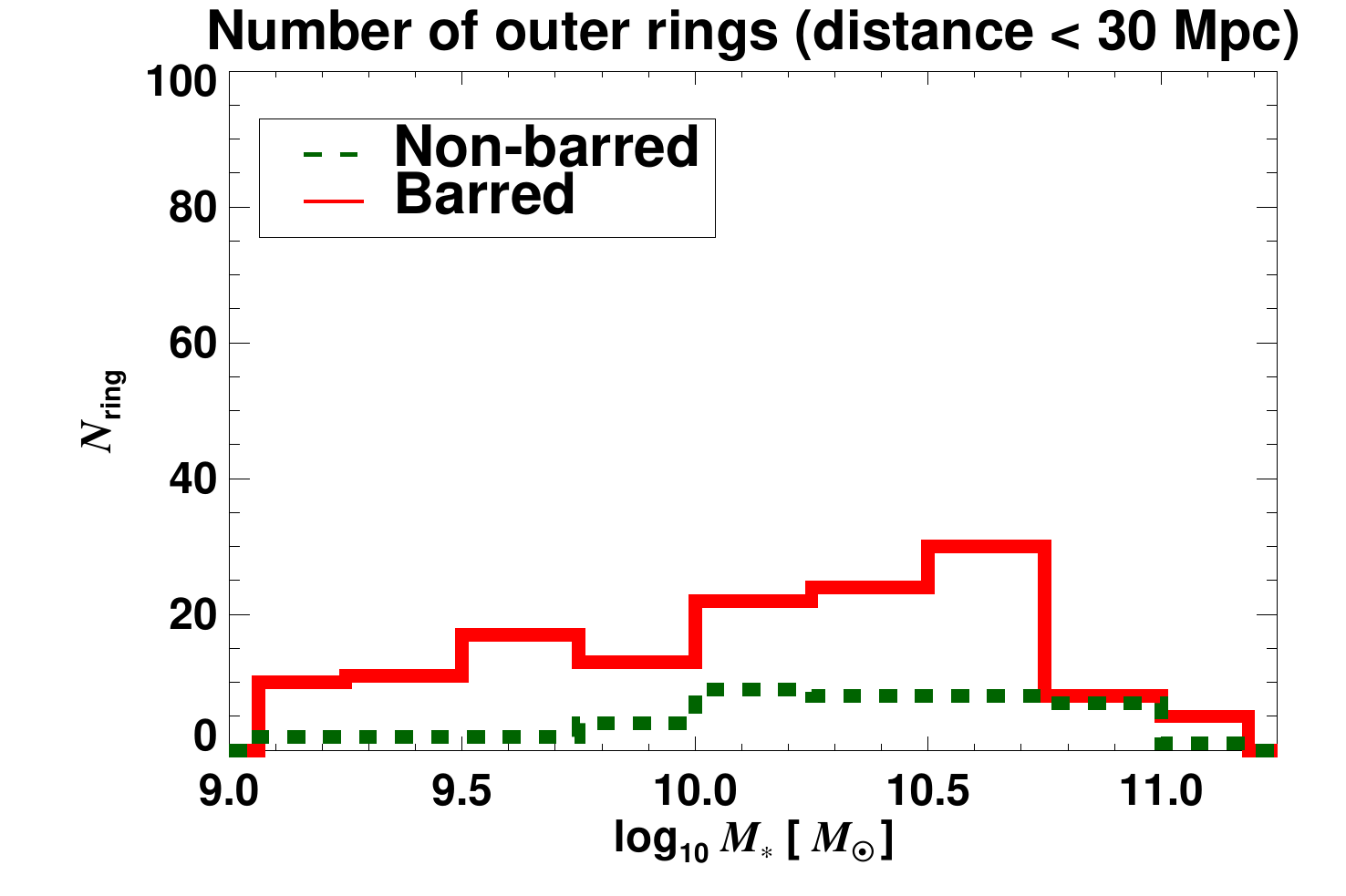}\\
\includegraphics[width=0.485\textwidth]{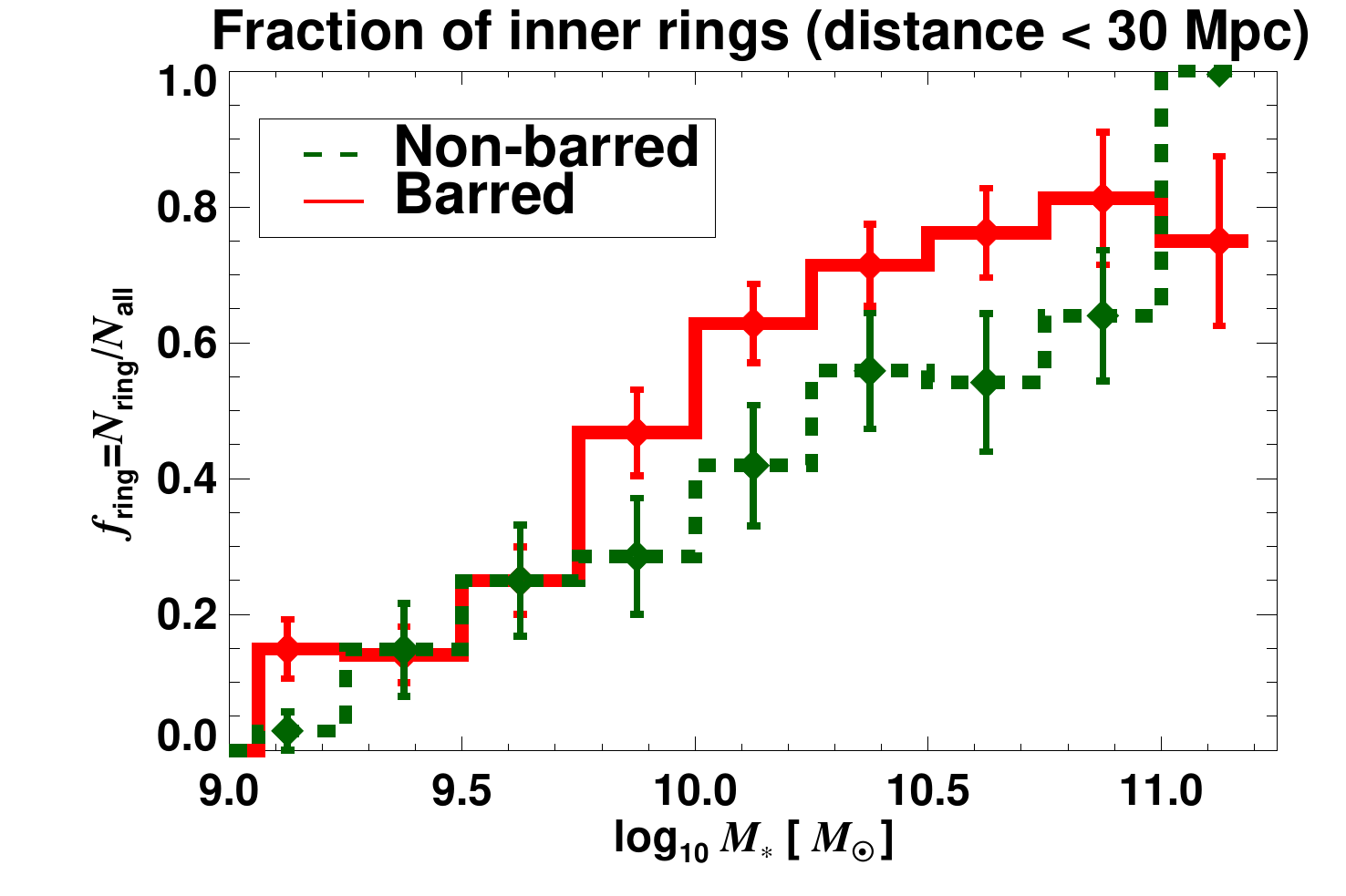}
\includegraphics[width=0.485\textwidth]{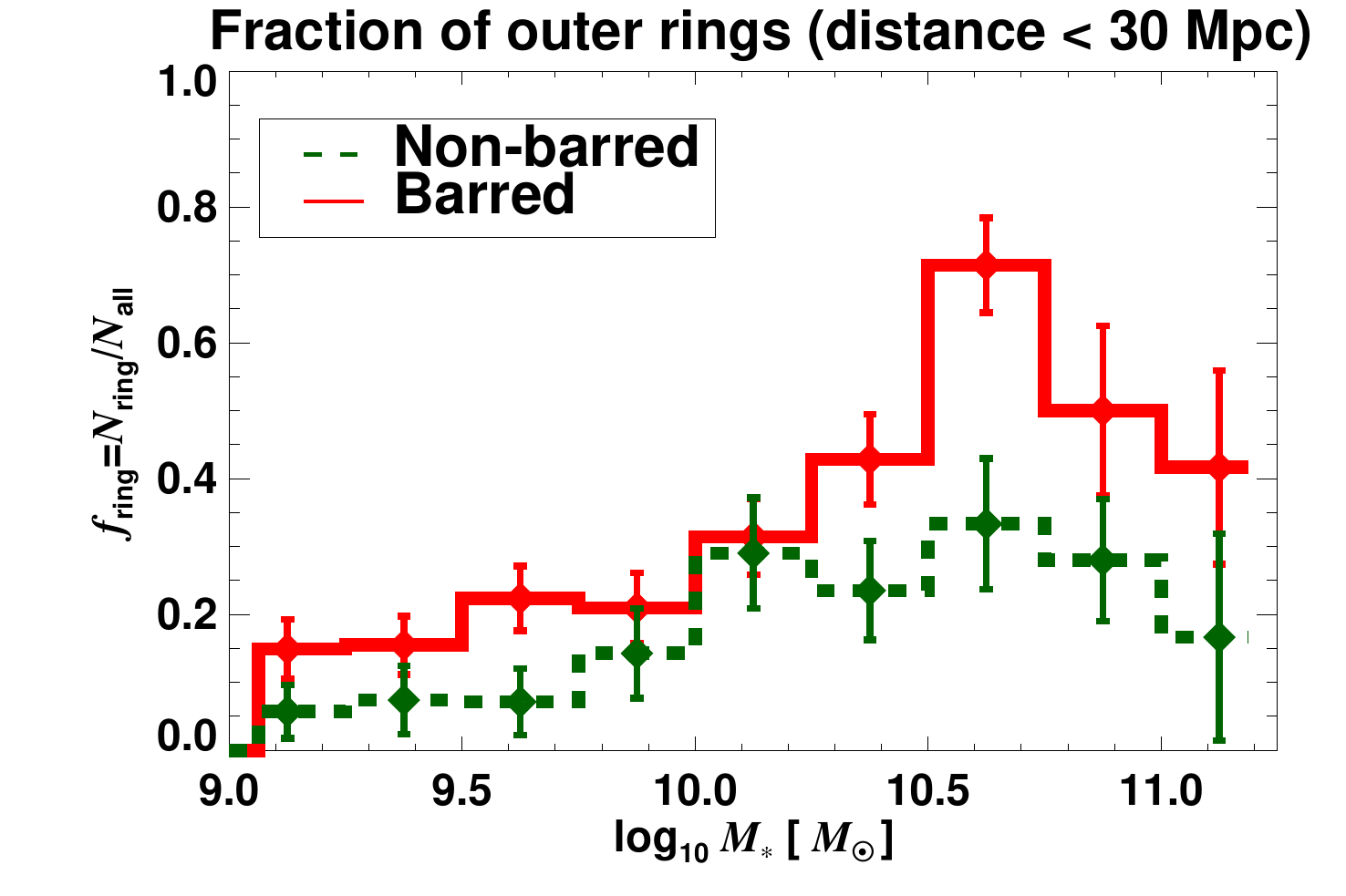}\\
\caption{
As Fig.~\ref{ring_frec_vs_mass} but for a more complete volume-limited sub-sample after imposing a distance limit d $<$ 30 Mpc.
}
\label{ring_frec_vs_mass_volume}
\end{figure*}

The trends for $f_{\rm ring}$ are practically the same as 
reported in Sect.~\ref{intrinsic_rings_mass} when a more complete volume-limited sub-sample is used (Fig.~\ref{ring_frec_vs_mass_volume}).
Ring fractions are listed in Tables~\ref{inner_ring_type_volume_limited} and~\ref{outer_ring_type_volume_limited}. 
Specifically, $18.1 \pm 2.5 \%$ of the non-barred galaxies in this new sub-sample host outer rings, in contrast to the $29.8 \pm 2.1 \%$ of barred ones. 
Barred galaxies host inner rings in $43.6 \pm 2.3\%$ of the cases, which is larger than the $36.6 \pm 3.1 \%$ of non-barred galaxies hosting them. 
Thus, the fraction of inner (outer) rings in barred galaxies is $1.19 \pm 0.12$ ($1.65 \pm 0.26$) larger than in their non-barred counterparts.
%
%
%
%
\begin{table*}
\centering
\small
\begin{tabular}{|| c| c| c| c||}
\hline
\multicolumn{1}{||c}{\centering Inner ring fraction (d < 30 Mpc):} & \multicolumn{1}{|c}{\centering All} &
\multicolumn{1}{|c}{\centering Barred} & \multicolumn{1}{|c||}{\centering Non-barred} \tabularnewline
\hline
\hline
${\rm log_{10}}(M_{\ast})\in[9,9.5)$&  12.5 $\pm$   2.3 $\%$ (200)&  14.5 $\pm$   3.0 $\%$ (138)&   8.1 $\pm$   3.5 $\%$ ( 62) \\
${\rm log_{10}}(M_{\ast})\in[9.5,10)$&  32.6 $\pm$   3.4 $\%$ (193)&  35.0 $\pm$   4.1 $\%$ (137)&  26.8 $\pm$   5.9 $\%$ ( 56) \\
${\rm log_{10}}(M_{\ast})\in[10,10.5)$&  60.7 $\pm$   3.5 $\%$ (191)&  66.7 $\pm$   4.2 $\%$ (126)&  49.2 $\pm$   6.2 $\%$ ( 65) \\
${\rm log_{10}}(M_{\ast})\in[10.5,11)$&  69.2 $\pm$   4.5 $\%$ (107)&  77.6 $\pm$   5.5 $\%$ ( 58)&  59.2 $\pm$   7.0 $\%$ ( 49) \\
${\rm log_{10}}(M_{\ast})\in[11,11.5)$&  78.9 $\pm$   9.4 $\%$ ( 19)&  69.2 $\pm$  12.8 $\%$ ( 13)& 100.0 $\pm$   0.0 $\%$ (  6) \\
\hline
\end{tabular}
\caption{
As in Table. \ref{inner_ring_type} but for a more complete volume-limited sample with distance d$< 30$ Mpc.}
\label{inner_ring_type_volume_limited}
\end{table*}

\begin{table*}
\centering
\small
\begin{tabular}{|| c| c| c| c||}
\hline
\multicolumn{1}{||c}{\centering Outer ring fraction (d < 30 Mpc):} & \multicolumn{1}{|c}{\centering All} &
\multicolumn{1}{|c}{\centering Barred} & \multicolumn{1}{|c||}{\centering Non-barred} \tabularnewline
\hline
\hline
${\rm log_{10}}(M_{\ast})\in[9,9.5)$&  12.5 $\pm$   2.3 $\%$ (200)&  15.2 $\pm$   3.1 $\%$ (138)&   6.5 $\pm$   3.1 $\%$ ( 62) \\
${\rm log_{10}}(M_{\ast})\in[9.5,10)$&  18.7 $\pm$   2.8 $\%$ (193)&  21.9 $\pm$   3.5 $\%$ (137)&  10.7 $\pm$   4.1 $\%$ ( 56) \\
${\rm log_{10}}(M_{\ast})\in[10,10.5)$&  33.0 $\pm$   3.4 $\%$ (191)&  36.5 $\pm$   4.3 $\%$ (126)&  26.2 $\pm$   5.5 $\%$ ( 65) \\
${\rm log_{10}}(M_{\ast})\in[10.5,11)$&  49.5 $\pm$   4.8 $\%$ (107)&  65.5 $\pm$   6.2 $\%$ ( 58)&  30.6 $\pm$   6.6 $\%$ ( 49) \\
${\rm log_{10}}(M_{\ast})\in[11,11.5)$&  36.8 $\pm$  11.1 $\%$ ( 19)&  46.2 $\pm$  13.8 $\%$ ( 13)&  16.7 $\pm$  15.2 $\%$ (  6) \\
\hline
\end{tabular}
\caption{
As in Table. \ref{inner_ring_type_volume_limited} but for outer rings.}
\label{outer_ring_type_volume_limited}
\end{table*}

%
%
\section{Ring dimensions as a function of galaxy mass}\label{app_inner}
\begin{figure}
\centering
\includegraphics[width=0.4\textwidth]{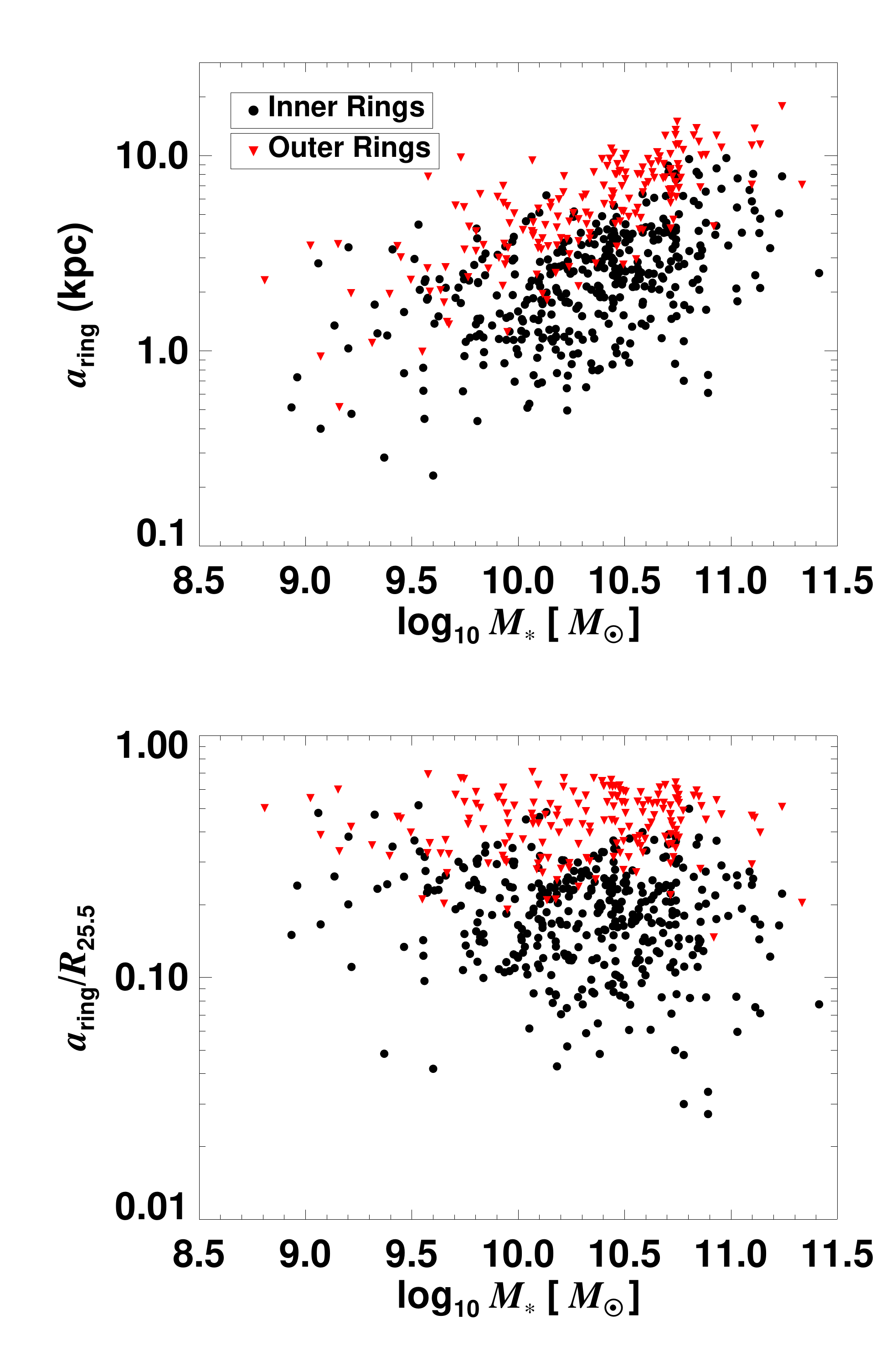}
\includegraphics[width=0.4\textwidth]{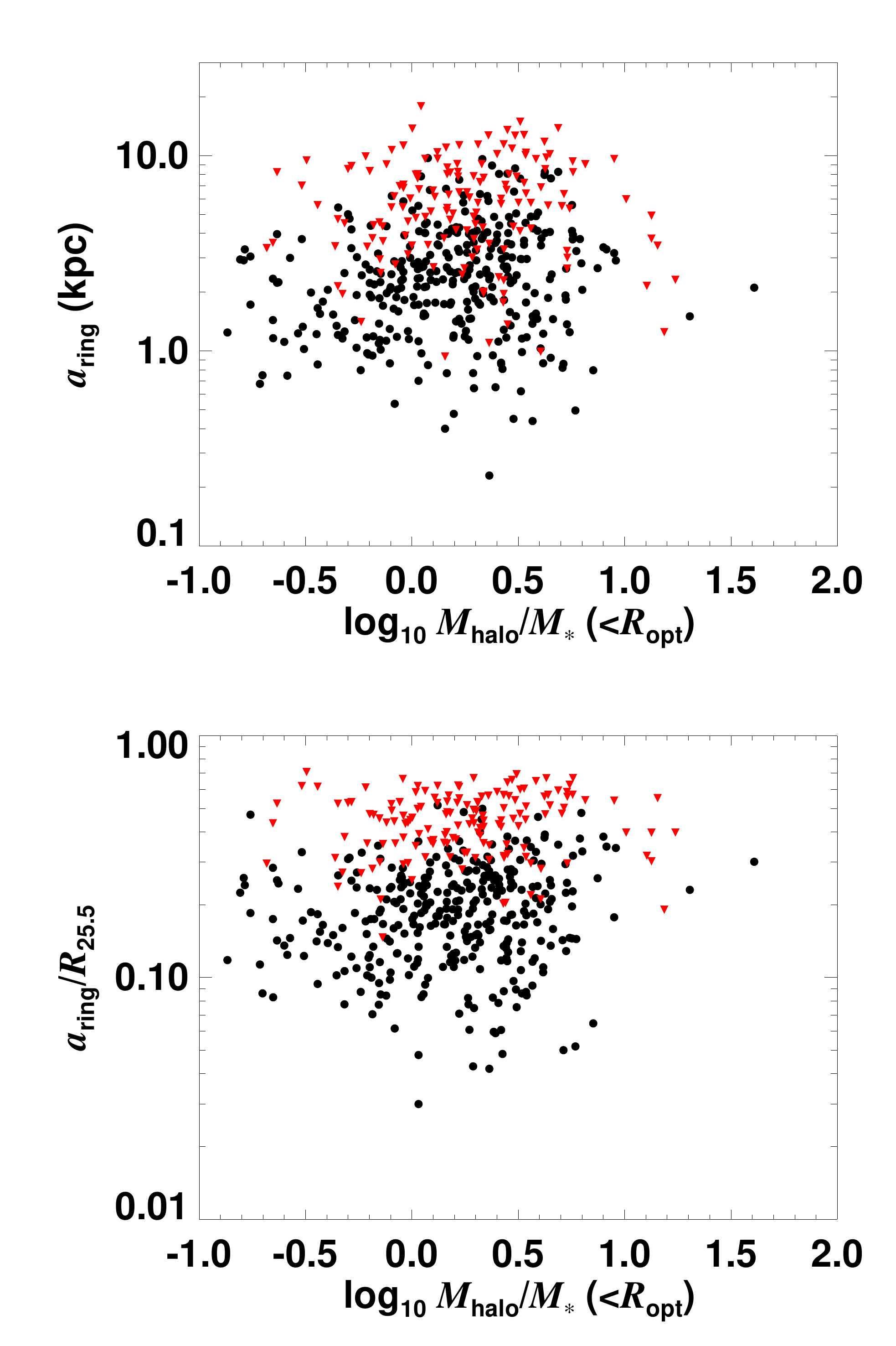}\\
\includegraphics[width=0.4\textwidth]{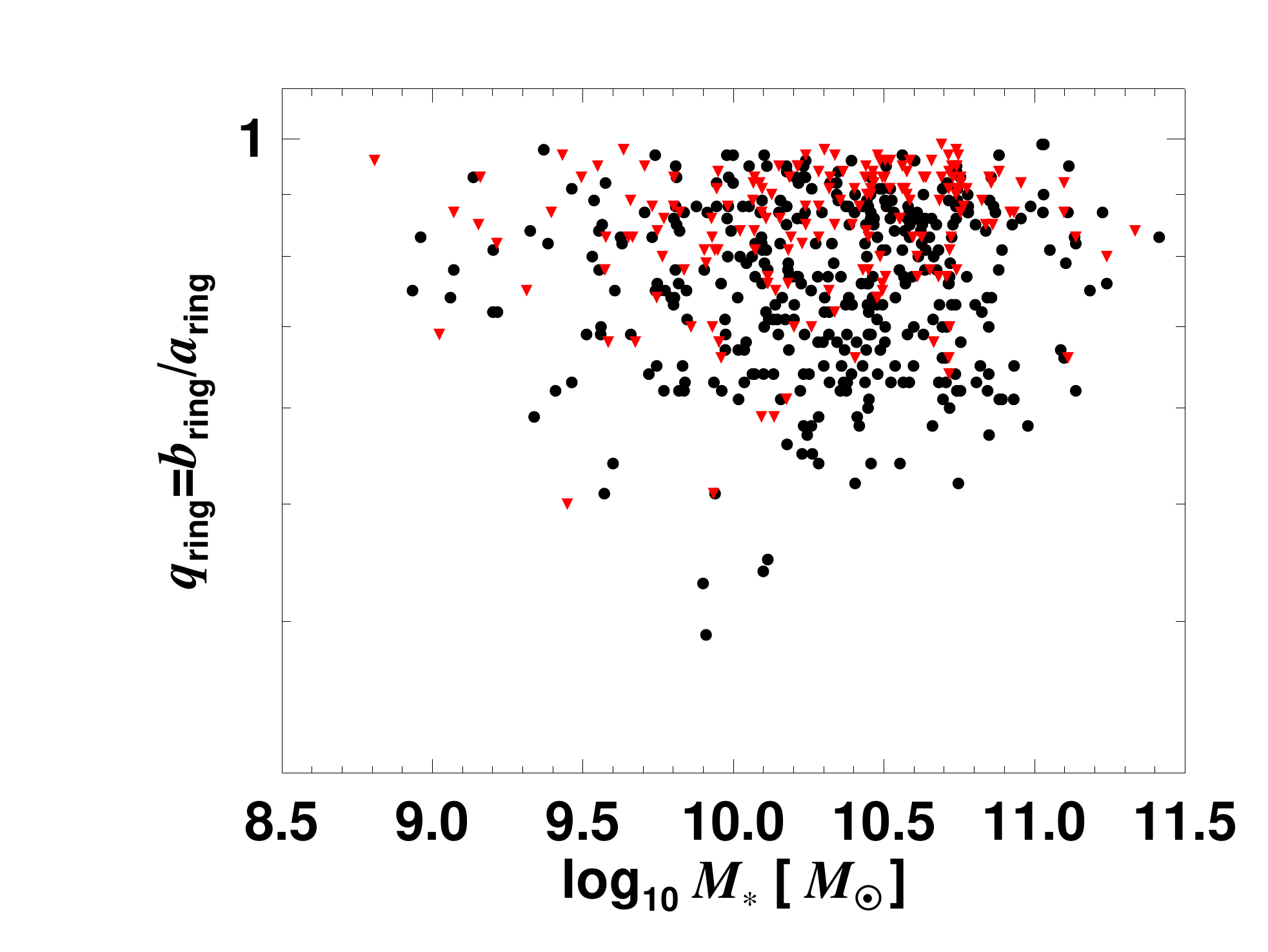}
\includegraphics[width=0.4\textwidth]{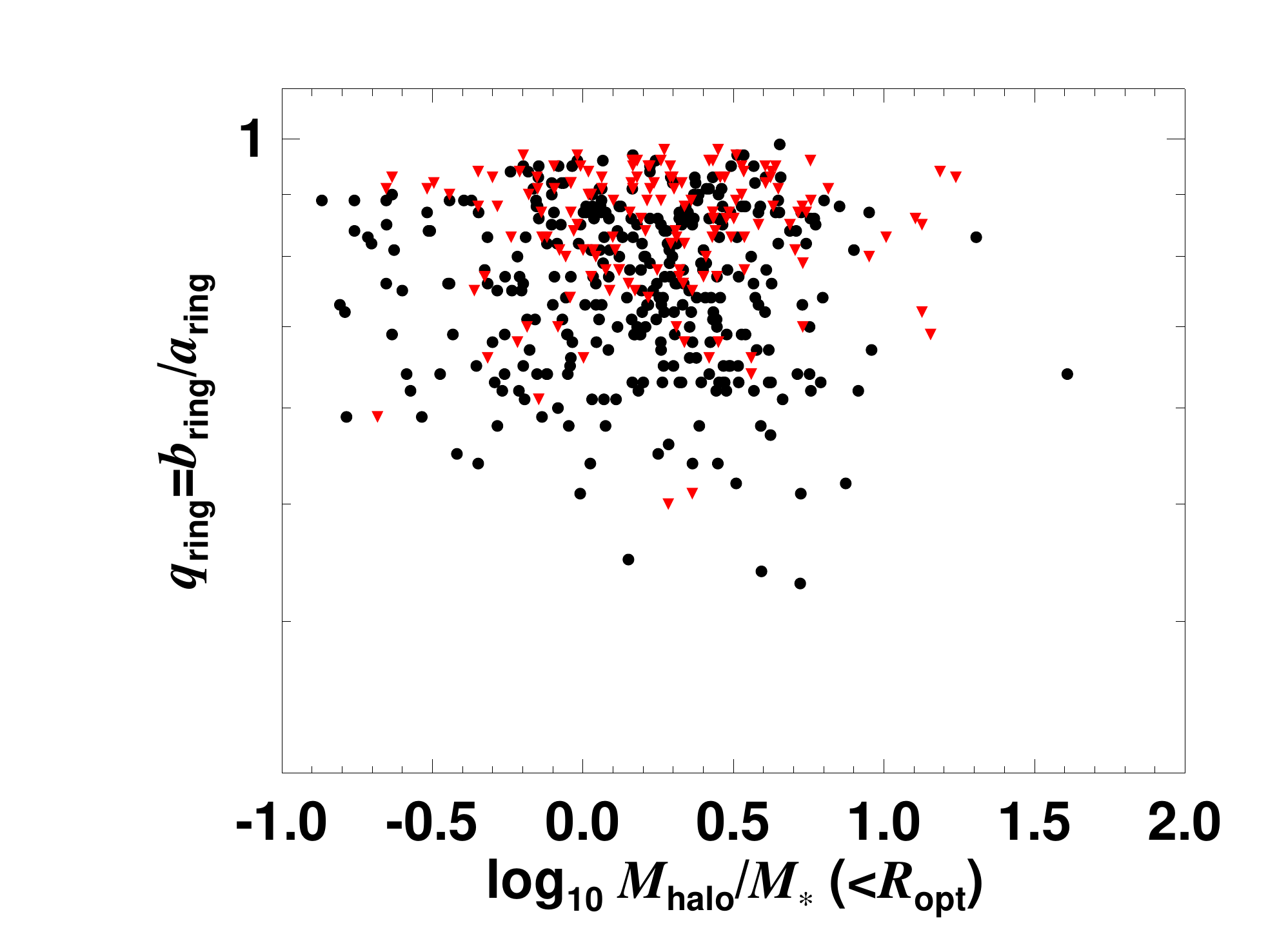}
\caption{
Ring sizes in kiloparsecs (upper panels), normalised to the disc size (middle panels), 
and ring intrinsic axial ratios (lower panels), 
as a function of the total stellar mass (left) 
and the halo-to-stellar mass ratio within the optical radius (see Eq.~\ref{halo-to-stellar-eq}) (right).
}
\label{inner_outer_mass}
\end{figure}

%
%
For a comparison of ring dimensions as a function of total stellar mass, 
the reader is referred to Fig.~12 in \citet[][]{2015A&A...582A..86H} and Fig.~\ref{inner_outer_mass} in this section. 
We also analyse ring dimensions versus $M_{\rm halo}/M_{\ast}(<R_{\rm opt})$ (right panels). 
As expected, ring sizes in physical size increase with increasing $M_{\ast}$ (upper left panel). 
Ring sizes normalised to disc size and axis ratios are not strongly controlled by the total stellar mass or 
the relative content of dark matter of the host galaxy. We encourage such results to be tested in simulations\footnote{For the 
normalisation of bar lengths by disc sizes in simulations, we note that $R_{25.5}$ roughly corresponds to the isophote tracing the stellar density 
$\sim 3.6$~$M_{\odot}$ pc$^{-2}$ \citep[][]{2015ApJS..219....3M}.}. 
%
%
%
%
\clearpage
%
\section{Frequency of outer rings as seen with SOM}\label{app_ml}
\begin{figure*}
\centering
\begin{tabular}{c c c c c c c c}
\includegraphics[width=0.24\textwidth]{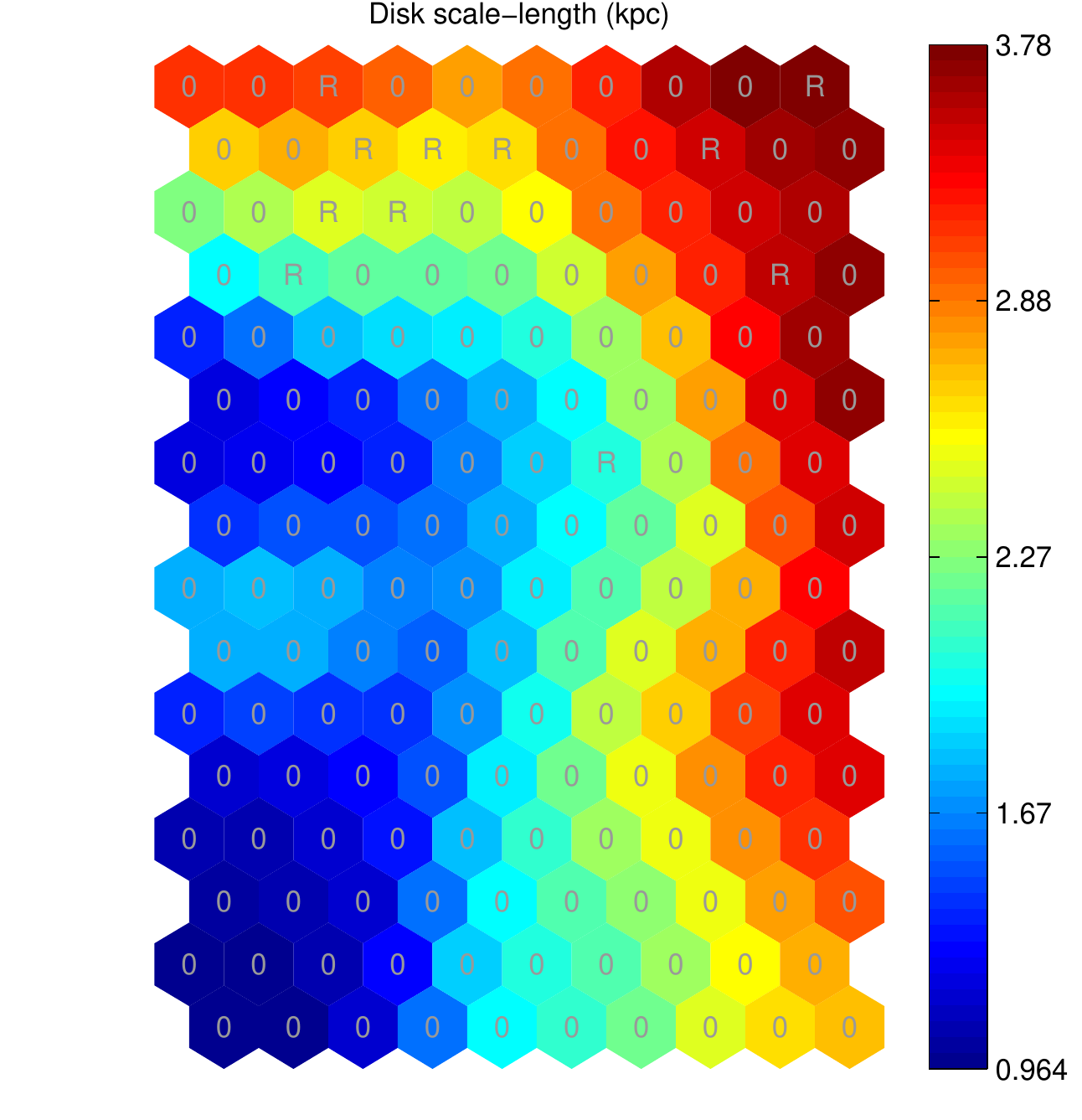}
\includegraphics[width=0.25\textwidth]{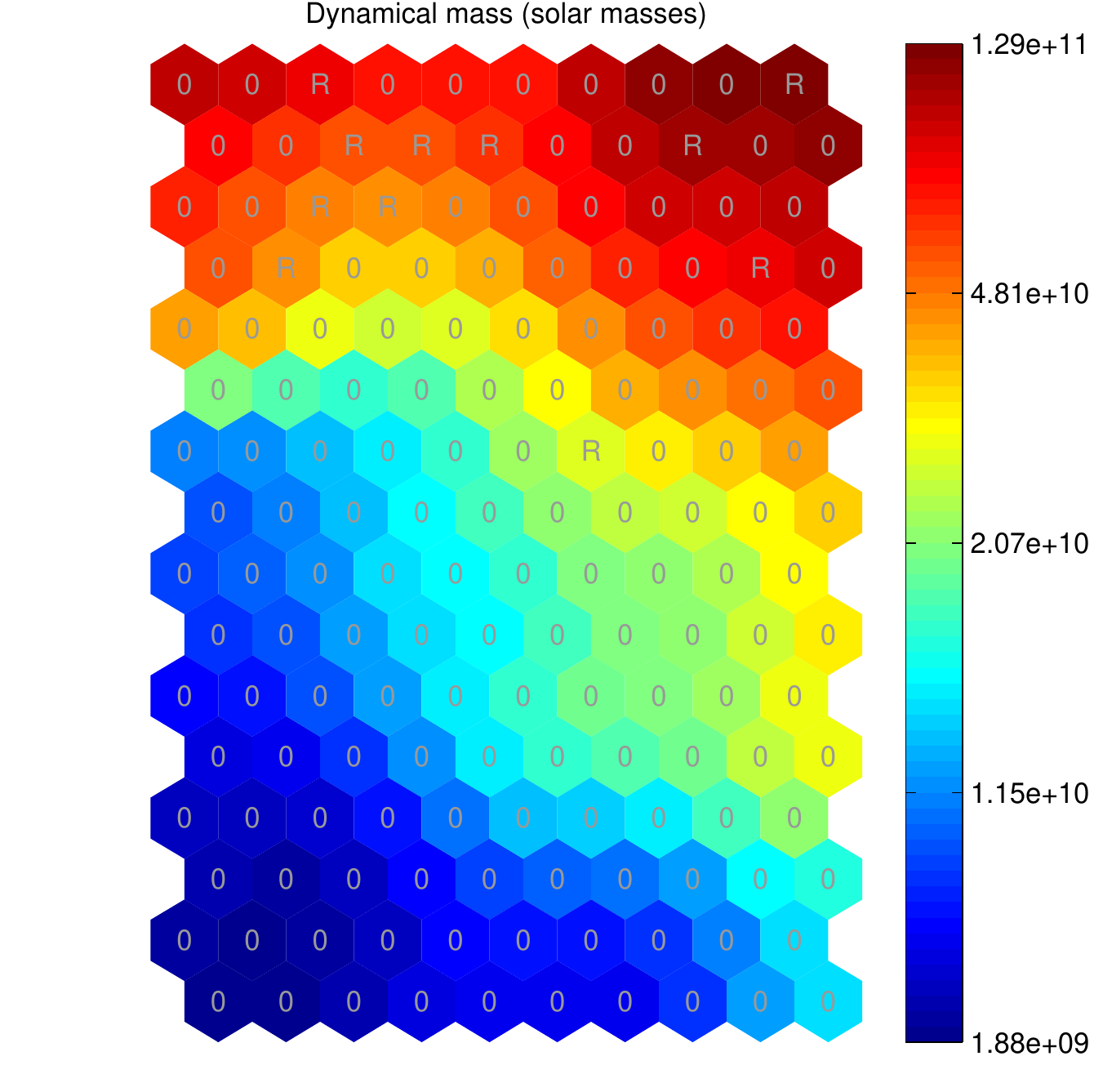}
\includegraphics[width=0.25\textwidth]{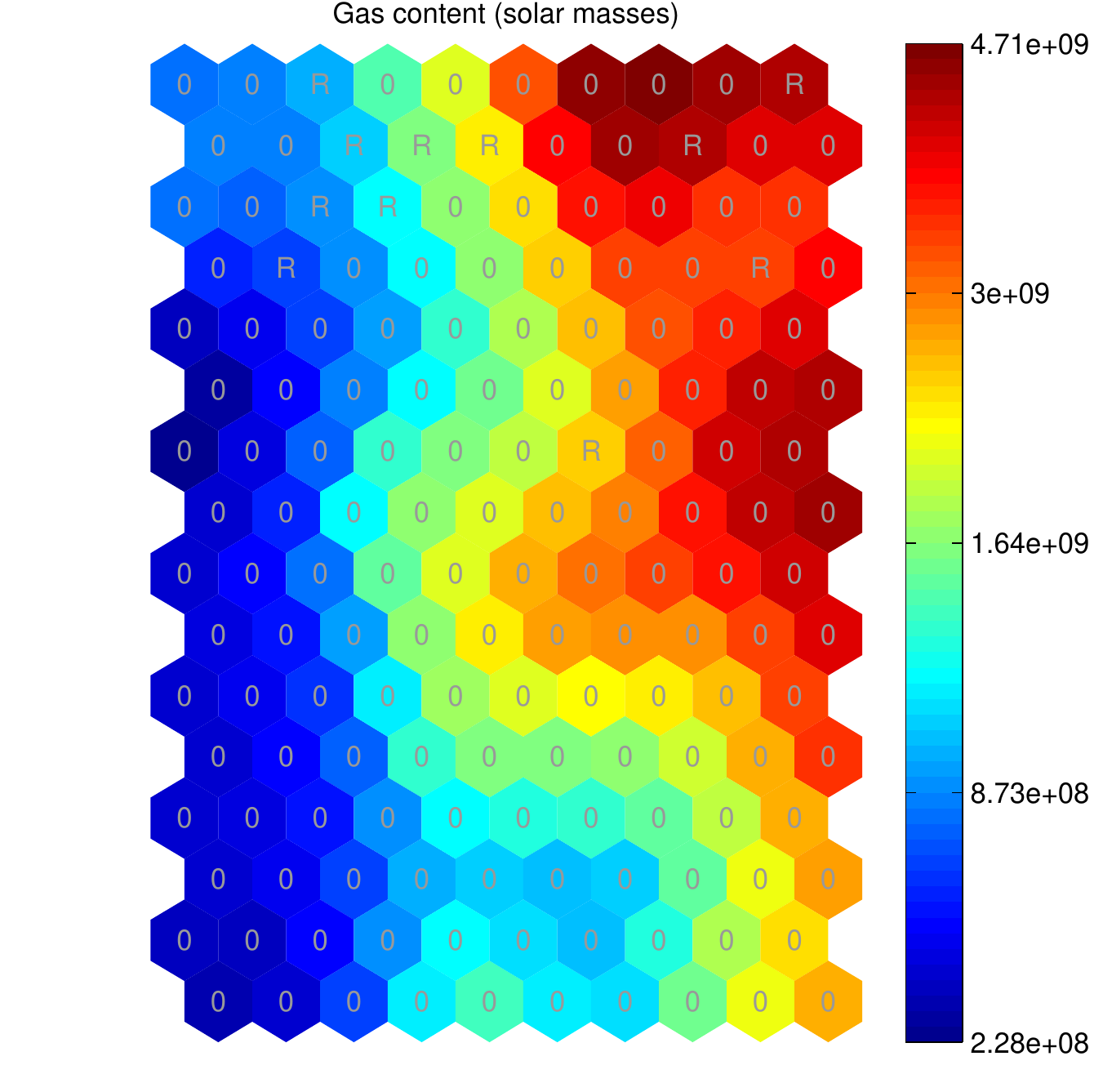}
\includegraphics[width=0.25\textwidth]{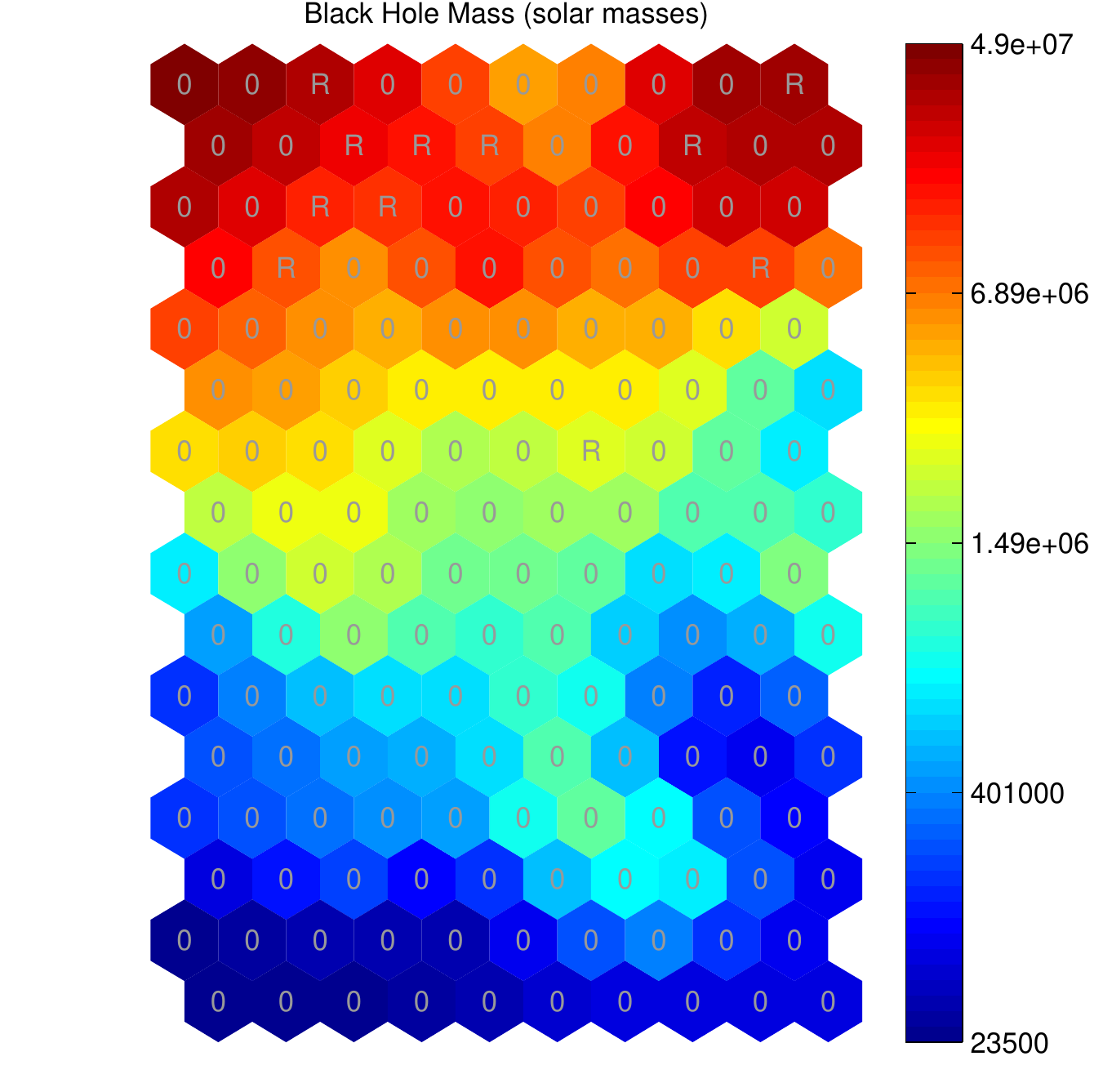}\\
\includegraphics[width=0.24\textwidth]{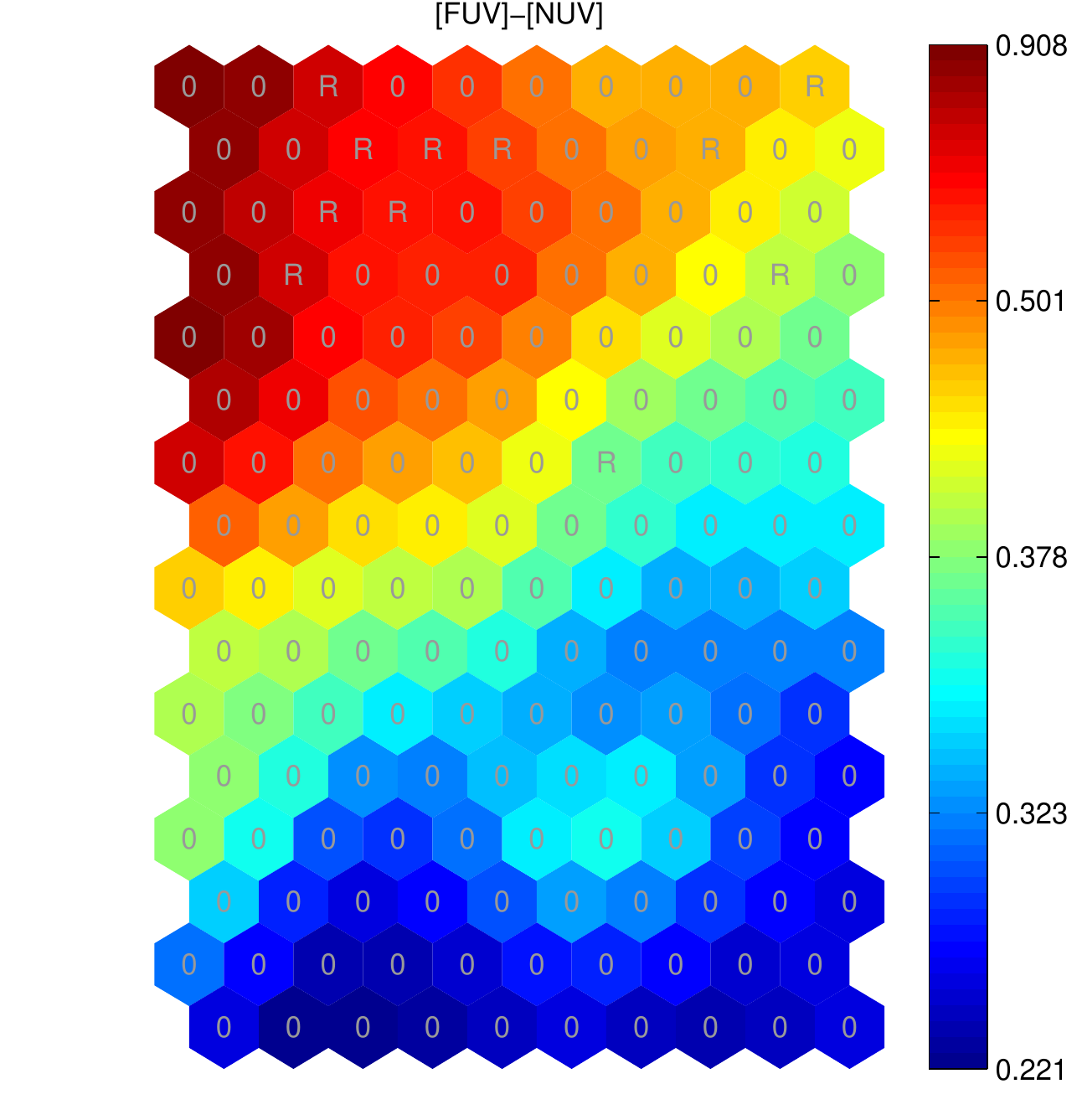}
\includegraphics[width=0.24\textwidth]{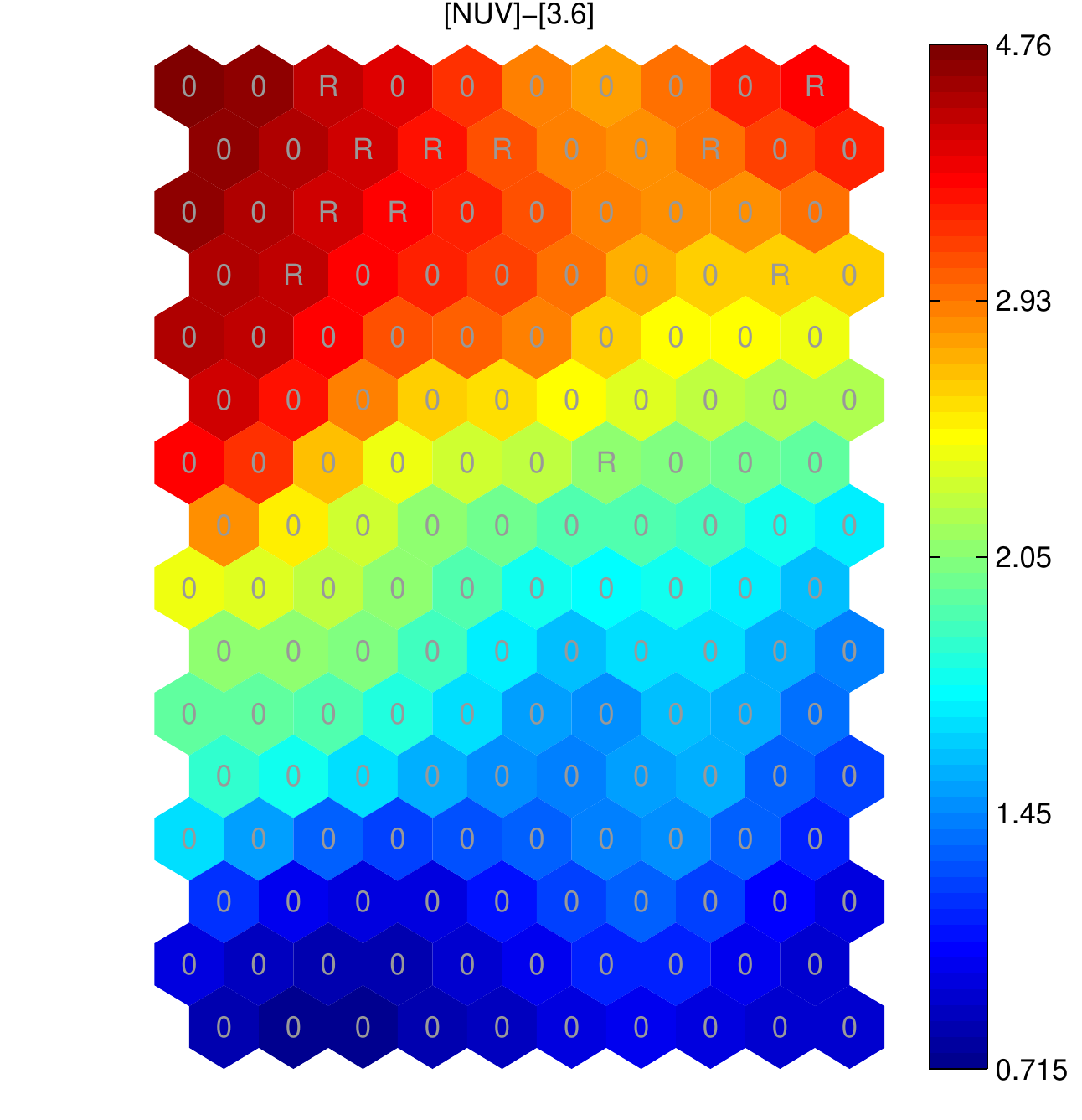}
\includegraphics[width=0.25\textwidth]{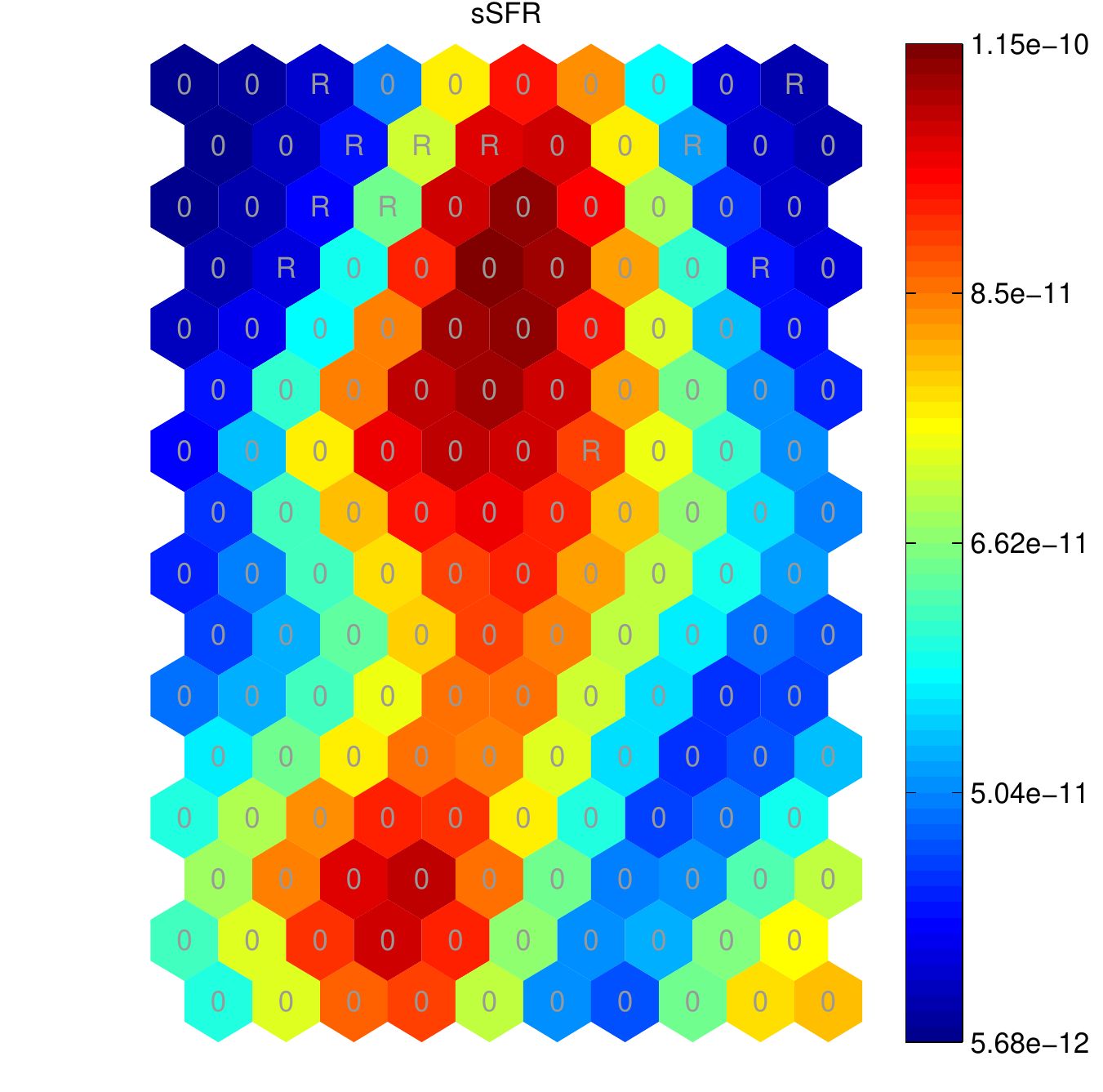}
\includegraphics[width=0.25\textwidth]{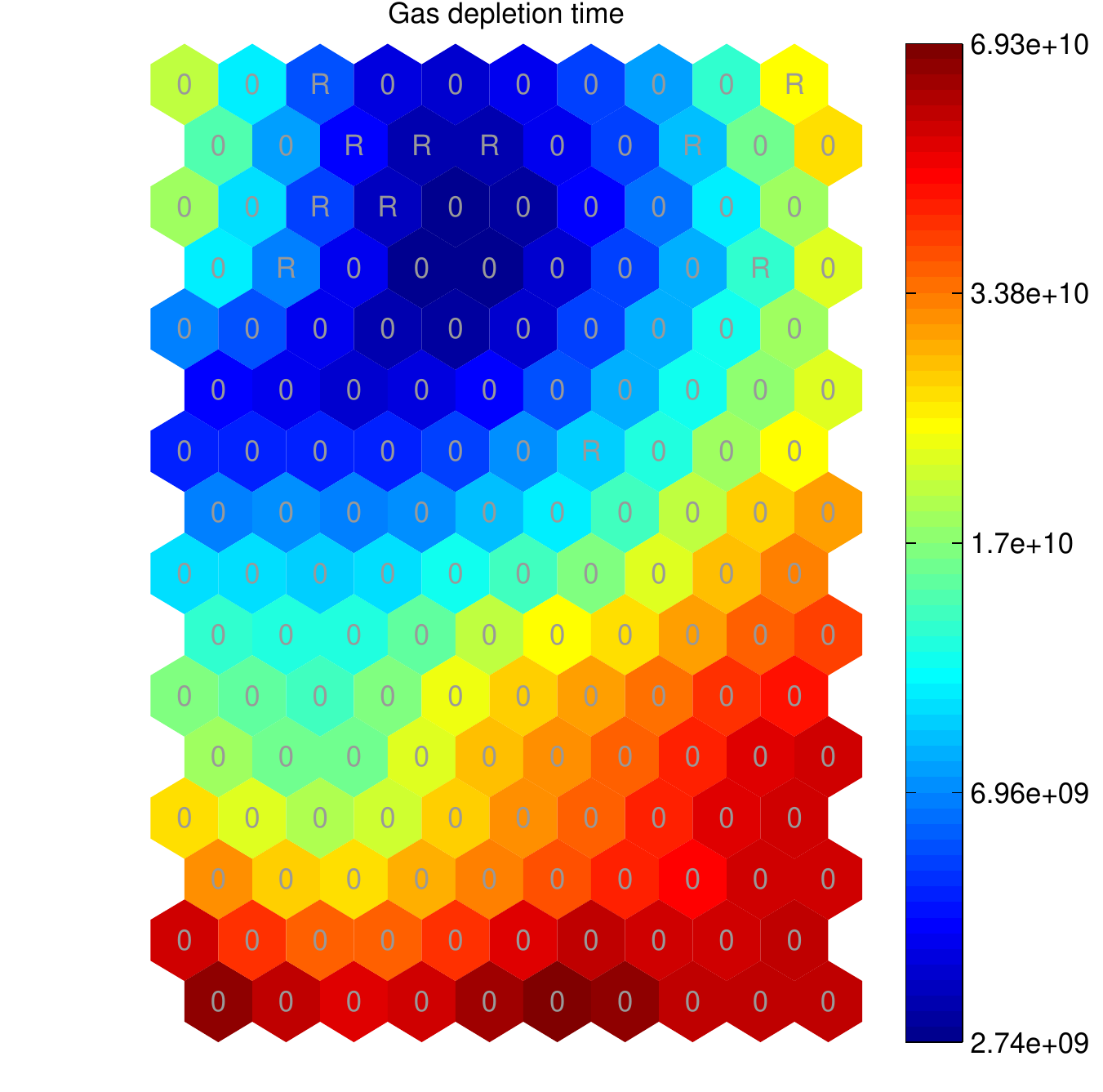}\\
\includegraphics[width=0.26\textwidth]{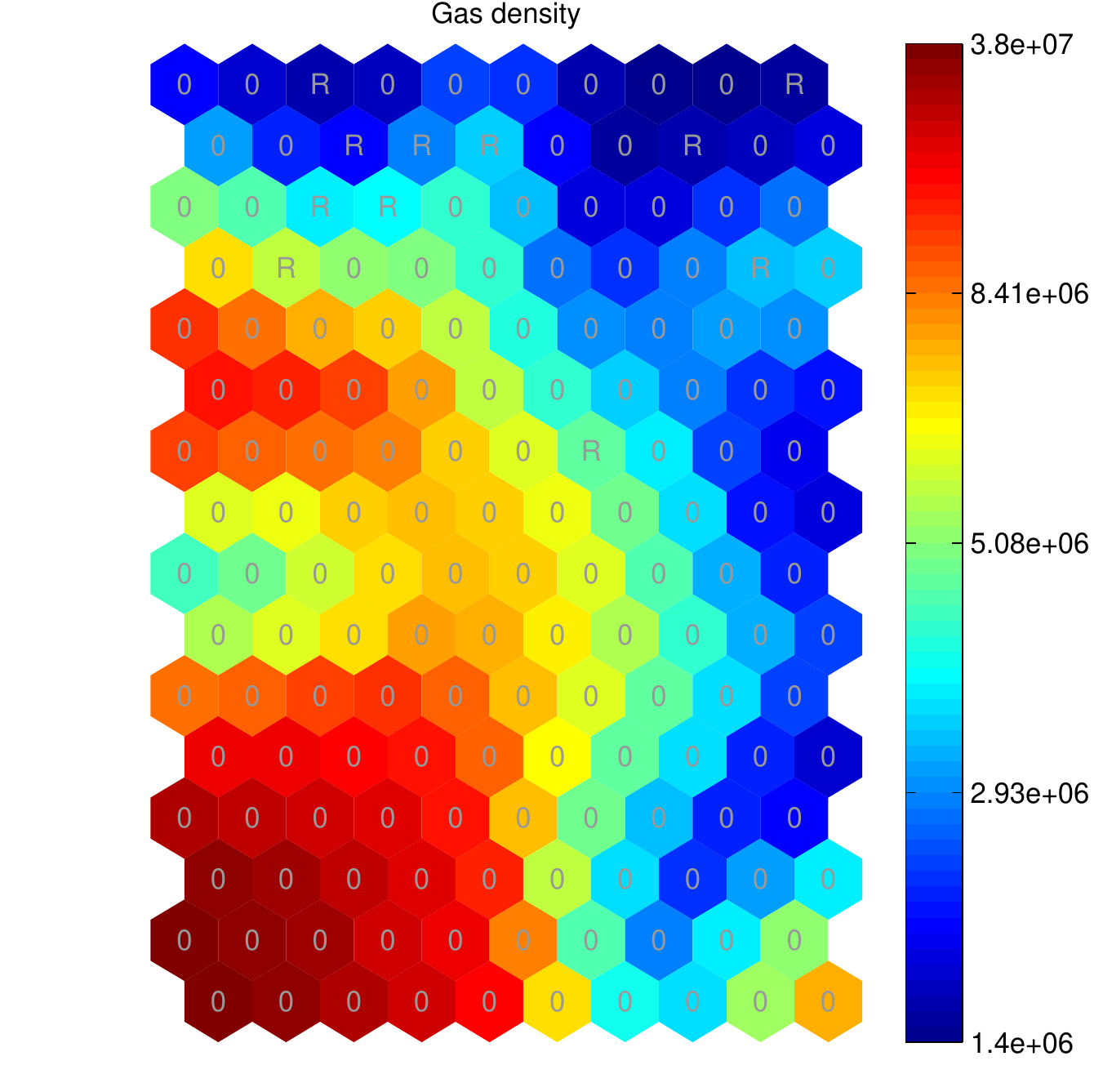}
\includegraphics[width=0.25\textwidth]{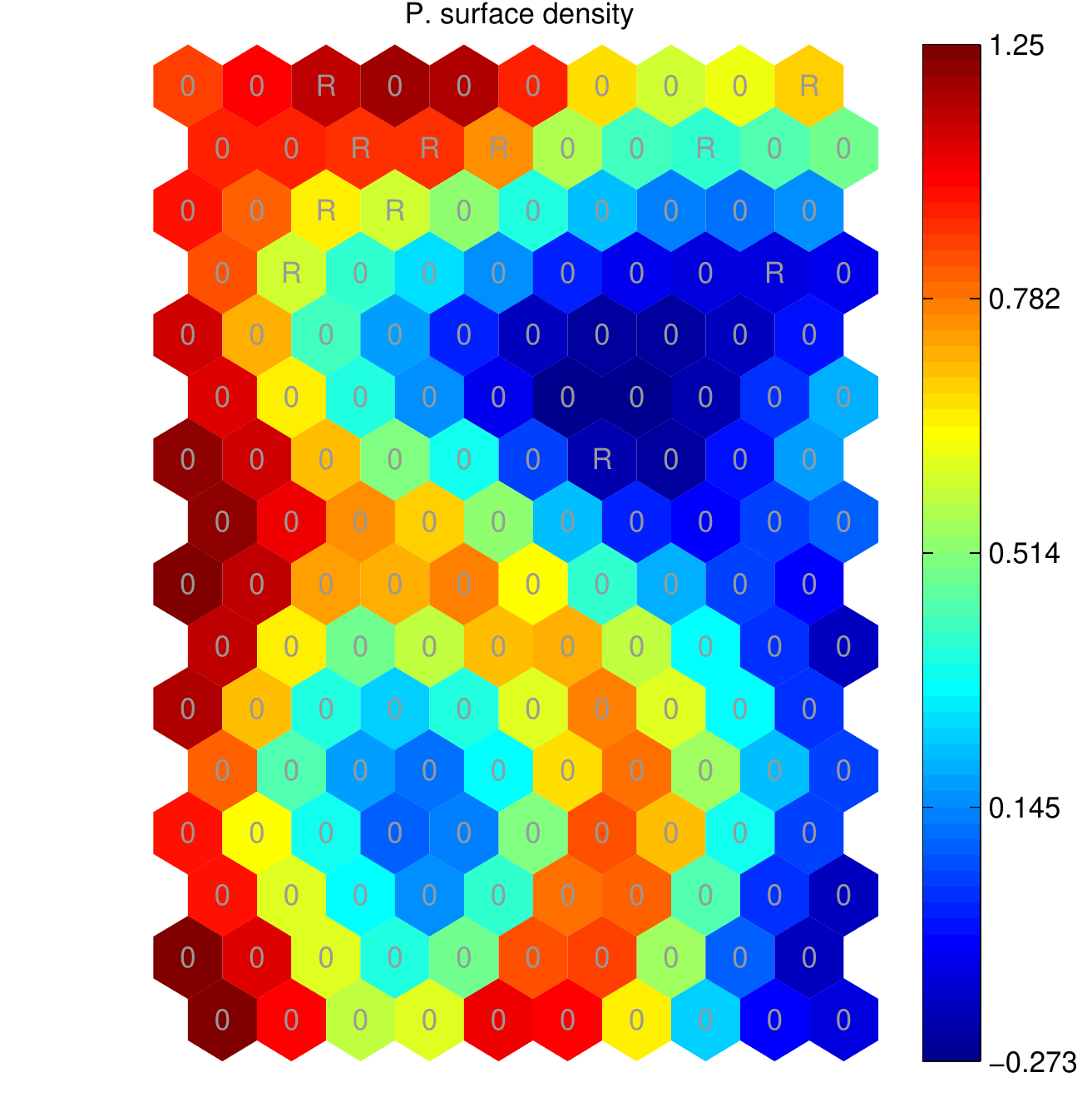}
\includegraphics[width=0.24\textwidth]{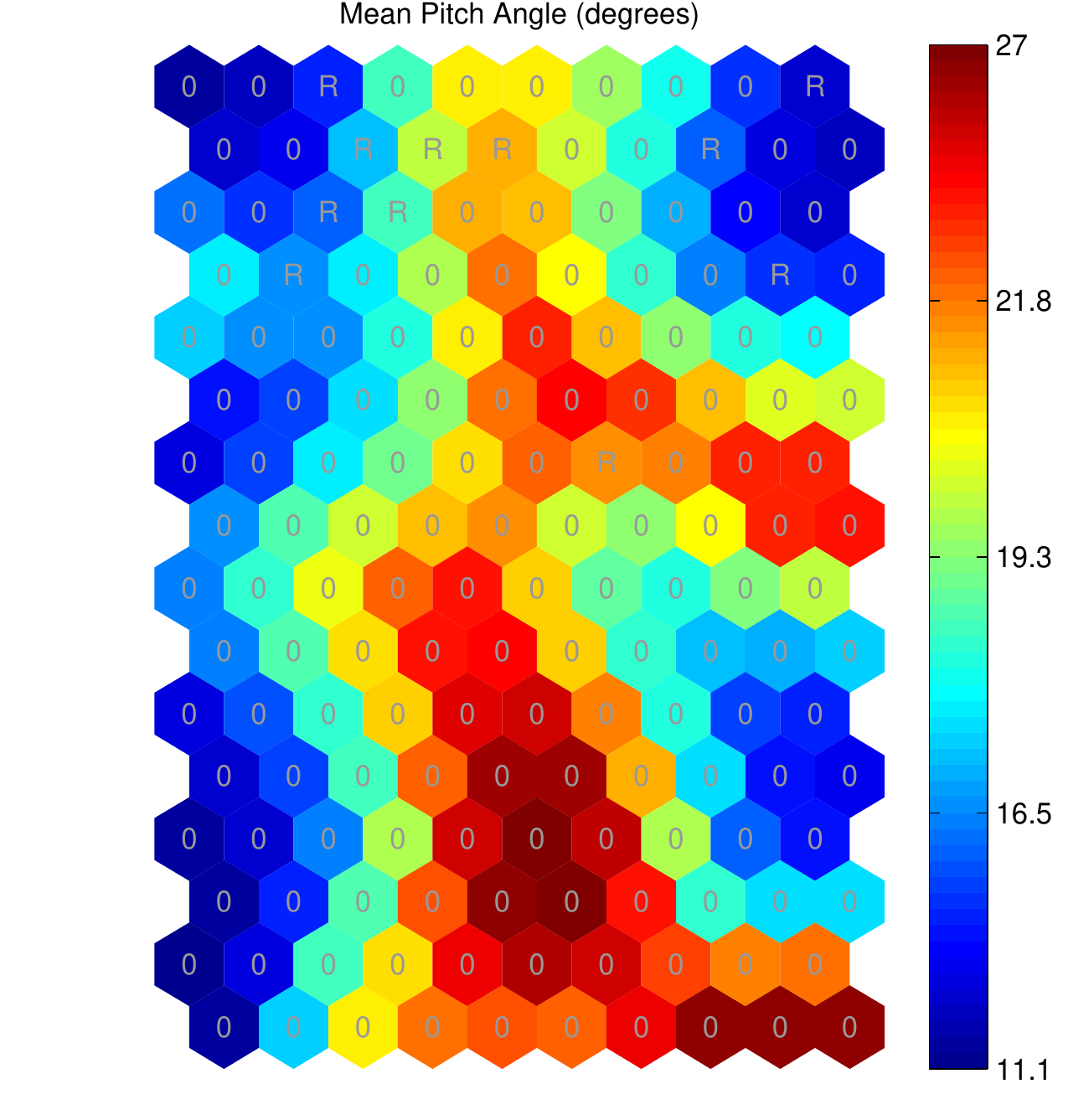}
\includegraphics[width=0.24\textwidth]{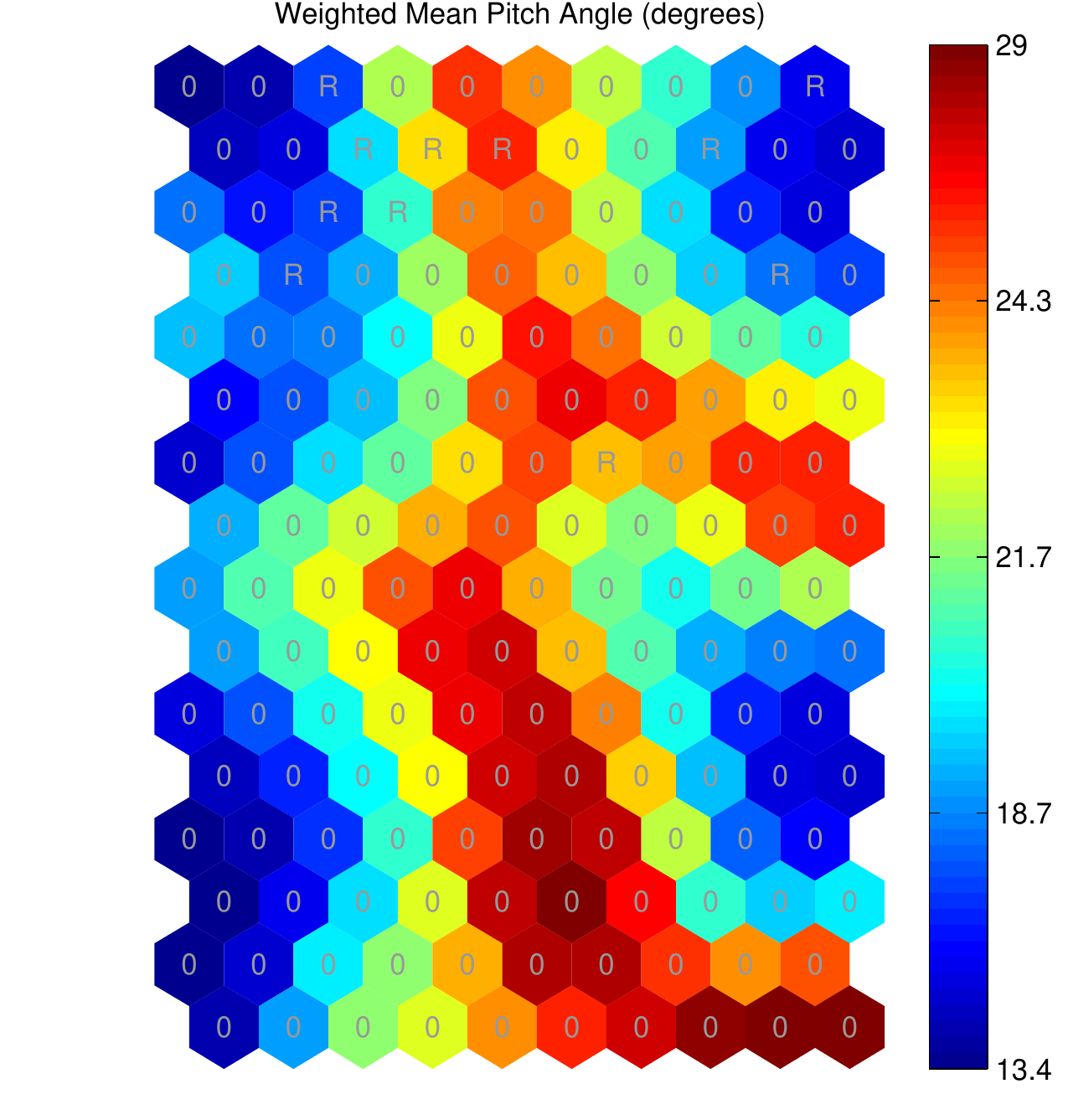}\\
\end{tabular}
\caption{
As Fig.~\ref{ml_properties_inner} but with the post-labelling done based on the presence ("R") or absence ("0") of outer rings. 
We show the remaining 12 components that were not displayed in Fig.~\ref{ml_properties_inner}.
}
\label{ml_properties_outer}
\end{figure*}

%
%
In Fig.~\ref{ml_properties_outer} we show the output of the SOM training on S$^4$G parameters, 
with the post-labelling done based on the presence ("R") or absence ("0") of outer rings 
(we show the remainder of the components that did not appear in Fig.~\ref{ml_properties_inner}; see Sect.~\ref{SOM_analysis} for further details). 
Only the uppermost neurons of the SOM, tracing the most massive or reddest galaxies, 
are typically best matching units for data of outer ringed galaxies.
%
%
\clearpage
%
%
\section{Post-labelling of SOM based on morphological types and on the presence of a bar}\label{ttype_ml}
\begin{figure}
\centering
\includegraphics[width=0.35\textwidth]{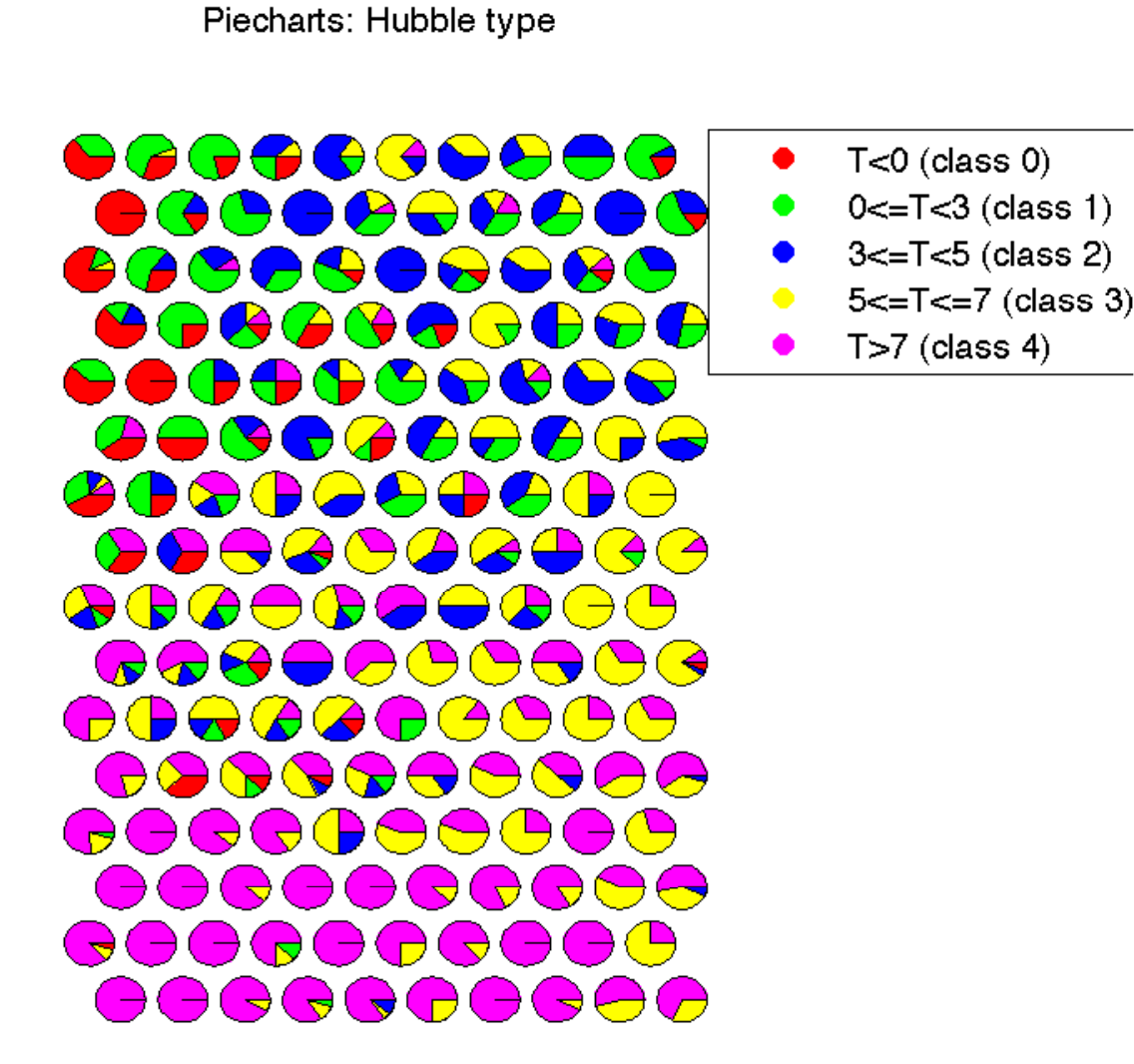}
\includegraphics[width=0.29\textwidth]{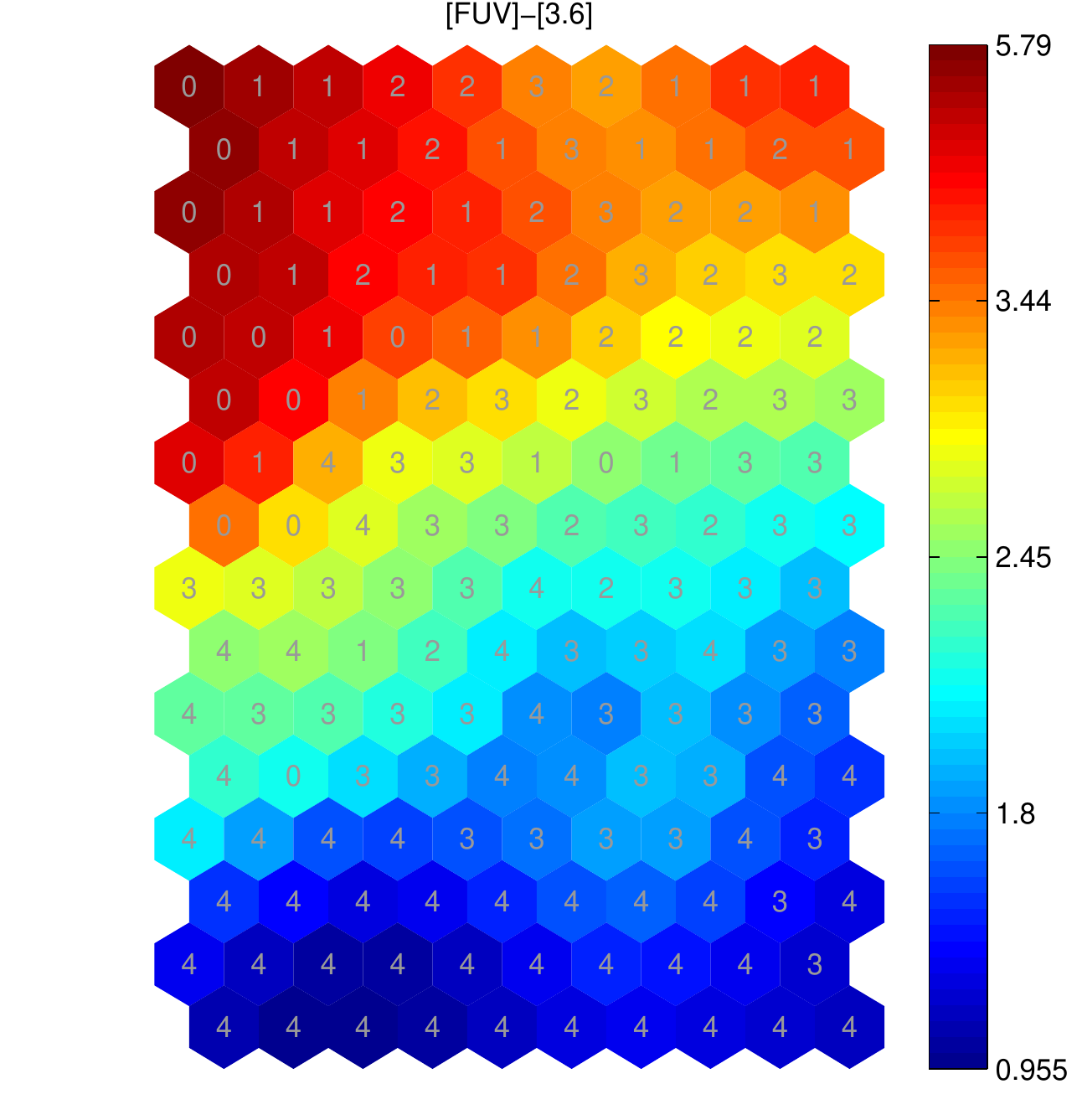}
\includegraphics[width=0.34\textwidth]{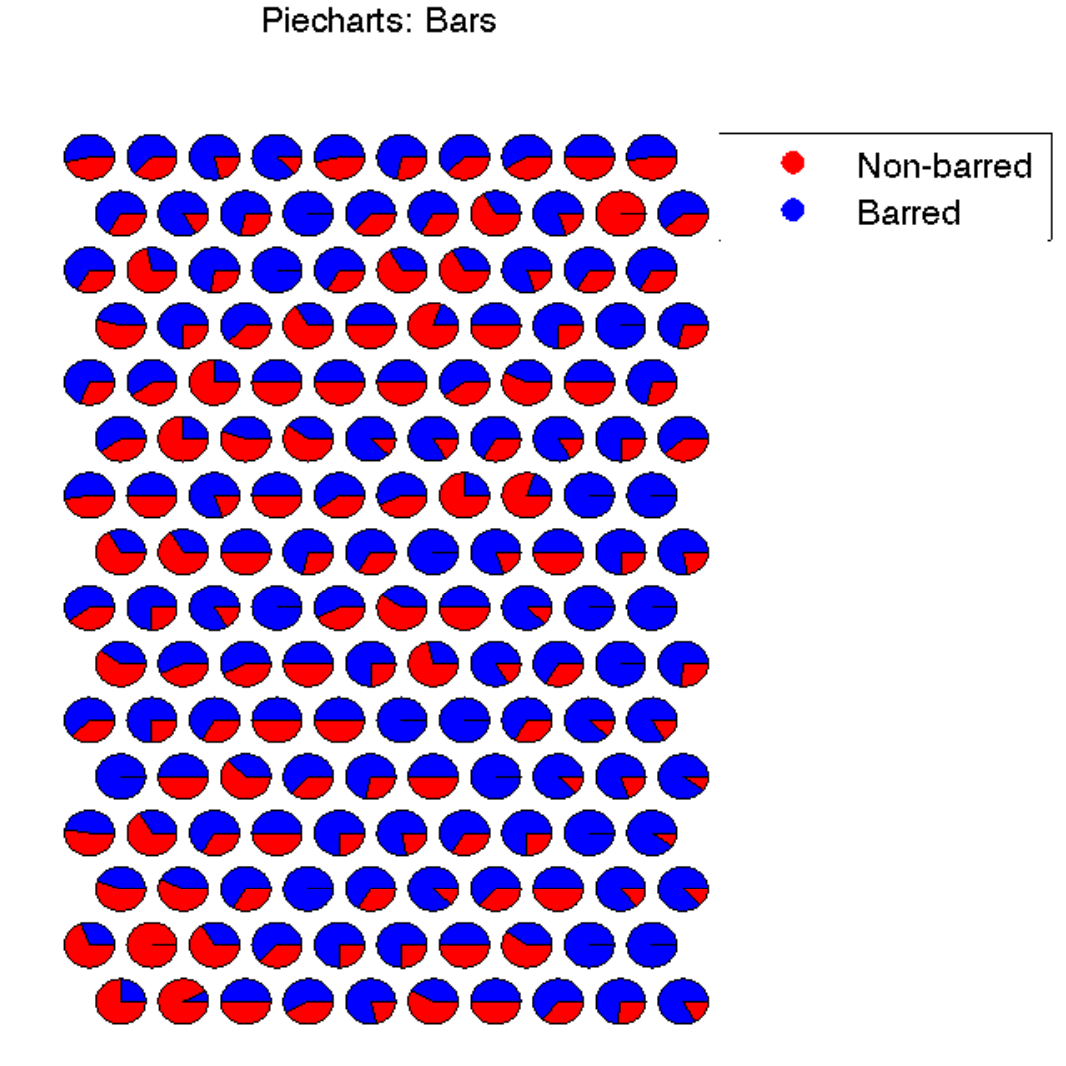}
\caption{
Left panel: Pie charts showing the hit histograms of the trained SOM (as in Fig.~\ref{pie_charts}) based on morphological types (see legend and text). 
Middle panel: [FUV]-[3.6] component plane of the trained SOM with the winning classes over-plotted (based on Hubble type). 
Right panel: Pie charts showing the hit histogram for barred (blue) and non-barred (red) galaxies. 
}
\label{morpho_type_SOM}
\end{figure}

%
%
In the left panel of Fig.~\ref{morpho_type_SOM} we show the output of the SOM training (same setup as in Fig.~\ref{ml_properties_inner}) 
with the post-labelling done based on the Hubble stages ($T$) determined by \citet[][]{2015ApJS..217...32B} 
(we note that $T$ was not used in the training). 
For a quick visualisation we also show the [FUV]-[3.6] component plane (middle panel). We consider the following families:
\begin{enumerate}
%
%
\item Class "0": Lenticular galaxies (S0s). $T < 0$.
\item Class "1": Early-type spirals. $0 \le T < 3$.
\item Class "2": Intermediate-type spirals. $3 \le T < 5$.
\item Class "3": Late-type spirals. $5 \le T \le 7$.
\item Class "4": Magellanic and irregulars galaxies. $T > 7$.
\end{enumerate}
%
%
Late-type spirals, Magellanic, and irregular galaxies (classes "3" and "4") dominate the lower part of the SOM 
(hitting blue colours and faint galaxies). Early-types galaxies ($T<5$) appear in the upper part, 
and S0s are labelled in the units with reddest colours (clustering in the upper left corner of the SOM). 
When the post-labelling is done based on the presence of bars (right panel in Fig.~\ref{morpho_type_SOM}), 
we find that barred and non-barred galaxies appear evenly distributed across the SOM.
%
%
%
\end{appendix}
%
%
\end{document}